\documentclass{article}
\usepackage{amssymb}\usepackage{amsmath,amsthm}
\usepackage{amsfonts}
\usepackage{appendix}
\begin{document}
\newcommand{\cc}{\mathbb{C}}
\newcommand{\rr}{\mathbb{R}}
\newcommand{\hh}{\mathbb{H}}
\newcommand{\te}{{\mathbb{T}}}
\newcommand{\tes}{{{\mathbb{T}}^L}}
\newcommand{\tesR}{{{\mathbb{T}}^R}}
\newcommand{\oo}{\mathbb{O}}
\newcommand{\BB}{\mathbb{B}}
\newcommand{\ip}{{\,{\rm i}\,}}
\newcommand{\ipp}{{\,{\rm J}\,}}
\newcommand{\im}{-{\rm i}}
\newcommand{\w}{{\rm w}}
\newcommand{\z}{\mathbb{Z}}
\newcommand{\op}{[\![}
\newcommand{\cl}{]\!]}
\newcommand{\nn}{\nonumber}
\newcommand{\cdote}{\!\cdot\! }
	
\newcommand{\bee}[1]{\renewcommand{\baselinestretch}{1}\begin{equation}\label{#1}}
\newcommand{\ee}{\end{equation}\renewcommand{\baselinestretch}{2}}
	
\newcommand{\ba}{\left[\begin{array}}
\newcommand{\ea}{\end{array}\right]}
\newcommand{\bac}{\begin{array}}
\newcommand{\eac}{\end{array}}
\newcommand{\eacb}{\end{array}\!\!\!\!\!\!}
\newcommand{\bc}{\begin{center}}
\newcommand{\ec}{\end{center}}
\newcommand{\beaa}{\renewcommand{\baselinestretch}{1}\begin{eqnarray}}
\newcommand{\eeaa}{\end{eqnarray}\renewcommand{\baselinestretch}{2}}
\newcommand{\Ra}{\Rightarrow}
\newcommand{\ra}{\rightarrow}
\newcommand{\sg}{\sigma}
\newcommand{\ev}{\equiv}
\newcommand{\mas}{\dagger}
\newcommand{\del}{\delta}
\newcommand{\numberthis}{\addtocounter{equation}{1}\tag{\theequation}}
\newcommand{\half}{\mbox{$\frac{1}{2}$}}

\pagestyle{myheadings} \markright{\hfill L. A. Wills-Toro: Mixing Higgs and Gauge Fields II\ \ \ \ \ \ \ }

\begin{titlepage}
\title{{\bf Poincar\'e--like extension Mixing Higgs and Gauge Fields in a U(1) symmetric model}}
	
\author{Luis Alberto {\bf Wills-Toro}$^{1,2,3}$\\
~\\
1. School of Mathematics, Universidad Nacional de Colombia,\\ Medell\'{\i}n,
Colombia, e--mail: lawillst@unal.edu.co\\
2. Departamento de F\'{\i}sica Te\'orica y del Cosmos, Universidad de Granada\\
3. CIMAT}
\date{}
	
\maketitle
	
\thispagestyle{empty}
	
\begin{abstract}
We continue the program by investigating symmetric structures underlying features of the Standard Model. We then expand the symmetry to encompass translations before contraction. A field theory model emerges with the goal of replicating a coupling to gravity before contraction. Then we obtain an expanded second-order gravity model after contraction that incorporates the abelian internal symmetry.
~
{\bf  }
\end{abstract}
\end{titlepage}

 \pagenumbering{arabic}
\section{Introduction}
In our previous work \cite{U1I}, we initiated a program to delve into the symmetries that underlie certain ad-hoc structures within the Standard Model (SM). For simplicity, we examined a model operating in 4D space-time with just U(1) internal symmetry, where $g_{\mu\nu}={\rm diag}(1,-1,-1,-1)$ and $\mu,\nu\in \{0,1,2,3\}$. We explored a symmetry capable of inducing a mixing effect among the Higgs field $H$, its conjugate $\bar{H}$, and the gauge field (the $U(1)$ connection) represented by $B_{\mu}$. Since all these fields have a naive energy dimension of 1, we introduced generators $T_{\mu}$ and $\hat{T}_{\mu}$ with  dimension 0, satisfying the relation:
\bee{THbar}
 \,[T_{\mu},B_{\nu}]=c'g_{\mu\nu}{H},\ \ \ [T_{\mu},\bar{H}]=b'B_{\mu}+\cdots,
\ee
where $b',c'$ are constants.
In \cite{U1I}, a dimensionless subalgebra was derived based on the following Ansatz (with constants $a$, $u$, $q'$, and $h$):
\bee{TTbar}
\,[T_{\mu},\hat{T}_{\,\,\nu}]=\ip aM_{\mu\nu}+u\mathcal{T}_{\mu\nu}-q' g_{\mu\nu} D+ (h/2) g_{\mu\nu}{T_{[o]}},
\ee
This expression combines the internal generator ${T_{[o]}}$ of the U(1) symmetry with the external (space-time) symmetries of the Lorentz subalgebra, featuring generators $M_{\mu\nu}$, a real symmetric tensor generator $\mathcal{T}_{\mu\nu}$ (whose role is not yet specified), and a scalar generator $D$ (possibly related to dilatations?). We now aim to extend this algebra to five dimensions (5D), leaving $g_{44}$ unspecified for now, and later consider contraction. This extension not only involves the Poincaré algebra explicitly but also provides a concrete method for obtaining invariant multiplets of generators and fields (connections).

It can be verified that the same algebra obtained in \cite{U1I} (equations (5)-(15) therein) closes in 5D without altering the arrays $\sigma(\mu\nu)_{\rho}^{\ \sigma}:=g_{\mu\rho}\delta_{\nu}^{\,\,\sigma}+g_{\nu\rho}\delta_{\mu}^{\,\,\sigma}-(1/2)g_{\mu\nu}\delta_{\rho}^{\,\,\sigma}$. Consequently, we can subsequently impose traceless conditions and separate degrees of freedom in 4 space-time dimensions. Extrapolating the contraction construction, for $\mu, \nu\in \{0,1,2,3\}$, we define a parameter $R$ with a naive energy dimension of -1 associated with a potential contraction:
\beaa
 \tilde{P}_{\mu}:=&-M_{4\mu}/R,& \tilde{Q}_{\mu}\ :=\ \ \mathcal{T}_{4\mu}/R\\
\hat{P}_{\mu}:=&(\sqrt{2})^{-1} (\tilde{P}_{\mu}+\ip \tilde{Q}_{\mu}),& \hat{P}_{\mu}^*\ :=\ \ (\sqrt{2})^{-1} (\tilde{P}_{\mu}-\ip \tilde{Q}_{\mu}).\\
\Phi:=&\hat{T}_4/R,&\  \bar{\Phi}\ :=\ \  {T}_4/R.
\eeaa
The combination of the Hermitian generators $\tilde{P}_{\mu}, \tilde{Q}_{\mu}$ into an adjoint pair $\hat{P}_{\mu}, \hat{P}_{\mu}^*$ will force that the associated transformation parameters are related by complex conjugation.
We must address the mismatch of traceless conditions through the following adjustments:
\bee{Traceless}
  \,\mathcal{T}:=\half g^{\mu\nu}\mathcal{T}_{\mu\nu},\ \  T_{\mu\nu}:=\mathcal{T}_{\mu\nu}-\half g_{\mu\nu}\mathcal{T},\ \ \ 
  S:=-\frac{2}{g_{44}}\mathcal{T}_{44}.
  \ee
The newly defined symmetric tensor $T_{\mu\nu}$ is traceless by construction. With this redefinition, it is observed that the scalars ${T_{[o]}}$ and $S-4\mathcal{T}$, as well as the scalars ${\hat{T}_{[o]}}\!-\!{T_{[o]}}$ and $5\mathcal{T}$, share identical nontrivial commutation relations with the remaining algebra generators. Hence, their difference serves as a central charge for the entire algebra. We can either identify generators with identical commutation relations or define a central charge $\mathcal{D}$ based on their difference.
We opt for the identification. $\mathcal{T}$ functions as a central charge for the Lorentz-like subalgebra. We thus adopt:
  \bee{Fix}
  {\hat{T}_{[o]}}:=(T_{[o]}+5\mathcal{T})=(5S\!-{T_{[o]}})/4,\ \ \ \mathcal{T}=(S-{T_{[o]}})/4=({\hat{T}_{[o]}}\!-\!{T_{[o]}})/5,\ \ \ 
  \ee
since these choices result in a sum of squares Casimir operator for the resulting algebra, without requiring the limit $R\to \infty$ inherent in the standard contraction procedure. With these Ansätze, the extension of the Lorentz algebra yields the following non-trivial relations:
   \beaa
  \,[T_{\mu},\hat{T}_{\,\,\nu}]&=&\epsilon(\ip M_{\nu\mu}+T_{\nu\mu}+(1/10) g_{\nu\mu} (4 {T_{[o]}}+{\hat{T}_{[o]}}))+2\epsilon g_{\nu\mu} T_{[o]},\label{Dim0Start}\\
  \,[{T_{[o]}},T_{\rho}]&=&T_{\rho},\ \ \ \ \ \ \ \ \ \  \,[{T_{[o]}},\hat{T}_{\rho}] =\ -\hat{T}_{\rho},\label{Eqn9}\\
  \,[{\hat{T}_{[o]}},T_{\rho}]&=&T_{\rho},\ \ \ \ \ \ \ \ \ \ \ \ \  \,[{\hat{T}_{[o]}},\hat{T}_{\rho}]=\ -\hat{T}_{\rho},\\
  \,[T_{\mu\nu},T_{\rho}]&=&\sigma(\mu\nu)_{\rho}^{\ \tau}T_{\tau},\ \  \,[T_{\mu\nu},\hat{T}_{\rho}] =\ -\sigma(\mu\nu)_{\rho}^{\ \tau}\hat{T}_{\tau},\\
  \,[M_{\mu\nu},T_{\rho}]&=&-\ip\sigma[\mu\nu]_{\rho}^{\ \tau}T_{\tau},\  \,[M_{\mu\nu},\hat{T}_{\rho}] =\ -\ip\sigma[\mu\nu]_{\rho}^{\ \tau}\hat{T}_{\tau},\label{Eqn12}\\
  \,[M_{\mu\nu},T_{\rho\sigma}]&=&-\ip(\sigma[\mu\nu]_{\rho}^{\,\,\tau}\delta_{\sigma}^{\,\,\kappa}+\sigma[\mu\nu]_{\sigma}^{\,\,\kappa}\delta_{\rho}^{\tau})T_{\tau\kappa},\label{MinExStart}\\
  \,[M_{\mu\nu},M_{\rho\sigma}]&=&-\ip(\sigma[\mu\nu]_{\rho}^{\,\,\tau}\delta_{\sigma}^{\,\,\kappa}+\sigma[\mu\nu]_{\sigma}^{\,\,\kappa}\delta_{\rho}^{\tau})M_{\tau\kappa},\\
  \,[T_{\mu\nu},T_{\rho\sigma}]&=&-\ip(\sigma(\mu\nu)_{\rho}^{\,\,\tau}\delta_{\sigma}^{\,\,\kappa}-\sigma(\mu\nu)_{\sigma}^{\,\,\kappa}\delta_{\rho}^{\tau})M_{\tau\kappa},\ \  \rm{with} \label{TTM}\\
  \sigma[\mu\nu]_{\rho}^{\ \tau}&:=&g_{\mu\rho}\delta_{\nu}^{\,\,\tau}-g_{\nu\rho}\delta_{\mu}^{\,\,\tau},\\
  \sigma(\mu\nu)_{\rho}^{\ \tau}&:=&g_{\mu\rho}\delta_{\nu}^{\,\,\tau}+g_{\nu\rho}\delta_{\mu}^{\,\,\tau}-(1/2)g_{\mu\nu}\delta_{\rho}^{\,\,\tau},\label{MinExEnd}\\
 {\hat{T}_{[o]}}&:=&(5S\!-{T_{[o]}})/4,\ \  \mathcal{T}=\!(S-{T_{[o]}})/4=\!({\hat{T}_{[o]}}-{T_{[o]}})/5,\ \ \label{Dim0End} 
  \eeaa
where $\epsilon\in\{+1,-1\}$, and $\mu,\nu,\rho,\sigma,\tau,\kappa\in \{0,1,2,3\}$. This elucidates the origin of scalar generators in the coinciding 4D Lorentz extension obtained in \cite{U1I}. The combination $\mathcal{T}=(S-{T_{[o]}})/4=({\hat{T}_{[o]}}-{T_{[o]}})/5$, which serves as a central charge for the subalgebra (\ref{Dim0Start})--(\ref{MinExEnd}), is not central for the full extension. The further relations of the closed extension of the Poincaré algebra are:
 \beaa
 \,[{T_{[o]}},\hat{P}_{\rho}]&=&0,\hspace{2cm}[{T_{[o]}},\hat{P}^*_{\rho}] =0,\\
 \,[{\hat{T}_{[o]}},\hat{P}_{\rho}]&=&5 \hat{P}_{\rho},\hspace{1.6cm}[{\hat{T}_{[o]}},\hat{P}^*_{\rho}] =-5 \hat{P}^*_{\rho},\\
  \,[{T_{[o]}},\Phi]&=&-\Phi,\hspace{1.8cm}[{T_{[o]}},\bar{\Phi}] =+\bar{\Phi},\label{ToPhi}\\
   \,[{\hat{T}_{[o]}},\Phi]&=&+4\Phi,\hspace{1.6cm} [{\hat{T}_{[o]}},\bar{\Phi}] =-4\bar{\Phi},\label{TohatPhi}\\
    \,[M_{\rho\sigma},\hat{P}_{\tau}]&=&-\ip \sigma[\rho\sigma]_{\tau}^{\,\,\kappa}\hat{P}_{\kappa},\  [M_{\rho\sigma},\hat{P}^*_{\rho}] =-\ip \sigma[\rho\sigma]_{\tau}^{\,\,\kappa}\hat{P}^*_{\kappa},\\
      \,[T_{\rho\sigma},\hat{P}_{\tau}]&=& \sigma(\rho\sigma)_{\tau}^{\,\,\kappa}\hat{P}_{\kappa},\hspace{.5cm} \,[T_{\rho\sigma},\hat{P}^*_{\rho}] =- \sigma(\rho\sigma)_{\tau}^{\,\,\kappa}\hat{P}^*_{\kappa},
        \eeaa
   \beaa
     [\hat{T}_{\rho},\hat{P}_{\tau}]&=&\sqrt{2}\ip g_{\rho\tau}{\Phi},   ,\hspace{.7cm}\,[T_{\rho},\hat{P}^*_{\tau}]=\sqrt{2}\ip g_{\rho\tau}\bar{\Phi},\label{Eqn25}\\
         \,[T_{\rho},\hat{P}_{\tau}]&=&0,\hspace{2.2cm} [\hat{T}_{\rho},\hat{P}^*_{\tau}]=0,\\
       \,[T_{\rho},{\Phi}]&=&-\sqrt{2}\ip\epsilon \hat{P}_{\rho},\hspace{1.1cm} [\hat{T}_{\rho},\bar{\Phi}]=-\sqrt{2}\ip \epsilon \hat{P}^*_{\rho},\\
     \,[\hat{T}_{\rho},\Phi]&=&0,\hspace{2.4cm} [{T}_{\rho},\bar{\Phi}]=0,\label{Eqn28}\\
      \,[\Phi,\Phi]&=&0,\hspace{2.5cm}[\bar{\Phi},\bar{\Phi}]=0,\\
    \,\![\hat{P}_{\rho},\hat{P}_{\sigma}]&=&0,\hspace{2.2cm}[\hat{P}^*_{\rho},\hat{P}^*_{\sigma}]=0,\\ 
   \,[\Phi,\hat{P}_{\rho}]&=&0,
   \hspace{2.4cm}[\bar{\Phi},\hat{P}^*_{\rho}]=0, \\
    \,[\Phi,\hat{P}^*_{\rho}]&=&-\frac{\sqrt{2}\ip  g_{44}}{ R^2}\hat{T}_{\rho},
   \hspace{.6cm}[\bar{\Phi},\hat{P}_{\rho}]=-\frac{\sqrt{2}\ip g_{44}}{ R^2}T_{\rho},\label{PhiPT}
     \eeaa
     \beaa   
        \,[\Phi,\bar{\Phi}]&=&-
       \frac{ g_{44}}{ R^2} \epsilon\left(\left(\frac{2}{5}( {T_{[o]}}-{\hat{T}_{[o]}})\right)+2{T_{[o]}}\right),\label{PhiPhiTs}\\
        \,[\hat{P}_{\rho},\hat{P}^*_{\sigma}]&=&\!\!\!\frac{ g_{44}}{ R^2}\!\left(\!\left(\ip M_{\sigma\rho}\!+\!T_{\sigma\rho}\!+\!\frac{1}{10} g_{\sigma\rho} (4 {T_{[o]}}\!+\!{\hat{T}_{[o]}})\right)\right.\nn\\
       &&\left. -g_{\sigma\rho}\!\left(\!\frac{2}{5}({T}_{[o]}\!-\! {\hat{T}_{[o]}})\!\right)\right),\label{PPMTs}
 \eeaa

The dimension 0 subalgebra (\ref{Dim0Start})--(\ref{Dim0End}) constitutes the Lorentz-like Extension (LE), unaffected by the contraction procedure. Restricting the generators to $T_{\mu}$, $\bar{T}_{\mu}$, $M_{\mu\nu}$, $T_{\mu\nu}$, and $(24 {T_{[o]}}+{\hat{T}_{[o]}})$ forms a dimension $24=5^2-1$ subalgebra, excluding the LE central charge $({\hat{T}_{[o]}}-{T_{[o]}})/5$. The complete LE possesses a dimension of $25=5^2$, suggesting a potential connection to the Lie algebras ${\mathfrak{su}}(t_2,s_2)$ and ${\mathfrak{u}}(t_2,s_2)$ with $s_2,t_2>0$ and $s_2+t_2=5$, as explored in \cite{U1I}.

The Non-contracted Poincaré-like Extension (NPE) in (\ref{Dim0Start})--(\ref{PPMTs}) introduces generators $\hat{P}_{\mu}$, $\hat{P}^*_{\mu}$ (or $\tilde{P}_{\mu}$, $\tilde{Q}_{\mu}$), $\Phi$, and $\bar{\Phi}$, completing $35=6^2-1$ generators. This suggests a potential relationship to the pseudo-unitary special Lie algebra ${\mathfrak{su}}(t_3,s_3)$ with $s_3,t_3>0$ and $s_3+t_3=6$. Meanwhile, the Contracted Poincaré Extension (CPE) could be associated with an inhomogeneous pseudo-unitary ${\mathfrak{iu}}(t_2,s_2)$ with $s_2,t_2>0$ and $s_2+t_2=5$.

We could introduce, alternatively, a real scalar central charge $\mathcal{D}$ to this algebra, as discussed earlier. The new generators $\mathcal{D}$, $T_{[o]N}$, and $\hat{T}_{[o]N}$ are incorporated into the algebra using the replacements:
\bee{CCharge} T_{[o]}=T_{[o]N}-(1/6)\mathcal{D},\ \ \hat{T}_{[o]}=\hat{T}_{[o]N}+4\mathcal{D}.\ee
The impact of these changes is evident in the  alternative relations:
\beaa 
\,[\Phi,\bar{\Phi}]&=&
\frac{ g_{44}}{10 R^2} \epsilon(24 {T_{[o]N}}-4{\hat{T}_{[o]N}}-20\mathcal{D}),\\
\,[\hat{P}_{\rho},\hat{P}^*_{\sigma}]&=&\frac{ g_{44}}{ R^2}(\ip M_{\sigma\rho}+T_{\sigma\rho}+( g_{\sigma\rho}/2) ({\hat{T}_{[o]N}}+4\mathcal{D})).
\eeaa
The resulting algebra possesses $36=6^2$ generators, potentially linked to the Lie algebra ${\mathfrak{u}}(t_3,s_3)$ with $s_3,t_3>0$ and $s_3+t_3=6$ (while ${\mathfrak{so}}(t_4,s_4)$ with $s_4,t_4>0$ and $s_4+t_4=9$ has dimension 36, it lacks a central charge). In this study, we will sporadically address the inclusion of $\mathcal{D}$. Relations (\ref{Dim0Start}), (\ref{TTM}), (\ref{PhiPT}--\ref{PPMTs}) suggest a potential interpretation of this extension as a generalized conformal symmetry.

The algebra (\ref{Dim0Start})--(\ref{PPMTs}) in the limit $R\rightarrow \infty$ is referred to as the Contracted Poincaré-like extension (CPE). The CPE exhibits nontrivial ideals. The abelian subalgebras generated by $\{\hat{P}_{\mu},\Phi\}$, $\{\hat{P}^*_{\mu},\bar{\Phi}\}$, and $\{\hat{P}_{\mu},\Phi, \hat{P}^*_{\mu},\bar{\Phi}\}$ remain invariant in the CPE. This suggests the possibility of introducing chirality to the electroweak sector.

The natural inclusion of complex vectors in this extension hints at a scenario involving complex space-time dimensions. Symmetry breaking mechanisms must be devised to transition to the four real space-time dimensions observed at low energies. The complex phases associated with complex space-time could potentially be related to the origin of the $SU(3)$ symmetry of strong interactions, although this remains unclear at present. The algebra implies a possible relation to a 10-dimensional space-time associated with the 10 translations $\hat{P}_{\mu},\hat{P}^*_{\mu}$ (or $\tilde{P}_{\mu}, \tilde{Q}_{\mu}$), $\Phi, \bar{\Phi}$. Further investigations might shed light on these aspects. As discussed in \cite{U1I}, advancements towards the necessity of 10-dimensional complex space-times at very high energy realms have been suggested in string theory, quantum gravity, and noncommutative geometry. Relevant references can be found therein.

For the NPE (\ref{Dim0Start})--(\ref{PPMTs}), we find a quadratic Casimir operator $\mathbb{I}_{(2)}$ ($\{A,B\}:=AB+BA$):
\beaa
\mathbb{I}_{(2)}&=&  2R^2(  g_{44}^{-1}\{\hat{P}^{\mu},\hat{P}^*_{\mu}\}+ g_{44}^{-1}\epsilon^{-1} \{\Phi, \bar{\Phi}\})\nn\\
&&+2\epsilon^{-1}\{\hat{T}^{\kappa},T_{\kappa}\}+M^{\rho\sigma}M_{\rho\sigma}+T^{\rho\sigma}T_{\rho\sigma}+\frac{24}{5}{T_{[o]}}^2+\frac{1}{5}{\hat{T}_{[o]}}^2.\label{CasimirPEx1}
   \eeaa
\beaa
\mathbb{I}_{(2)}&=&[\hat{P}^{\mu}\ \hat{P}^{*\mu}\ \Phi\  \bar{\Phi}\ T^{\kappa}\ \hat{T}^{\kappa}\ M^{\rho\tau}\ T ^{\rho\tau}\ {T_{[o]}}\ {\hat{T}_{[o]}}]\nn\\
&&\cdot M_1\cdot [\hat{P}^*_{\mu}\ \hat{P}_{\mu}\  \bar{\Phi}\ {\Phi}\ \hat{T}_{\kappa}\ T_{\kappa}\ M_{\rho\tau}\ T_{\rho\tau}\ {T_{[o]}}\ {\hat{T}_{[o]}}]^{tr}\!\!,\label{CasimirPEx}\\
{\rm with}&&\!\!M_1:={\rm diag}\left(\frac{ 2R^2}{ g_{44}},\frac{ 2R^2}{ g_{44}},\frac{2 R^2}{\epsilon g_{44}},\frac{2 R^2}{\epsilon g_{44}}, \frac{2}{\epsilon},\ \frac{2}{\epsilon},\ {1},\ 
{1},\ \frac{24}{5},\ \frac{1}{5}\right).\label{CasimirPExM1}
\eeaa
This Casimir operator further supports the suggestion of a connection to a special pseudo-unitary group. Moreover, it fosters the expectation that it might lead to a Lagrangian where the squared field strengths associated with point-dependent translation invariance carry the dimensionful coupling of gravity. Subsequent developments should provide substantial evidence for the transition to an internal $\oplus$ external symmetry paradigm, ultimately leading to the emergence of the Standard Model coupling to gravity.

The Contracted Poincaré-like Extension (CPE) [(\ref{Dim0Start})--(\ref{PPMTs}) in the limit $R\to \infty$] exhibits non-trivial ideals, indicating the presence of at least two Casimir operators, as it is a non-simple algebra:
\beaa
J&=&2g_{44}^{-1}(\{\hat{P}^{\mu},\hat{P}^*_{\mu}\}+2\epsilon^{-1} \{\Phi, \bar{\Phi}\})\nn\\
&=&[\hat{P}^{\mu}\ \hat{P}^{*\mu}\ \Phi\  \bar{\Phi}]
\cdot M_2\cdot [\hat{P}^*_{\mu}\ \hat{P}_{\mu}\  \bar{\Phi}\ {\Phi}]^{tr}\!\!,\label{CasPEx2}
\eeaa
\beaa
{\rm with}&&\!\!M_2:={\rm diag}\left(\frac{ 2}{ g_{44}},\frac{ 2}{ g_{44}},\frac{2}{\epsilon g_{44}},\frac{2 }{\epsilon g_{44}}\right).\label{CasimirPExM2}
\eeaa
Any constant multiple of $M_1$ and $M_2$ produces an invariant in (\ref{CasimirPEx}) and (\ref{CasPEx2}).

In \cite{U1I}, certain subalgebras of the Lorentz-like Extension (LE) were discussed. In that context, a potential origin of the Higgs field associated with the generators $T_\mu, \hat{T}_\mu$ required $\epsilon=-1$. We now turn our attention to some interesting subalgebras of the Non-contracted Poincaré-like Extension (NPE). Utilizing the following generators:
\beaa
\tilde{\cal P}_\mu=&(R/\sqrt{2})(\hat{P}_\mu+\hat{P}^*_\mu) +(T_\mu+\hat{T}_\mu)/\sqrt{2},\nn\\
\tilde{\cal K}_\nu=&(R/\sqrt{2})(\hat{P}_\nu+\hat{P}^*_\nu) -(T_\nu+\hat{T}_\nu)/\sqrt{2},\nn\\
\tilde{\cal D}=&-(R/(\sqrt{2}))(\Phi+\bar{\Phi}),\hspace{2.5cm}\label{ConfAlg}\\
\,[\tilde{\cal P}_\mu,\tilde{\cal K}_\nu]=&(g_{44}-\epsilon)\ip M_{\mu\nu}-2\ip g_{\mu\nu}\tilde{\cal D}.\hspace{1cm}\label{ConfAlgRel}
\eeaa
We verify that $<\!\{\tilde{\cal P}_\mu,\tilde{\cal K}_\nu, M_{\rho\sigma},\tilde{\cal D} \}\!>$ is a conformal algebra   for $\epsilon=-1=-g_{44}$. 

Another conformal algebra is generated for  $\epsilon=-1=-g_{44}$ with the choice: 
\beaa
\tilde{\cal P}_\mu=&(R/\sqrt{2})(\hat{P}_\mu+\hat{P}^*_\mu) +\ip(T_\mu-\hat{T}_\mu)/\sqrt{2},\nn\\
\tilde{\cal K}_\nu=&(R/\sqrt{2})(\hat{P}_\nu+\hat{P}^*_\nu) -\ip(T_\nu-\hat{T}_\nu)/\sqrt{2},\nn\\
\tilde{\cal D}=&(R/(\sqrt{2}))\ip(\Phi-\bar{\Phi}).\label{ConfAlg2}\hspace{2.7cm}
\eeaa
 Further subalgebras are easily found, together with their casimirs.
\section{Differential Representation}
We delve into the representability of the Poincaré-like extension (\ref{Dim0Start}--\ref{PPMTs}) in terms of differential operators, building upon those obtained for the Lorentz-like extension in \cite{U1I}. For each symmetry generator ${\cal O}$, we assign a differential operator $\ip\delta_{{\cal O}}$ that obeys to the same algebraic structure. Once again, there are (at least) two ways to represent this algebra as differential operators. The various generators can be expressed either in terms of Hermitian (H) and Anti-Hermitian (AH) operators, or in terms of Hermitian (H) and Adjoint Pairs (AP) of generators. Here is a description of the found differential representations:
\begin{table}[h]
	${\footnotesize
		\bac{c|cccccccccc}\hline{\stackrel{{\rm Generator} \rightarrow}
			{\downarrow{\rm Diff. Rep.}}} 
		&\  {\hat{P}}_{\rho}\  &\ {\hat{P}^*}_{\rho}\  
		&\  {\Phi}\  &\ {\bar{\Phi}}\ 
		&\  {T}_{\rho}\  &\ \hat{T}_{\rho}\  & \  M_{\rho\tau}\  & \ T_{\rho\tau}\  & \ T[o] \ &\ \hat{T}[o]\ \\
		\hline
		5D&{\rm H} &{\rm H}&{\rm H} &{\rm H}&{\rm H} &{\rm H}&{\rm H}&{\rm AH}&{\rm AH}&{\rm AH}\\
		\hline
		10D&{\rm AP} &{\rm AP}&{\rm AP} &{\rm AP}&{\rm AP} &{\rm AP}&{\rm H}&{\rm H}&{\rm H}&{\rm H}\\
		\hline
		\eac}$
	\caption{Differential Representations of Ext. Poincar\'e Alg.; $\rho,\tau=0,1,2,3.$}\label{DFull}
\end{table}

For the 5D representation in terms of differential operators, we introduce, in addition to the 4D space-time coordinates $\chi_{\rho}$ with $\partial_{\rho}=\partial/\partial \chi^{\rho},\ \rho=0,1,2,3$, new variables $\chi_5$ and $\chi_4$. We define $\partial_5=\partial/\partial \chi^5$, $\partial_5 (\chi_5)=\epsilon=g_{55}$, $\chi^5=\epsilon^{-1}\chi_5$, and $\partial_4=\partial/\partial \chi^4$, $\partial_4 (\chi_4)=g_{44}$, $\chi^4=g_{44}^{-1}\chi_4$. The variable $\chi^5$ is associated with the symmetry aiming to combine the Higgs and gauge fields, while $\chi_4$ is the variable that leads to the Poincaré algebra after an optional contraction.

The 5D differential representation devised for the Lorentz-like extension is expanded to the 5D representation of the Poincaré-like extension:
\beaa
-\ip \delta_{M_{\mu\nu}}&=&-\ip(\chi_{\mu}\partial_{\nu}-\chi_{\nu}\partial_{\mu}),\label{Rep5DStart}\\
-\ip\delta_{T_{\mu\nu}}&=&(\chi_{\mu}\partial_{\nu}+\chi_{\nu}\partial_{\mu}-\half g_{\mu\nu} \chi^{\rho}\partial_{\rho}),\\
-\ip\delta_{{T_{[o]}}}&=&-\chi^5\partial_5+\frac{1}{6}(\chi^5\partial_5+\chi^4\partial_4+\chi^{\rho}\partial_{\rho}),\\
-\ip\delta_{{\hat{T}_{[o]}}}&=&+(4\chi^5\partial_5+5\chi^{\rho}\partial_{\rho})-4(\chi^5\partial_5+\chi^4\partial_4+\chi^{\rho}\partial_{\rho}),\\
-\ip\delta_{T_{\mu}}&=&+\sqrt{2}\ip \chi_5\partial_{\mu},\ \ -\ip\delta_{\hat{T}_{\nu}}=-\sqrt{2}\ip \chi_{\nu}\partial_{5},
\eeaa
\beaa
-\ip\delta_{\hat{P}_{\mu}}&=&-\frac{{\sqrt{2}}\ip}{R} \chi_4\partial_{\mu},\ \ -\ip\delta_{\hat{P}^*_{\nu}}=\frac{{\sqrt{2}}\ip}{R} \chi_{\nu}\partial_{4},\\
-\ip\delta_{\Phi}&=&\frac{\sqrt{2}\ip}{R} \chi_{4}\partial_{5},\ \ -\ip\delta_{\bar{\Phi}}=-\frac{\sqrt{2}\ip}{R} \chi_5\partial_{4}.\label{Rep5DEnd}
\eeaa
It is straightforward to verify that the quantity ${\cal I}_o=\chi^{\rho}\chi_{\rho}+g_{44}\chi^4\chi^4+\epsilon\chi^5\chi^5$ is not invariant under the Anti-Hermitian operators, precluding a simple geometric interpretation with metric ${\rm diag(1,-1,-1,-1,g_{44},\epsilon)}$. 
However, ${\cal I}_o$ is invariant under the subalgebra $<\{M_{\mu\nu},T_{\rho}+\hat{T}_{\rho},\hat{P}_{\sigma}+{\hat{P}^*}_{\sigma},\Phi+\bar{ \Phi}\}>$ generated by (\ref{ConfAlg}).

The quantity ${\cal I}=(\chi^{\mu}\chi_{\mu})^4(\chi^{4}\chi_{4})(\chi^{5}\chi_{5})$ is invariant under $\delta_{M_{\mu\nu}}, \delta_{{T_{[o]}}}$, and $\delta_{{\hat{T}_{[o]}}}$. It is also invariant under the constrained action under $ -\ip\delta_{T_{\mu}}-\ip\delta_{\hat{T}_{\mu}}$ adopting parameters $\bar{\xi}^{\mu}={\xi}^{\mu}$ real (or under $+\ip\delta_{T_{\mu}}-\ip\delta_{\hat{T}_{\mu}}$ adopting $\bar{\xi}^{\mu}=-{\xi}^{\mu}$ imaginary) and the condition $\chi^{5}\chi_{5}=(1/4)\chi^{\mu}\chi_{\mu}$, resembling a brane condition.
 ${\cal I}$ is not invariant under $\xi^{\mu\nu}\delta_{T_{\mu\nu}}$ for arbitrary parameter $\xi^{\mu\nu}$, but, ${\cal I}$ is invariant under $ -\ip\delta_{\hat{P}_{\mu}}-\ip \delta_{\hat{P}^*_{\mu}}$ for parameters $\bar{X}^{\mu}=X^{\mu}$ (or under $ \ip\delta_{\hat{P}_{\mu}}-\ip \delta_{\hat{P}^*_{\mu}}$ for parameters $\bar{X}^{\mu}=-X^{\mu}$) and the condition $\chi^{4}\chi_{4}=1/4(\chi^{\mu}\chi_{\mu})$, resembling a brane condition. Also, ${\cal I}$ is invariant under $ -\ip\delta_{\Phi}-\ip \delta_{\bar{\Phi}}$ for parameters $\bar{X}=X$ (or under $ +\ip\delta_{\Phi}-\ip \delta_{\bar{\Phi}}$ with $\bar{X}=-X$) and the condition $\chi^{4}\chi_{4}=\chi^{5}\chi_{5}$.

If we aim to include the central charge $\mathcal{D}$ using (\ref{CCharge}) in this 5D representation:
\beaa
-\ip\delta_{\mathcal{D}}&=&\chi^5\partial_5+\chi^4\partial_4+\chi^{\rho}\partial_{\rho},\nn\\
-\ip\delta_{{T_{[o]N}}}&=&\chi^5\partial_5,\ \ 
-\ip\delta_{{\hat{T}_{[o]N}}}=-(4\chi^5\partial_5+5\chi^{\rho}\partial_{\rho}).\label{CCharge5D}
\eeaa

We naively expect that $T_{\mu}$ and $\hat{T}_{\mu}$ (as the Higgs field) — as well as $\hat{P}_{\mu}$ and $\hat{P}_{\mu}^*$ — form an adjoint pair, as hinted by relations (\ref{THbar}), (\ref{Eqn9}--\ref{Eqn12}), (\ref{Eqn25}--\ref{Eqn28}). However, this expectation is not substantiated under this representation. This 5D representation might be instrumental either as a starting model, despite its limitations in providing a geometric framework with full invariance, or as an intermediate stage when the symmetry is partially broken.

The contraction procedure is adjusted to provide a differential representation. We employ $R\rightarrow\infty$, $\chi_4/R\rightarrow (1/\sqrt{2})$, $g_{44}^{-1}\chi_4\partial_4\rightarrow -\epsilon^{-1}\chi_{5}\partial_5-\chi^\rho\partial_{\rho}+3$, and $(1/R)\partial_4\rightarrow 0$. This leaves us with no nontrivial representation of $\hat{P}_{\mu}^*$, $\bar{\Phi}$.

As a digression, we contemplate a representation of the NPE without $\chi^4$, say with $-\ip\delta_{\hat{P}_{\mu}}=-\ip\partial_{\mu}$ and $-\ip\delta_{\Phi}=\ip\partial_5$. In analogy to conformal representations, we set $-\ip\delta_{\hat{P}_{\mu}^*}=(-2\sqrt{2}\ip\,g_{44}/R^2)\chi_\mu(\chi^\rho\partial_\rho+\chi^5\partial_5-3))$ and $-\ip\delta_{\bar{\Phi}}=-(\ip 2 \sqrt{2}g_{44}/R^2)\chi_5(\chi^\rho\partial_\rho+\chi^5\partial_5-3)$. However, this attempt fails to represent the algebra.

We also complete the 10D representation presented in \cite{U1I} to the extended Poincaré algebra by complexifying the coordinates. Let $z_{a}=\chi_{a}+\ip y_{a}$ and $\bar{z}_a=\chi_{a}-\ip y_{a}$ for $a=0,1,2,3,4,5$. This leads to a representation with
\beaa \bar{z}^{a}z_{a}&=&g_{\mu\nu}\bar{z}^{\mu}{z}^{\nu}+g_{44}\bar{z}^{4}{z}^{4}+\epsilon\bar{z}^{5}{z}^{5}\nn\\
&=&g_{\mu\nu}{\chi}^{\mu}{\chi}^{\nu}+g_{\mu\nu}{y}^{\mu}{y}^{\nu}+g_{44}{\chi}^{4}{\chi}^{4}+g_{44}{y}^{4}{y}^{4}+\epsilon{\chi}^{5}{\chi}^{5}+\epsilon{y}^{5}{y}^{5}
\eeaa 
invariant, establishing the isomorphism to an algebra $su(s_2,t_2)$ with $s_2,t_2>0$ and $s_2+t_2=6$, as previously conjectured. Denoting $\partial_a=\partial/\partial \chi^a$ and $\hat{\partial}_a=\partial/\partial y^a$, the 10D representation with such invariance is:
\beaa
-\ip \delta_{M_{\mu\nu}}\!\!\!\!&=&\!\!\!\!-\ip(\chi_{\mu}\partial_{\nu}-\chi_{\nu}\partial_{\mu})-\ip(y_{\mu}\hat{\partial}_{\nu}-y_{\nu}\hat{\partial}_{\mu}),\label{Rep10DStart}\\
-\ip\delta_{T_{\mu\nu}}\!\!\!\!&=&\!\!\!\!\!-\ip(\chi_{\mu}\hat{\partial}_{\nu}+\chi_{\nu}\hat{\partial}_{\mu}-\!\half g_{\mu\nu} \chi^{\rho}\hat{\partial}_{\rho})\!+\!\ip(y_{\mu}\partial_{\nu}+y_{\nu}\partial_{\mu}-\!\half g_{\mu\nu} y^{\rho}\partial_{\rho}),\\
-\ip\delta_{{T_{[o]}}}\!\!\!\!&=&\!\!\!\!\!-\frac{5\ip}{6}\epsilon^{\!-1}(y_5\partial_5\!-\!\chi_5\hat{\partial}_5)+\!\frac{\ip}{6}g_{44}^{-1}(y_4\partial_4\!-\!\chi_4\hat{\partial}_4)\!+\!\frac{\ip}{6}(y^{\rho}\partial_{\rho}\!-\!\chi^{\rho}\hat{\partial}_{\rho}),\\
-\ip\delta_{{\hat{T}_{[o]}}}\!\!\!\!&=&\!\!\!\!-4\ip g_{44}^{-1}(y_4\partial_4-\!\chi_4\hat{\partial}_4)+\ip(y^{\rho}\partial_{\rho}-\!\chi^{\rho}\hat{\partial}_{\rho}),\\
-\ip\delta_{T_{\mu}}\!\!\!\!&=&\!\!\!\!-\frac{1}{\sqrt{2}}\left(\! -\ip(\chi_5+\!\ip y_5)(\partial_{\mu}-\!\ip\hat{\partial}_{\mu})+\ip(\chi_{\mu}-\!\ip y_{\mu})(\partial_5+\!\ip\hat{\partial}_5)\right)\!,\\
-\ip\delta_{\hat{T}_{\nu}}\!\!\!\!&=&\!\!\!\!+\frac{1}{\sqrt{2}}\left(\!-\ip(\chi_{\mu}+\!\ip y_{\mu})(\partial_5-\!\ip\hat{\partial}_5) +\ip(\chi_5-\!\ip y_5)(\partial_{\mu}+\!\ip\hat{\partial}_{\mu})\right)\!,
\eeaa\beaa
-\ip\delta_{\hat{P}_{\mu}}\!\!\!\!&=&\!\!\!\!+\frac{1}{\sqrt{2}R}\left(\! -\ip(\chi_4+\!\ip y_4)(\partial_{\mu}-\!\ip\hat{\partial}_{\mu})+\ip(\chi_{\mu}-\!\ip y_{\mu})(\partial_4+\!\ip\hat{\partial}_4)\right)\!,\\
-\ip\delta_{\hat{P}^*_{\mu}}\!\!\!\!&=&\!\!\!\!-\frac{1}{\sqrt{2}R}\left(\!-\ip(\chi_{\mu}+\!\ip y_{\mu})(\partial_4-\!\ip\hat{\partial}_4) +\ip(\chi_4-\!\ip y_4)(\partial_{\mu}+\!\ip\hat{\partial}_{\mu})\right)\!,\\
-\ip\delta_{{\Phi}}\!\!\!\!&=&\!\!\!\!-\frac{1}{\sqrt{2}R}\left(\! -\ip(\chi_4+\!\ip y_4)(\partial_{5}-\!\ip\hat{\partial}_{5})+\ip(\chi_{5}-\!\ip y_{5})(\partial_4+\!\ip\hat{\partial}_4)\right)\!,\\
-\ip\delta_{\bar{\Phi}}\!\!\!\!&=&\!\!\!\!\frac{1}{\sqrt{2}R}\left(\! -\ip(\chi_{5}+\!\ip y_{5})(\partial_4-\!\ip\hat{\partial}_4)+\ip(\chi_4-\!\ip y_4)(\partial_{5}+\!\ip\hat{\partial}_{5})\right)\!.\label{Rep10DEnd}
\eeaa
If we  include the central charge $\mathcal{D}$ using (\ref{CCharge}) in this 10D representation, the central charge does not look like a dilatation:
\beaa
-\ip\delta_{\mathcal{D}}&=&-\ip(y^5\partial_5-\!\chi^5\hat{\partial}_5)-\ip(y^4\partial_4-\!\chi^4\hat{\partial}_4)-\ip(y^{\rho}\partial_{\rho}-\!\chi^{\rho}\hat{\partial}_{\rho}),\nn\\
-\ip\delta_{{T_{[o]N}}}&=&-\ip(y^5\partial_5-\!\chi^5\hat{\partial}_5),\nn\\ 
-\ip\delta_{{\hat{T}_{[o]N}}}&=&+4\ip(y^5\partial_5-\!\chi^5\hat{\partial}_5)+5\ip(y^{\rho}\partial_{\rho}-\!\chi^{\rho}\hat{\partial}_{\rho}).\label{CCharge10D}
\eeaa
The contraction procedure is adjusted to provide a differential representation. Employ $R\rightarrow\infty$, $\chi_4/R\rightarrow 1/\sqrt{2}$, $y_4\partial_4-\!\chi_4\hat{\partial}_4\rightarrow -g_{44}\epsilon^{-1}(y_5\partial_5-\!\chi_5\hat{\partial}_5)-g_{44}(y^\rho\partial_\rho-\!\chi^\rho\hat{\partial}_\rho)$, $y_4/R\rightarrow 0$, $\partial_4/R\rightarrow 0$, and $\hat{\partial_4}/R\rightarrow 0$. This contraction does not make any generator trivial. In the contracted 10D representation, $-\ip\delta_{\hat{P}_\mu}=-\frac{\ip}{2}(\partial_\mu-\ip\hat{\partial}_\mu)$, $-\ip\delta_{\hat{P}^*_\mu}=-\frac{\ip}{2}(\partial_\mu+\ip\hat{\partial}_\mu)$, $-\ip\delta_{\Phi}=+\frac{\ip}{2}(\partial_5-\ip\hat{\partial}_5)$, $-\ip\delta_{\bar{\Phi}}=+\frac{\ip}{2}(\partial_5+\ip\hat{\partial}_5)$. Possible mechanisms for the transition (before or after contraction) from the 10D to 5D representation will be addressed elsewhere.

We  consider now two  bases $\tilde{N}_{AB}$ and $\tilde{M}_{AB}$ of operators for the 10D representation, using $\partial_{Z^A}=(1/2)(\partial_{x^A}-\ip\partial_{y^A}),\ \ 
\partial_{\bar{Z}^A}=(1/2)(\partial_{x^A}+\ip\partial_{y^A})$:
  \beaa
 -\ip\delta_{\tilde{N}_{AB}}&=&-\ip(z_A\partial_{z^B}-\bar{z}_B\partial_{\bar{z}^A}),\\
 -\ip\delta_{\tilde{M}_{AB}}&=& -\ip\delta_{\tilde{N}_{AB}}-(g_{AB}/6)(-\ip\delta_{\tilde{N}^C_{\ C}}),\ \  A,B=0,1,2,3,4,5.
 \eeaa 
Lowering and rising of capital labels via $g_{AB}={\rm diag(1,-1,-1,-1,g_{44},\epsilon)}$. Such differential operators represent the algebra with:
\beaa
\tilde{M}_{AB}&=& \tilde{N}_{AB}-(g_{AB}/6)\tilde{N}^C_{\ C},\\
\,[\tilde{N}_{AB},\tilde{N}_{CD}]&=&-\ip(g_{AD}\tilde{N}_{CB}-g_{BC}\tilde{N}_{AD}),\\
\,[\tilde{M}_{AB},\tilde{M}_{CD}]&=&-\ip(g_{AD}\tilde{M}_{CB}-g_{BC}\tilde{M}_{AD}).
\eeaa
Repeated up and down indices $B,C$ are summed over $\{\mu,4,5\}\!=\!\{0,1,2,3,4,5\}$. The $\tilde{N}_{AB}$ generate a ${\mathfrak{u}}(t_3,s_3)$ algebra, while the $\tilde{M}_{AB}$ generate a ${\mathfrak{su}}(t_3,s_3)$ with $s_3>2,t_3>0$ and $s_3+t_3=6$.
Using (\ref{Rep10DStart})--(\ref{Rep10DEnd}), we find the conversion between diverse generator bases:
\beaa
\tilde{N}^{C}_{\ C} &=&\tilde{N}^\rho_{\ \rho}+\tilde{N}^4_{\ 4}+\tilde{N}^5_{\ 5}=\frac{6\ip}{20}( 4T_{[o]}+\hat{T}_{[o]}-{5\ip}\tilde{N}^\rho_{\ \rho}),\ \tilde{M}^{C}_{\ C}=0,\ \label{TildeConvStart}\\
\tilde{N}^4_{\ 4} &=&\frac{\ip }{4}\hat{T}_{[o]}+\frac{1 }{4}\tilde{N}^\rho_{\ \rho},\hspace{1.9cm} \tilde{M}^4_{\ 4}=-\frac{\ip}{5}(T_{[o]}-\hat{T}_{[o]}),\\ 
\tilde{N}^5_{\ 5} \!&=\!&\!\!\!\!\frac{6\ip}{5} T_{[o]}\!+\!\frac{\ip}{20}\hat{T}_{[o]}\!+\!\frac{1}{4}\tilde{N}^\rho_{\ \rho},\  \tilde{M}^5_{\ 5}=\ip T_{[o]},\ 
\tilde{M}^\rho_{\ \rho}=\!-\frac{\ip}{5}(4T_{[o]}\!+\!\hat{T}_{[o]}),\ \ \\
\tilde{N}_{\mu\nu}&=&-\frac{\ip}{2}(\ip M_{\mu\nu}\!+\! T_{\mu\nu})\!+\!\frac{1}{4}g_{\mu\nu}\tilde{N}^\rho_{\ \rho},\\
\tilde{M}_{\mu\nu}&=&-\frac{\ip}{2}(\!\ip M_{\mu\nu}\!+\! T_{\mu\nu}\!+\!\frac{g_{\mu\nu}}{10}(4T_{[o]}\!+\!\hat{T}_{[o]})),\\
\tilde{N}_{5\mu}&=&-\sqrt{2}\,T_\mu/2=\tilde{M}_{5\mu},\hspace{1.9cm}
\tilde{N}_{\mu 5}=\sqrt{2}\,\bar{T}_\mu/2=\tilde{M}_{\mu 5},\\
\tilde{N}_{4\mu}&=&\sqrt{2}R\, \hat{P}_\mu/2=\tilde{M}_{4\mu},\hspace{1.6cm}
\tilde{N}_{\mu 4}=-\sqrt{2}R\,\hat{P}^*_\mu/2=\tilde{M}_{\mu 4},\\
\tilde{N}_{45}&=&-\sqrt{2}\,R\, \Phi/2=\tilde{M}_{45},\hspace{1.5cm}
\tilde{N}_{5 4}=\sqrt{2}\,R\,\bar{\Phi}/2=\tilde{M}_{5 4}.
\label{TildeConvEnd}
\eeaa
The corresponding central charge is $\tilde{N}^{C}_{\ C}$ in (\ref{TildeConvStart}). The expression for $-2\ip\tilde{M}_{\mu\nu}$ appears frequently and has extraordinary properties. We will use it in place of the corresponding expression. The quadratic Casimir $\mathbb{I}_{(2)}$ in (\ref{CasimirPExM1}) corresponds to the standard Casimir for ${\mathfrak{su}}(t_3,s_3)$:
\beaa
-\frac{1}{4}\mathbb{I}_{(2)}&=&\tilde{M}^{AB}\tilde{M}_{BA}=\tilde{N}^{AB}\tilde{N}_{BA}-\frac{1}{6}(\tilde{N}^{C}_{\ C})^2.
\eeaa
A quartic Casimir $\mathbb{I}_{(4)}$ for the NPE consists of six anticommutators or squares of quadratics, using nine quadratics which build a multiplet:
\beaa
\!\!&&\!\!-\frac{1}{16}\mathbb{I}_{(4)}=\!\tilde{M}^{AB}\tilde{M}_{BC}\tilde{M}^{CD}\tilde{M}_{DA}=\nn\\
\!\!&&\!\half\{\tilde{M}^{\rho A}\tilde{M}_{A\sigma},\tilde{M}^{\sigma A}\tilde{M}_{A\rho}\}\!+\!\{\tilde{M}^{\rho A}\tilde{M}_{A4},\tilde{M}^{4 B}\tilde{M}_{B\rho}\}\!+\!\{\tilde{M}^{\rho A}\tilde{M}_{A5},\tilde{M}^{5 B}\tilde{M}_{B\rho}\}\nn\\
\!\!&&\!+(\tilde{M}^{4A}\tilde{M}_{A4})^2
+\{\tilde{M}^{4A}\tilde{M}_{A5},\tilde{M}^{5 B}\tilde{M}_{B4}\}+(\tilde{M}^{5A}\tilde{M}_{A5})^2.
\eeaa

From NPE $\simeq{\mathfrak{su}}(t_3,s_3)$, we determine the nature of the CPE-algebra, which is, as conjectured, an inhomogeneous pseudo-unitary ${\mathfrak{iu}}(t_2,s_2)$ with $s_2>2,t_2>0$ and $s_2+t_2=5$.  Determination of a quartic invariant (extrapolating the  squared Pauli-Luba\'nski pseudo--vector invariant of the Poincar\'e group) is done in Appendix A.

\section{Field Representation \& Local Invariance}
We expect the 4D space--time translations to be an effective outcome of this symmetry, so we propose covariant derivatives ${\mathbb{D}}_{\cal O}$ where ${\cal O}$ is a Lie algebra generator of NPE or of some invariant subalgebra of the CPE. The covariant derivatives involve one of the differential representations $-\ip\delta_{\cal O}$ in terms of coordinates $(\chi_0,\chi_1,\chi_2,\chi_3,\chi_4,\chi_5)$, or $(z_0,z_1,z_2,z_3,z_4,z_5)$ --excluding $\chi_4$ or $z_4$ if contracted-- and fields $\lambda^{\hat{\cal O}}{\cal A}_{{\cal O}}^{\ \hat{\cal O}}(x)$ that are functions thereof. The factor $\lambda^{\hat{\cal O}}$ is a coupling constant assigned to each type of generators. Coupling constant indices do not participate in the repeated index summation convention. The covariant derivative ${\mathbb{D}}_{{\cal O}}$ and the field strengths $\lambda^{\hat{\cal O}} {\mathbb F}_{\ {\cal O}{\cal O}'}^{\hat{\cal O}}$ are defined as follows:
\beaa
&&{\mathbb{D}}_{{\cal O}}:= \ip(-\ip\delta_{{\cal O}})\otimes Id+\sum_{\hat{\cal O}}\ip\lambda^{\hat{\cal O}}{\cal A}_{{\cal O}}^{\ \hat{\cal O}}\otimes {\hat{\cal O}},\\
&&\sum_{\hat{\cal O}}\ip\lambda^{\hat{\cal O}} {\mathbb F}_{\ {\cal O}{\cal O}'}^{\hat{\cal O}}\otimes {\hat{\cal O}}:= \,[{\mathbb{D}}_{{\cal O}},{\mathbb{D}}_{{\cal O}'}]+\ip {\mathbb{D}}_{[{\cal O},{\cal O}']},\label{FieldStrengthsOp}\\
&&\lambda^{\hat{\cal O}} {\mathbb F}_{\ {\cal O}{\cal O}'}^{\hat{\cal O}}\!:=
\delta_{{\cal O}}(\lambda^{\hat{\cal O}}{\cal A}_{{\cal O}'}^{\ \hat{\cal O}})\! -\! \delta_{{\cal O}'}(\lambda^{\hat{\cal O}}{\cal A}_{{\cal O}}^{\ \hat{\cal O}})\nn\\
&&\hspace{2cm}+{\lambda^{\hat{\cal O}}}{\cal F}_{\ {\cal O}{\cal O}'}^{\hat{\cal O}}+\ip(\ip C_{\ {\cal O}{\cal O}'}^{{\cal O}''})(\lambda^{\hat{\cal O}}{\cal A}_{{\cal O}''}^{\ \hat{\cal O}}),\ \ \label{FSNew}\\
&&{\lambda^{\hat{\cal O}}}{\cal F}_{\ {\cal O}{\cal O}'}^{\hat{\cal O}}:= {\ip}(\lambda^{\tilde{\cal O}}{\cal A}_{{\cal O}}^{\ \tilde{\cal O}})\ip C_{\ \tilde{\cal O}\tilde{\cal O}'}^{\hat{\cal O}}
(\lambda^{\tilde{\cal O}'}{\cal A}_{{\cal O}'}^{\ \tilde{\cal O}'}).\label{QuadTerms}
\eeaa
where $[{\cal O}'',{\cal O}']=\ip\,C_{\ {\cal O}''{\cal O}'}^{{\cal O}}\,{\cal O}$, and $Id$ is the universally commuting identity operator. The calligraphic  ${\lambda^{\hat{\cal O}}}{\cal F}_{\ {\cal O}{\cal O}'}^{\hat{\cal O}}$ corresponds to standard quadratic terms due to non--abelianity. The labels of standard covariant derivatives build the abelian subalgebra of translations. Here, we have a full invariant subalgebra {\bf which is not necessarily commutative}. The linear term in (\ref{FieldStrengthsOp}) accounts for the non-commutativity geometric nature of the diverse labels, and it is the way in which non-commutative geometry is involved here.

Both fields, and field strengths are left--modules over the extended Poincar\'e--like Lie algebra, transforming only with respect to their upper index:
\beaa \ [\tilde{{\cal O}},\lambda^{\hat{\cal O}}{\cal A}_{{\cal O}}^{\ \hat{\cal O}}(x)]_1&=-\ip\, C_{\ \tilde{{\cal O}}{\cal O}''}^{\hat{{\cal O}}}\lambda^{{\cal O}''}{\cal A}_{{\cal O}}^{\ {\cal O}''}(x),&\label{LeftModField}\\
\ [\tilde{{\cal O}},\lambda^{\hat{\cal O}} {\mathbb F}_{\ {\cal O}{\cal O}'}^{\hat{\cal O}}(x)]_1&=-\ip\, C_{\ \tilde{{\cal O}}{\cal O}''}^{\hat{\cal O}}\lambda^{{\cal O}''} {\mathbb F}_{\ {\cal O}{\cal O}'}^{{\cal O}''}(x).&\label{LeftModNew}
\eeaa
We can use a matrix notation for the covariant derivative,
\beaa
\,[D_{\cal O}]_{\hat{\cal O}'}^{\ \ \hat{\cal O}}=\delta_{\cal O}[Id]_{\hat{\cal O}'}^{\ \ \hat{\cal O}}+\ip (\lambda^{\tilde{\cal O}}{\cal A}_{\cal O}^{\ \tilde{\cal O}})[R_{\tilde{\cal O}}]_{\hat{\cal O}'}^{\ \ \hat{\cal O}},\ \ 
{\rm with}\ \ [R_{\cal O}]_{\hat{\cal O}'}^{\ \ \hat{\cal O}}=-\ip C_{\ {\cal O}\hat{\cal O}'}^{\hat{\cal O}},
\eeaa
from which  (\ref{FSNew}) follows. $R_{\cal O}$ is the well known adjoint representation.

We assign to the generic generator ${\cal O}'$ a parameter ${\varsigma}^{{\cal O}'}(x)$. The first order transformation of the field $\lambda^{{\cal O}}{\cal A}_{a}^{{\cal O}}$ (together with its coupling) is adopted to be:
\beaa 
(\lambda^{\hat{{\cal O}}}\!{{\cal A}_{{\cal O}}^{\ \hat{\cal O}}})\!\mapsto\!(\lambda^{\hat{{\cal O}}}\!{{\cal A}_{{\cal O}}^{\ \hat{\cal O}}})'\!=\!(\lambda^{\hat{{\cal O}}}\!{{\cal A}_{{\cal O}}^{\ \hat{\cal O}}})\!+\!\sum_{\tilde{{\cal O}}}\!\ip{\varsigma}^{\tilde{{\cal O}}}\!(x) [\tilde{{\cal O}},\lambda^{\hat{\cal O}}\!{\cal A}_{{\cal O}}^{\ \hat{{\cal O}}}]_1\!+\!
\ip(-\ip\delta_{{\cal O}}\varsigma^{\hat{{\cal O}}}\!(x)).\ \ \label{NewFOTField}
\eeaa
From this, the field strengths end up transforming homogeneously (as a curvature), as they would with global (rigid) transformations:
\beaa
(\lambda^{\hat{\cal O}} {\mathbb {F}}_{\ {\cal O}{\cal O}'}^{\hat{\cal O}})\mapsto (\lambda^{\hat{\cal O}} {\mathbb {F}}_{\ {\cal O}{\cal O}'}^{\hat{\cal O}})'= (\lambda^{\hat{\cal O}} {\mathbb {F}}_{\ {\cal O}{\cal O}'}^{\hat{\cal O}})+\sum_{\tilde{{\cal O}}}\ip{\varsigma}^{\tilde{{\cal O}}}(x) [\tilde{{\cal O}},\lambda^{\hat{\cal O}} {\mathbb {F}}_{\ {\cal O}{\cal O}'}^{\hat{\cal O}}]_1.\ \ \label{NewFOTFieldSt}
\eeaa
Notice that the action of $-\ip\delta_{{\cal O}''}(\cdot)$  commutes with the $[{\cal O}',\cdot]_1$ actions.

The indices in ${\cal O}, \hat{\cal O}$ of the fields ${\cal A}_{{\cal O}}^{\ \hat{\cal O}}$ play distinct roles. The $\hat{\cal O}$ acts as a generator of the Non-Contracted Poincaré-like Extension (NPE) (\ref{Dim0Start})--(\ref{PPMTs}) or resides in some invariant subalgebra of the Contracted Poincaré-like Extension (CPE). Its related indices are Lorentz--like indices linked to tetrads and similar fields, aiding in the local definition of a Lorentz basis. These upper indices undergo transformation according to (\ref{LeftModField}--\ref{LeftModNew}) as part of the left module of the NPE, or as part of the CPE module if contracted. 

On the other hand, the lower indices involved in ${\cal O}$ correspond to those of the differential representation expressed in terms of manifold coordinates and their derivatives involving space--time coordinates. They are thus Einstein--like indices.

In introducing the coupling constants, we transition the upper indices to an interaction basis with generators: ${T_{[o]}}_{\text{int}} := h{T_{[o]}}, {\hat{T}_{[o]\text{int}}} := {\hat{h}}{\hat{T}_{[o]}}, T_{\tau\text{int}} := \ell\, T_{\tau}, \hat{T}_{\tau\text{int}} := \ell\, \hat{T}_{\tau}, M_{\rho\sigma\text{int}} := {r'}M_{\rho\sigma}, T_{\rho\sigma\text{int}} := r T_{\rho\sigma}, \hat{P}_{\tau\text{int}} = s\hat{P}_{\tau}, \hat{P}_{\tau\text{int}}^{*} = {\bar{s}}\hat{P}_{\tau}^{*}$,
$\Phi_{\text{int}} = q\Phi, \hat{\Phi}_{\text{int}} = \bar{q}\hat{\Phi}$. We maintain the use of the original renormalized algebra (\ref{Dim0Start})--(\ref{PPMTs}) and keep track of the coupling constants to signal an interaction basis by simply appending them to the fields and field strengths. 

For each generator ${\cal O}, {\hat{\cal O}}$ with naive dimensions dim${\cal O}$, dim${\hat{\cal O}}$ respectively, we associated a field (connection) $\lambda^{\hat{\cal O}}{\cal A}_{{\cal O}}^{\ \hat{\cal O}}(x)$ with a naive dimension of dim${\cal O} -$dim${\hat{\cal O}}$. The notation of these fields depends on their upper index, as follows:
\beaa
{\mathbb{D}}_{\cal O}\!:\!\!&\!=\!&\!\!\ip(-\ip\delta_{{\cal O}})\otimes Id+\ip \ell (\hat{\phi}_{\cal O}^{\, \, \tau}\otimes {T_{\tau}}+\phi_{\cal O}^{\, \, \tau}\otimes {\hat{T}_{\tau}})\nn\\
\!\!&\!\!+&\!\!\ip h B_{\cal O}\!\otimes {T_{[o]}}+\ip \hat{h} \hat{B}_{\cal O}\!
\otimes {\hat{T}_{[o]}}+\ip r' B^{\, \, [\rho\sigma]}_{\cal O}\!\otimes \!{M_{\rho\sigma}}+\ip r B^{\, \, (\rho\sigma)}_{\cal O}\!\otimes\! {T_{\rho\sigma}}\nn\\
\!\!&\!\!+&\!\!\ip {\rm s}\, \bar{{\mathcal{E}}}_{\cal O}^{\   \tau}\!\otimes {\hat{P}_{\tau}}+
\ip \bar{\rm s}\, {{\mathcal{E}}}_{\cal O}^{\ \tau}\!\otimes {\hat{P}^{*}_{\tau}}+\ip {\rm q} \bar{{\mathcal{V}}}_{\cal O}\!\otimes {{\Phi}}+
\ip \bar{\rm q} {{\mathcal{V}}}_{\cal O}\otimes {\bar{ \Phi}},\label{CovDerNPE}
\eeaa
with $hB_{\cal O},\hat{h}\hat{B}_{\cal O},\ell\hat{ \phi}_{\cal O}^{\,\, \tau},\ell\phi_{\cal O}^{\,\,\tau},r'B^{\,\,[\rho\sigma]}_{\cal O},rB^{\,\,(\rho\sigma)}_{\cal O}, \bar{s}{{\mathcal{E}}}_{\cal O}^{\ \tau}, s\bar{{\mathcal{E}}}_{\cal O}^{\ \tau}, \bar{q}{{\mathcal{V}}}_{\cal O},q\bar{{\mathcal{V}}}_{\cal O}$ gauge fields --connections--  corresponding  to generators ${T_{[o]}},{\hat{T}_{[o]}},T_{\tau},\hat{T}_{\tau},M_{\rho\sigma},T_{\rho\sigma},\hat{P}^{*}_{\tau}\!,\hat{P}_{\tau}\!, \hat{\Phi},{\Phi}$, respectively,  with  coupling constants
$h,\hat{h} ,\ell,\ell, r',r, \bar{\rm s}, {\rm s}, \bar{\rm q}, {\rm q}$, respectively.
We compute the formal commutator $[{\mathbb{D}}_{{\cal O}},{\mathbb{D}}_{{\cal O}'}]$ in (\ref{FieldStrengthsOp}) to obtain the field  strengths:
\beaa
&&\,\!\!\![{\mathbb{D}}_{{\cal O}},{\mathbb{D}}_{{\cal O}'}]+\ip {\mathbb{D}}_{[{\cal O},{\cal O}']}\!=\!\ip \ell(\hat{\mathbb {F}}_{\ {\cal O}{\cal O}'}^{ \rho}\otimes {T_{\rho}}\!+\!{\mathbb {F}}_{\ {\cal O}{\cal O}'}^{\rho}\otimes {\hat{T}_{\rho}}) \nn\\
&&\ +\ip  h{\mathbb {F} }_{{\cal O}{\cal O}'}\otimes {T_{[o]}}+ \ip  \hat{h}\hat{\mathbb {F}}_{{\cal O}{\cal O}'}\otimes {\hat{T}_{[o]}}+\ip r'{\mathbb {G}}_{\ \ {\cal O}{\cal O}'}^{[\tau\kappa]}\otimes {M_{\tau\kappa}}\!+\!\ip r{\mathbb {H}}_{\ \ {\cal O}{\cal O}'}^{(\tau\kappa)}\otimes {T_{\tau\kappa}}\nn\\
&&\ +\ip {\rm s}\bar{{\mathbb {R}}}_{\ {\cal O}{\cal O}'}^{\kappa}\otimes {\hat{P}_{\kappa}}
\!+\!\ip \bar{\rm s}{{\mathbb {R}}}_{\ {\cal O}{\cal O}'}^{\kappa}\otimes {\hat{P}^{*}}_{\kappa}+\ip {\rm q} \bar{{\mathbb {Q}}}_{{\cal O}{\cal O}'}\otimes {{\Phi}}+
\ip  \bar{\rm q}{{\mathbb {Q}}}_{{\cal O}{\cal O}'}\otimes {\bar{ \Phi}};\ \ \label{FieldStrengthsOpPoin}
\eeaa
\beaa
\,\hat{\mathbb {F}}_{\ {\cal O}{\cal O}'}^{ \rho}:\!\!&=&\!\!\delta_{{\cal O}}(\hat{\phi}_{{\cal O}'}^{\, \, \tau})-\delta_{{\cal O}'}(\hat{\phi}_{{\cal O}}^{\, \, \tau})+\hat{{\cal F}}_{\ {\cal O}{\cal O}'}^{\rho}+\ip(\ip C_{\ {\cal O}{\cal O}'}^{{\cal O}''})\hat{\phi}_{{\cal O}''}^{\, \, \tau},\\
\,{\mathbb {F}}_{\ {\cal O}{\cal O}'}^{ \rho}:\!\!&=&\!\!\delta_{{\cal O}}({\phi}_{{\cal O}'}^{\, \, \tau})-\delta_{{\cal O}'}({\phi}_{{\cal O}}^{\, \, \tau})+{{\cal F}}_{\ {\cal O}{\cal O}'}^{\rho}+\ip(\ip C_{\ {\cal O}{\cal O}'}^{{\cal O}''}){\phi}_{{\cal O}''}^{\, \, \tau},\\
{\mathbb {F}}_{{\cal O}{\cal O}'}:\!\!&=&\!\!\delta_{{\cal O}}({B}_{{\cal O}'})-\delta_{{\cal O}'}({B}_{{\cal O}})+{{\cal F}}_{ {\cal O}{\cal O}'}+\ip(\ip C_{\ {\cal O}{\cal O}'}^{{\cal O}''}){B}_{{\cal O}''},\label{FSTo}\\
\hat{\mathbb {F}}_{{\cal O}{\cal O}'}:\!\!&=&\!\!\delta_{{\cal O}}(\hat{B}_{{\cal O}'})-\delta_{{\cal O}'}(\hat{B}_{{\cal O}})+{\hat{\cal  F}}_{{\cal O}{\cal O}'}+\ip(\ip C_{\ {\cal O}{\cal O}'}^{{\cal O}''})\hat{B}_{{\cal O}''},\label{FSThato}\\
{\mathbb {G}}_{\ \ {\cal O}{\cal O}'}^{[\tau\kappa]}:\!\!&=&\!\!\delta_{{\cal O}}({B}_{{\cal O}'}^{\ [\rho\sigma]})\!-\!\delta_{{\cal O}'}({B}_{{\cal O}'}^{\ [\rho\sigma]})\!+\!{\hat{\cal G}}_{\ \ {\cal O}{\cal O}'}^{[\rho\sigma]}\!+\!\ip(\ip C_{\ {\cal O}{\cal O}'}^{{\cal O}''}){B}_{{\cal O}''}^{\ [\rho\sigma]}, \\
\eeaa
\beaa
{\mathbb {H}}_{\ \ {\cal O}{\cal O}'}^{(\tau\kappa)}:\!\!&=&\!\!\delta_{{\cal O}}({B}_{{\cal O}'}^{\ (\rho\sigma)})\!-\!\delta_{{\cal O}'}({B}_{{\cal O}'}^{\ (\rho\sigma)})\!+\!{\hat{\cal H}}_{\ \ {\cal O}{\cal O}'}^{(\rho\sigma)}\!+\!\ip(\ip C_{\ {\cal O}{\cal O}'}^{{\cal O}''}){B}_{{\cal O}''}^{\ (\rho\sigma)},\ \ \\
\,\bar{{\mathbb {R}}}_{\ {\cal O}{\cal O}'}^{\rho}:\!\!&=&\!\!\delta_{{\cal O}}(\bar{{\mathcal{E}}}_{{\cal O}'}^{\, \, \rho})-\delta_{{\cal O}'}(\bar{{\mathcal{E}}}_{{\cal O}}^{\, \, \rho})+\bar{{\cal R}}_{\ {\cal O}{\cal O}'}^{\rho}+\ip(\ip C_{\ {\cal O}{\cal O}'}^{{\cal O}''})\bar{{\mathcal{E}}}_{{\cal O}''}^{\, \, \rho},\\
\,{{\mathbb {R}}}_{\ {\cal O}{\cal O}'}^{\rho}:\!\!&=&\!\!\delta_{{\cal O}}({{\mathcal{E}}}_{{\cal O}'}^{\, \, \rho})-\delta_{{\cal O}'}({{\mathcal{E}}}_{{\cal O}}^{\, \, \rho})+{{\cal R}}_{\ {\cal O}{\cal O}'}^{\rho}+\ip(\ip C_{\ {\cal O}{\cal O}'}^{{\cal O}''}){{\mathcal{E}}}_{{\cal O}''}^{\, \, \rho},\\
\bar{\mathbb {Q}}_{{\cal O}{\cal O}'}:\!\!&=&\!\!\delta_{{\cal O}}(\bar{\mathcal{V}}_{{\cal O}'})-\delta_{{\cal O}'}(\bar{\mathcal{V}}_{{\cal O}})+{\bar{\cal Q}}_{{\cal O}{\cal O}'}+\ip(\ip C_{\ {\cal O}{\cal O}'}^{{\cal O}''})\bar{\mathcal{V}}_{{\cal O}''},\label{FSPhi}\\
{\mathbb {Q}}_{{\cal O}{\cal O}'}:\!\!&=&\!\!\delta_{{\cal O}}({\mathcal{V}}_{{\cal O}'})-\delta_{{\cal O}'}({\mathcal{V}}_{{\cal O}})+{{\cal Q}}_{ {\cal O}{\cal O}'}+\ip(\ip C_{\ {\cal O}{\cal O}'}^{{\cal O}''}){\mathcal{V}}_{{\cal O}''}. \label{FSPhibar}
\eeaa
The quadratic part  ${\cal F}_{\ {\cal O}{\cal O}'}^{\hat{\cal O}}$ of the generic field strengths ${\mathbb F}_{\ {\cal O}{\cal O}'}^{\hat{\cal O}}$ is given by
\beaa
\hat{{\cal F}}_{\ {\cal O}{\cal O}'}^{ \rho}\!\!\!&=&\!\!\!\!-\ip B_{{\cal O}}(-h)\hat{\phi}_{{\cal O}'}^{\, \, \rho}-\ip \hat{ \phi}_{{\cal O}}^{\, \, \rho}(+h) B_{{\cal O}'}-\ip \hat{B}_{{\cal O}}(-\hat{h})\hat{\phi}_{{\cal O}'}^{\, \, \rho}-\ip \hat{\phi}_{{\cal O}}^{\, \, \rho}(+\hat{h}) \hat{B}_{{\cal O}'}\nn\\
&&\hspace{1.1cm} -\ip B_{{\cal O}}^{\, \, [\tau\kappa]}(+\ip r'\sigma[\tau\kappa]_{\sigma}^{\,\,\rho})\hat{\phi}_{{\cal O}'}^{\, \, \sigma}-\ip \hat{\phi}_{{\cal O}}^{\, \, \sigma}(-\ip r'\sigma[\tau\kappa]_{\sigma}^{\,\,\rho}) B_{{\cal O}'}^{\, \, [\tau\kappa]}\nn\\
&&\hspace{1.1cm} -\ip B_{{\cal O}}^{\, \, (\tau\kappa)}(- r\sigma(\tau\kappa)_{\sigma}^{\,\,\rho})\hat{\phi}_{{\cal O}'}^{\, \, \sigma}-\ip \hat{\phi}_{{\cal O}}^{\, \, \sigma}(+ r\sigma(\tau\kappa)_{\sigma}^{\,\,\rho}) B_{{\cal O}'}^{\, \, (\tau\kappa)}\nn\\
&&\hspace{1.1cm}-\!\ip \bar{{\mathcal{E}}}_{{\cal O}}^{\ \rho}\!\left(\!-\frac{\sqrt{2}\ip s  \bar{q} g_{44}}{\ell R^2}\!\right)\!{{\mathcal{V}}}_{{\cal O}'}\!-\!\ip {{\mathcal{V}}}_{{\cal O}}\!\left(\!+\frac{\sqrt{2}\ip s \bar{q}
	g_{44}}{\ell R^2}\!\right)\!\bar{{\mathcal{E}}}_{{\cal O}'}^{\ \rho},\\
{\cal F}_{\ {\cal O}{\cal O}'}^{ \rho}\!\!\!&=&\!\!\!\!-\ip B_{{\cal O}}(+h)\phi_{{\cal O}'}^{\, \, \rho}-\ip \phi_{{\cal O}}^{\, \, \rho}(-h) B_{{\cal O}'}-\ip \hat{B}_{{\cal O}}(+\hat{h})\phi_{{\cal O}'}^{\, \, \rho}-\ip \phi_{{\cal O}}^{\, \, \rho}(-\hat{h}) \hat{B}_{{\cal O}'}\nn\\
&&\hspace{1.1cm} -\ip B_{{\cal O}}^{\, \, [\tau\kappa]}(+\ip r'\sigma[\tau\kappa]_{\sigma}^{\,\,\rho})\phi_{{\cal O}'}^{\, \, \sigma}-\ip \phi_{{\cal O}}^{\, \, \sigma}(-\ip r'\sigma[\tau\kappa]_{\sigma}^{\,\,\rho}) B_{{\cal O}'}^{\, \, [\tau\kappa]}\nn\\
&&\hspace{1.1cm} -\ip B_{{\cal O}}^{\, \, (\tau\kappa)}(+ r\sigma(\tau\kappa)_{\sigma}^{\,\,\rho})\phi_{{\cal O}'}^{\, \, \sigma}-\ip \phi_{{\cal O}}^{\, \, \sigma}(- r\sigma(\tau\kappa)_{\sigma}^{\,\,\rho}) B_{{\cal O}'}^{\, \, (\tau\kappa)}\nn\\
&&\hspace{1.1cm}-\!\ip {\mathcal{E}}_{{\cal O}}^{\ \rho}\!\left(\!-\frac{\sqrt{2}\ip \bar{s}{q}g_{44}}{\ell R^2}\!\right)\!\bar{\mathcal{V}}_{{\cal O}'}\!-\!\ip \bar{\mathcal{V}}_{{\cal O}}\!\left(\!+\frac{\sqrt{2}\ip \bar{s}{q}g_{44}}{\ell R^2}\!\right)\!{\mathcal{E}}_{{\cal O}'}^{\ \rho},
\eeaa
\beaa
{\cal F}_{{\cal O}{\cal O}'}\!\!\!&=&\!\!\!\!-\!\ip \hat{\phi}_{{\cal O}}^{\, \, \rho}\!\left(\!-\frac{12\epsilon \ell^2}{5h}g_{\rho\sigma}\!\right)\!\phi_{{\cal O}'}^{\, \, \sigma}\!-\ip\phi_{{\cal O}}^{\, \, \rho}\!\left(\!+\frac{12\epsilon \ell^2}{5h}g_{\rho\sigma}\!\right)\!\hat{\phi}_{{\cal O}'}^{\, \, \sigma}\!\nn\\
&&-\!\ip {{\mathcal{V}}}_{{\cal O}}\!\left(\!-\frac{12\epsilon q\bar{q}g_{44}}{5{h}R^2}\!\right)\!\bar{\mathcal{V}}_{{\cal O}'}\!-\!\ip \bar{\mathcal{V}}_{{\cal O}}\!\left(\!+\frac{12\epsilon q\bar{q}g_{44}}{5{h}R^2}\!\right)\!{{\mathcal{V}}}_{{\cal O}'},\label{FOO'}\\
\hat{{\cal F}}_{{\cal O}{\cal O}'}\!\!\!&=&\!\!\!\!-\!\ip \hat{\phi}_{{\cal O}}^{\, \, \rho}\!\left(\!-\frac{\epsilon \ell^2}{10\hat{h}}g_{\rho\sigma}\!\right)\!\phi_{{\cal O}'}^{\, \, \sigma}-\!\ip\phi_{{\cal O}}^{\, \, \rho}\!\left(\!+\frac{\epsilon \ell^2}{10\hat{h}}g_{\rho\sigma}\!\right)\!\hat{\phi}_{{\cal O}'}^{\, \, \sigma}\nn\\
&&-\!\ip \bar{{\mathcal{E}}}_{{\cal O}}^{\ \tau}\!\left(\!-\frac{ s\bar{s}g_{44}}{2\hat{h}R^2}g_{\tau\kappa}\!\right)\!{\mathcal{E}}_{{\cal O}'}^{\ \kappa}\!-\!\ip {\mathcal{E}}_{{\cal O}}^{\ \kappa}\!\left(\!+\frac{ s\bar{s}g_{44}}{2\hat{h}R^2}g_{\tau\kappa}\!\right)\!\bar{{\mathcal{E}}}_{{\cal O}'}^{\ \tau}\nn\\
&&-\!\ip {{\mathcal{V}}}_{{\cal O}}\!\left(\!+\frac{2\epsilon q\bar{q}g_{44}}{5\hat{h}R^2}\!\right)\!\bar{\mathcal{V}}_{{\cal O}'}\!-\!\ip \bar{\mathcal{V}}_{{\cal O}}\!\left(\!-\frac{2\epsilon q\bar{q}g_{44}}{5\hat{h}R^2}\!\right)\!{{\mathcal{V}}}_{{\cal O}'},\ \ \label{FhatOO'}\\
\!\!\!\!{\cal G}_{\ \ {\cal O}{\cal O}'}^{ [\tau\kappa]}\!\!\!&=&\!\!\!\!
-\ip B_{{\cal O}}^{\, \, [\rho\sigma]}(+\ip r'\left(\sigma[\rho\sigma]_{\gamma}^{\,\,\tau}\delta_{\delta}^{\,\,\kappa}+\sigma[\rho\sigma]_{\delta}^{\,\,\kappa}\delta_{\gamma}^{\,\,\tau}\right))B_{{\cal O}'}^{\, \, [\gamma\delta]}\nn\\
 &&\!\!\!-\ip B_{{\cal O}}^{\, \, (\rho\sigma)}\left(+ \frac{\ip r^2}{r'}\left(\sigma(\rho\sigma)_{\gamma}^{\,\,\tau}\delta_{\delta}^{\,\,\kappa}-\sigma(\rho\sigma)_{\delta}^{\,\,\kappa}\delta_{\gamma}^{\,\,\tau}\right)\right) B_{{\cal O}'}^{\, \, (\gamma\delta)}\ \ \\
&&\!\!\!-\!\ip \hat{\phi}_{{\cal O}}^{\, \, \gamma}\!\left(\!+\frac{\ip \epsilon \ell^2}{r'}(\sigma[\gamma\delta]_{\rho}^{\,\,\kappa} g^{\rho\tau})\!\right)\!\phi_{{\cal O}'}^{\, \, \delta}
\!-\ip\phi_{{\cal O}}^{\, \, \gamma}\!\left(\!+\frac{\ip \epsilon \ell^2}{2r'}(\sigma[\gamma\delta]_{\rho}^{\,\,\kappa} g^{\rho\tau})\!\right)\!\hat{\phi}_{{\cal O}'}^{\, \, \delta}\!\ \ \nn\\
&&\!\!\!-\!\ip \bar{{\mathcal{E}}}_{{\cal O}}^{\ \gamma}\!\left(\!+\frac{ \ip s\bar{s}g_{44}}{2r' R^2}(\sigma[\gamma\delta]_{\rho}^{\,\,\kappa} g^{\rho\tau})\!\right)\!{\mathcal{E}}_{{\cal O}'}^{\ \delta}\!-\!\ip {\mathcal{E}}_{{\cal O}}^{\ \gamma}\!\left(\!+\frac{\ip s\bar{s}g_{44}}{2r' R^2}(\sigma[\gamma\delta]_{\rho}^{\,\,\kappa} g^{\rho\tau})\!\right)\!\bar{{\mathcal{E}}}_{{\cal O}'}^{\ \delta},\ \ \ \nn\\
\!\!\!\!{\cal H}_{\ \ {\cal O}{\cal O}'}^{ (\tau\kappa)}\!\!\!&=&\!\!\!\!
-\ip B_{{\cal O}}^{\, \, [\rho\sigma]}(+\ip r'\left(\sigma[\rho\sigma]_{\gamma}^{\,\,\tau}\delta_{\delta}^{\,\,\kappa}+\sigma[\rho\sigma]_{\delta}^{\,\,\kappa}\delta_{\gamma}^{\,\,\tau}\right))B_{{\cal O}'}^{\, \, (\gamma\delta)}\hspace{2.5cm}\nn\\
&& -\ip B_{{\cal O}}^{\, \, (\gamma\delta)}(-\ip r'\left(\sigma[\rho\sigma]_{\gamma}^{\,\,\tau}\delta_{\delta}^{\,\,\kappa}+\sigma[\rho\sigma]_{\delta}^{\,\,\kappa}\delta_{\gamma}^{\,\,\tau}\right)) B_{{\cal O}'}^{\, \, (\rho\sigma)}\hspace{2.8cm}\\
&&-\!\ip \hat{\phi}_{{\cal O}}^{\, \, \gamma}\!\left(\!-\frac{ \epsilon \ell^2}{2r}
\left(\sigma(\gamma\delta)_{\rho}^{\,\,\kappa}g^{\rho\tau}
\right)\!\right)\!\phi_{{\cal O}'}^{\, \, \delta}-\ip\phi_{{\cal O}}^{\, \, \gamma}\!\left(\!+\frac{ \epsilon \ell^2}{2r}\left(\sigma(\gamma\delta)_{\rho}^{\,\,\kappa}g^{\rho\tau}\right)\!\right)\!\hat{\phi}_{{\cal O}'}^{\, \, \delta}\nn\\
&&\!\!-\!\ip \bar{{\mathcal{E}}}_{{\cal O}}^{\ \gamma}\!\left(\!-\frac{ s\bar{s}g_{44}}{2r R^2}\left(\sigma(\gamma\delta)_{\rho}^{\,\,\kappa}g^{\rho\tau}\right)\!\right)\!{\mathcal{E}}_{{\cal O}'}^{\ \delta}
\!-\!\ip {\mathcal{E}}_{{\cal O}}^{\ \delta}\!\left(\!+\frac{s\bar{s}g_{44}}{2r R^2}\left(\sigma(\gamma\delta)_{\rho}^{\,\,\kappa}g^{\rho\tau}\right)\!\right)\!\bar{{\mathcal{E}}}_{{\cal O}'}^{\ \gamma}.\nn\\
\!\!\bar{{\cal R}}_{\ {\cal O}{\cal O}'}^{\kappa}\!\!\!&=&\!\!\!\!-\ip \hat{B}_{{\cal O}}(-5\hat{h})\bar{{\mathcal{E}}}_{{\cal O}'}^{\ \kappa}-\ip \bar{{\mathcal{E}}}_{{\cal O}}^{\ \kappa}(+5\hat{h}) \hat{B}_{{\cal O}'}\nn\\
&&-\ip B_{{\cal O}}^{\, \, [\rho\sigma]}\left(+\ip r'\sigma[\rho\sigma]^{\ \kappa}_{\tau}\right)\bar{{\mathcal{E}}}_{{\cal O}'}^{\ \tau}-\ip \bar{{\mathcal{E}}}_{{\cal O}}^{\ \tau} \left(-\ip r'\sigma[\rho\sigma]^{\ \kappa}_{\tau}\right)B_{{\cal O}'}^{\, \, [\rho\sigma]}\nn\\
&&-\ip B_{{\cal O}}^{\, \, (\rho\sigma)}\left(- r\sigma(\rho\sigma)^{\ \kappa}_{\tau}\right)\bar{{\mathcal{E}}}_{{\cal O}'}^{\ \tau}-\ip \bar{{\mathcal{E}}}_{{\cal O}}^{\ \tau} \left(+ r \sigma(\rho\sigma)^{\ \kappa}_{\tau}\right)B_{{\cal O}'}^{\, \, (\rho\sigma)}\nn\\
&&\!\!\!-\!\ip \hat{\phi}_{{\cal O}}^{\, \, \kappa}\!\left(\!+\frac{\sqrt{2}\ip \epsilon \ell {q}}{{s}}\!\right)\!\bar{\mathcal{V}}_{{\cal O}'}
-\!\ip \!\bar{\mathcal{V}}_{{\cal O}}\left(\!-\frac{\sqrt{2}\ip \epsilon \ell {q}}{{s}}\!\right)\!\hat{\phi}_{{\cal O}'}^{\, \, \kappa},\\
\!\!{\cal R}_{\ {\cal O}{\cal O}'}^{\kappa}\!\!\!&=&\!\!\!\!-\ip \hat{B}_{{\cal O}}(+5\hat{h}){{\mathcal{E}}}_{{\cal O}'}^{\ \kappa}-\ip {{\mathcal{E}}}_{{\cal O}}^{\ \kappa}(-5\hat{h}) \hat{B}_{{\cal O}'}\nn\\
&&-\ip B_{{\cal O}}^{\, \, [\rho\sigma]}\left(+\ip r'\sigma[\rho\sigma]^{\ \kappa}_{\tau}\right){{\mathcal{E}}}_{{\cal O}'}^{\ \tau}-\ip {{\mathcal{E}}}_{{\cal O}}^{\ \tau}\left(-\ip r'\sigma[\rho\sigma]^{\ \kappa}_{\tau}\right)B_{{\cal O}'}^{\, \, [\rho\sigma]}\nn\\
&&-\ip B_{{\cal O}}^{\, \, (\rho\sigma)}\left(+ r\sigma(\rho\sigma)^{\ \kappa}_{\tau}\right){{\mathcal{E}}}_{{\cal O}'}^{\ \tau}-\ip {{\mathcal{E}}}_{{\cal O}}^{\ \tau} \left(- r\sigma(\rho\sigma)^{\ \kappa}_{\tau}\right)B_{{\cal O}'}^{\, \, (\rho\sigma)}\nn\\
&&\!\!\!-\!\ip {\phi}_{{\cal O}}^{\, \, \kappa}\!\left(\!+\frac{\sqrt{2}\ip \epsilon \ell \bar{q}}{\bar{s}}\!\right)\!{\mathcal{V}}_{{\cal O}'}
-\!\ip \!{\mathcal{V}}_{{\cal O}}\left(\!-\frac{\sqrt{2}\ip \epsilon \ell \bar{q}}{\bar{s}}\!\right)\!{\phi}_{{\cal O}'}^{\, \, \kappa},
\eeaa
\beaa
\!\!\bar{\cal Q}_{{\cal O}{\cal O}'}\!\!\!&=&\!\!\!\!-\ip {B}_{{\cal O}}(+{h})\bar{{\mathcal{V}}}_{{\cal O}'}-\ip \bar{{\mathcal{V}}}_{{\cal O}}(-{h}) {B}_{{\cal O}'}
-\ip \hat{B}_{{\cal O}}(-4\hat{h})\bar{{\mathcal{V}}}_{{\cal O}'}-\ip \bar{{\mathcal{V}}}_{{\cal O}}(+4\hat{h}) \hat{B}_{{\cal O}'}\nn\\
&&\!\!\!-\!\ip {\phi}_{{\cal O}}^{\, \, \tau}\!\left(\!-\frac{\sqrt{2}\ip  \ell {s}}{{q}}g_{\tau\kappa}\!\right)\!\bar{\mathcal{E}}_{{\cal O}'}^{\ \kappa}
-\!\ip \!\bar{\mathcal{E}}_{{\cal O}}^{\ \kappa}\left(\!+\frac{\sqrt{2}\ip  \ell {s}}{{q}}g_{\tau\kappa}\!\right)\!{\phi}_{{\cal O}'}^{\, \, \tau}.\label{SbarOO'}\\
\!\!{\cal Q}_{{\cal O}{\cal O}'}\!\!\!&=&\!\!\!\!-\ip {B}_{{\cal O}}(-{h}){{\mathcal{V}}}_{{\cal O}'}-\ip {{\mathcal{V}}}_{{\cal O}}(+{h}) {B}_{{\cal O}'}
-\ip \hat{B}_{{\cal O}}(+4\hat{h}){{\mathcal{V}}}_{{\cal O}'}-\ip {{\mathcal{V}}}_{{\cal O}}(-4\hat{h}) \hat{B}_{{\cal O}'}\nn\\
&&\!\!\!-\!\ip \hat{\phi}_{{\cal O}}^{\, \, \tau}\!\left(\!-\frac{\sqrt{2}\ip  \ell \bar{s}}{\bar{q}}g_{\tau\kappa}\!\right)\!{\mathcal{E}}_{{\cal O}'}^{\ \kappa}
-\!\ip \!{\mathcal{E}}_{{\cal O}}^{\ \kappa}\left(\!+\frac{\sqrt{2}\ip  \ell \bar{s}}{\bar{q}}g_{\tau\kappa}\!\right)\!\hat{\phi}_{{\cal O}'}^{\, \, \tau}\label{SOO'}.
\eeaa
Some terms ensuring tracelessness were added, inspired by the tracelessness of $T_{\tau\kappa}$. We can deduce the transformation properties of the fields from the quadratic terms (\ref{QuadTerms}) and the Ansatz (\ref{LeftModField}). First, dimensionless generators acting on fields associated with dimensionless generators remain as in \cite{U1I}:
\beaa
\,[ T_{\mu},hB_{{\cal O}}]_1&=&\!\left(\!-\frac{12\epsilon \ell}{5}g_{\mu\sigma}\!\right)\!\phi_{{\cal O}}^{\, \, \sigma},\hspace{.3cm}[\hat{T}_{\mu},hB_{{\cal O}}]_1\ =\ \!\left(\!+\frac{12\epsilon \ell}{5}g_{\mu\sigma}\!\right)\!\hat{\phi}_{{\cal O}}^{\, \, \sigma},\ \ \ \label{TmuBO}\\
\,[ T_{\mu},\hat{h}\hat{B}_{{\cal O}}]_1&=&\!\left(\!-\frac{\epsilon \ell}{10}g_{\mu\sigma}\!\right)\!\phi_{{\cal O}}^{\, \, \sigma},\hspace{.6cm}[\hat{T}_{\mu},\hat{h}\hat{B}_{{\cal O}}]_1\ =\ \!\left(\!+\frac{\epsilon \ell}{10}g_{\mu\sigma}\!\right)\!\hat{\phi}_{{\cal O}}^{\, \, \sigma},\label{TmuBhatO}\\
\,[T_{\mu},r'B_{{\cal O}}^{\, \, [\tau\kappa]}]_1&=&\!\left(\!+\frac{\ip \epsilon \ell}{2}(\sigma[\mu\delta]_{\rho}^{\,\,\kappa}g^{\rho\tau})\!\right)\!\phi_{{\cal O}}^{\, \, \delta},\nn\\
&&\!\!\![\hat{T}_{\mu},r'B_{{\cal O}}^{\, \, [\tau\kappa]}]_1\ =\ \!\!\left(\!+\frac{\ip \epsilon \ell}{2}(\sigma[\mu\delta]_{\rho}^{\,\,\kappa}g^{\rho\tau})\!\right)\!\hat{\phi}_{{\cal O}}^{\, \, \delta},\\
\,[T_{\mu},rB_{{\cal O}}^{\, \, (\tau\kappa)}]_1&=&\!\left(\!-\frac{\epsilon \ell}{2}
(\sigma(\mu\delta)_{\rho}^{\,\,\kappa}g^{\rho\tau})
\!\right)\!\phi_{{\cal O}}^{\, \, \delta},\nn\\
&&\!\!\![\bar{T}_{\mu},rB_{{\cal O}}^{\, \, (\tau\kappa)}]_1=\ \!\!\left(\!+\frac{ \epsilon \ell}{2}(\sigma(\mu\delta)_{\rho}^{\,\,\kappa}g^{\rho\tau})\!\right)\!\hat{\phi}_{{\cal O}}^{\, \, \delta},\ \ \ \\
\,[T_{\mu},\ell\,\phi_{{\cal O}}^{\, \, \tau}]&=&0,\hspace{2.9cm}[\hat{T}_{\mu},\ell\,\hat{\phi}_{{\cal O}}^{\, \, \tau}]_1\ =\ 0,\label{TmuPhiO}\\
\,[T_{\mu},\ell\,\hat{\phi}_{{\cal O}}^{\, \, \tau}]_1&=&\!h \delta_{\mu}^{\,\,\tau} B_{{\cal O}}+\hat{h} \delta_{\mu}^{\,\,\tau} \hat{B}_{{\cal O}}-\!\ip r'\sigma[\rho\sigma]_{\mu}^{\,\,\tau}B_{{\cal O}}^{\, \, [\rho\sigma]}+r\sigma(\rho\sigma)_{\mu}^{\,\,\tau}\!B_{{\cal O}}^{\, \, (\rho\sigma)}\!\!\!,\nn\\
\,[\hat{T}_{\mu},\ell\,\phi_{{\cal O}}^{\, \, \tau}]_1&=&\!\!\!\!\!-h \delta_{\mu}^{\,\,\tau} B_{{\cal O}}-\hat{h} \delta_{\mu}^{\,\,\tau} \hat{B}_{{\cal O}}-\!\ip r'\sigma[\rho\sigma]_{\mu}^{\,\,\tau}B_{{\cal O}}^{\, \, [\rho\sigma]}-r\sigma(\rho\sigma)_{\mu}^{\,\,\tau}\!B_{{\cal O}}^{\, \, (\rho\sigma)}\!\!\!,\hspace{1,5cm}\label{TmuPhibarO}\\
\,[{T_{[o]}},hB_{{\cal O}}]_1&=&0,\hspace{2.9cm}[{T_{[o]}},\hat{h}\hat{B}_{{\cal O}}]_1\ =\ 0,\\
\,[{T_{[o]}},r'B_{{\cal O}}^{\, \, [\tau\kappa]}]_1&=&0,\hspace{2.5cm}[{T_{[o]}},rB_{{\cal O}}^{\, \, (\tau\kappa)}]_1\ =\ 0,\\
\,[{T_{[o]}},\ell\,\phi_{{\cal O}}^{\, \, \tau}]_1&=&+\ell\,\phi_{{\cal O}}^{\, \, \tau},\hspace{2.0cm}[{T_{[o]}},\ell\,\hat{\phi}_{{\cal O}}^{\, \, \tau}]_1\ =-\ell\,\hat{\phi}_{{\cal O}}^{\, \, \tau},\\
\,[{\hat{T}_{[o]}},hB_{{\cal O}}]_1&=&0,\hspace{2.9cm}[{\hat{T}_{[o]}},\hat{h}\,\hat{B}_{{\cal O}}]_1\ =\ 0,\\
\!\!\![{\hat{T}_{[o]}},r'B_{{\cal O}}^{\, \, [\tau\kappa]}]_1&=&0,\hspace{2.5cm}[{\hat{T}_{[o]}},rB_{{\cal O}}^{\, \, (\tau\kappa)}]_1\ =\ 0,\\
\,[{\hat{T}_{[o]}},\ell\,\phi_{{\cal O}}^{\, \, \tau}]_1&=&+\ell\,\phi_{{\cal O}}^{\, \, \tau},\hspace{2.0cm}[{\hat{T}_{[o]}},\ell\,\hat{\phi}_{{\cal O}}^{\, \, \tau}]_1\ =-\ell\,\hat{\phi}_{{\cal O}}^{\, \, \tau},\\
\,[M_{\rho\sigma},hB_{{\cal O}}]_1&=&0,\hspace{2.8cm}[ M_{\rho\sigma},\hat{h}\hat{B}_{{\cal O}}]_1 =\ 0,\\
\!\!\![ M_{\rho\sigma},r'B_{{\cal O}}^{\, \, [\tau\kappa]}]_1&=&+\ip r'(\sigma[\rho\sigma]_{\gamma}^{\,\,\tau}\delta_{\delta}^{\,\,\kappa}+\sigma[\rho\sigma]_{\delta}^{\,\,\kappa}\delta_{\gamma}^{\tau})B_{{\cal O}}^{\, \, [\gamma\delta]},\\
\!\!\![ M_{\rho\sigma},rB_{{\cal O}}^{\, \, (\tau\kappa)}]_1&=&+\ip r(\sigma[\rho\sigma]_{\gamma}^{\,\,\tau}\delta_{\delta}^{\,\,\kappa}+\sigma[\rho\sigma]_{\delta}^{\,\,\kappa}\delta_{\gamma}^{\tau})B_{{\cal O}}^{\, \, (\gamma\delta)},\\
\,[ M_{\rho\sigma},\ell\,\phi_{{\cal O}}^{\, \, \tau}]_1&=&+\ip \ell\,\sigma[\rho\sigma]_{\gamma}^{\,\,\tau}\phi_{{\cal O}}^{\, \, \gamma},\hspace{0.7cm}
[ M_{\rho\sigma},\ell\,\hat{\phi}_{{\cal O}}^{\, \, \tau}]_1\ =\ +\ip \ell\,\sigma[\rho\sigma]_{\gamma}^{\,\,\tau}\hat{\phi}_{{\cal O}}^{\, \, \gamma},
\eeaa
\beaa
\!\![ T_{\rho\sigma},hB_{{\cal O}}]_1&=&0,\hspace{2.9cm}[T_{\rho\sigma},\hat{h}\hat{B}_{{\cal O}}]_1 =\ 0,\\
\,[T_{\rho\sigma},r'B_{{\cal O}}^{\, \, [\tau\kappa]}]_1&=&+\ip r(\sigma(\rho\sigma)_{\gamma}^{\,\,\tau}\delta_{\delta}^{\,\,\kappa}-\sigma(\rho\sigma)_{\delta}^{\,\,\kappa}\delta_{\gamma}^{\tau})B_{{\cal O}}^{\, \, (\gamma\delta)},\\
\!\![T_{\rho\sigma},rB_{{\cal O}}^{\, \, (\tau\kappa)}]_1&=&+\ip r'(\sigma(\rho\sigma)_{\gamma}^{\,\,\tau}\delta_{\delta}^{\,\,\kappa}-\sigma(\rho\sigma)_{\delta}^{\,\,\kappa}\delta_{\gamma}^{\,\,\tau})B_{{\cal O}}^{\, \, [\gamma\delta]},\\ 
\,[ T_{\rho\sigma},\ell\,\phi_{{\cal O}}^{\, \, \tau}]_1&=&+ \ell\,\sigma(\rho\sigma)_{\gamma}^{\,\,\tau}\phi_{{\cal O}}^{\, \, \gamma},\hspace{.9cm}
[ T_{\rho\sigma},\ell\,\hat{\phi}_{{\cal O}}^{\, \, \tau}]_1\ =\  -\ell\,\sigma(\rho\sigma)_{\gamma}^{\,\,\tau}\hat{\phi}_{{\cal O}}^{\, \, \gamma}.\hspace{1cm}
\eeaa
Dimensionless generators  acting on fields associated to dimension 1 generators:
\beaa
\,[{T_{[o]}},\bar{s}{{\mathcal{E}}}_{{\cal O}}^{\ \tau}]_1&=&0,\hspace{2.3cm}[{T_{[o]}},s\bar{{\mathcal{E}}}_{{\cal O}}^{\ \tau}]_1\ =\ 0,\hspace{1.5cm}\\
\,[{T_{[o]}},\bar{q}{{\mathcal{V}}}_{{\cal O}}]_1&=&-\bar{q}{{\mathcal{V}}}_{{\cal O}},\hspace{1.6cm}[{T_{[o]}},q\bar{{\mathcal{V}}}_{{\cal O}}]_1\ =\ +q\bar{{\mathcal{V}}}_{{\cal O}},\hspace{1.5cm}\\
\,[{\hat{T}_{[o]}},\bar{s}{{\mathcal{E}}}_{{\cal O}}^{\ \tau}]_1&=&+5\bar{s}{{\mathcal{E}}}_{{\cal O}}^{\ \tau},\hspace{1.4cm}[{\hat{T}_{[o]}},s\bar{{\mathcal{E}}}_{{\cal O}}^{\ \tau}]_1\ =\ -5s\bar{{\mathcal{E}}}_{{\cal O}}^{\ \tau},\hspace{1.8cm}\\
\,[{\hat{T}_{[o]}},\bar{q}{{\mathcal{V}}}_{{\cal O}}]_1&=&+4\bar{q}{{\mathcal{V}}}_{{\cal O}},\hspace{1.5cm}[{\hat{T}_{[o]}},q\bar{{\mathcal{V}}}_{{\cal O}}]_1\ =\ -4q\bar{{\mathcal{V}}}_{{\cal O}},\hspace{1.8cm}\\
\,[ M_{\rho\sigma},\bar{s}{{\mathcal{E}}}_{{\cal O}}^{\ \tau}]_1&=&+\ip \bar{s}\sigma[\rho\sigma]^{\ \tau}_{\kappa}{{\mathcal{E}}}_{{\cal O}}^{\ \kappa},\hspace{.1cm}
[ M_{\rho\sigma},s\bar{{\mathcal{E}}}_{{\cal O}}^{\ \tau}]_1\ =+\ip s\sigma[\rho\sigma]_{\kappa}^{\,\,\tau}\bar{{\mathcal{E}}}_{{\cal O}}^{\ \kappa},\ \\
\,[ M_{\rho\sigma},\bar{q}{{\mathcal{V}}}_{{\cal O}}]_1&=&0,\hspace{2.2cm}
[ M_{\rho\sigma},q\bar{{\mathcal{V}}}_{{\cal O}}]_1\ =\ 0,\hspace{2cm}\\
\,[ T_{\rho\sigma},\bar{s}{{\mathcal{E}}}_{{\cal O}}^{\ \tau}]_1&=&+ \bar{s}\sigma(\rho\sigma)_{\kappa}^{\,\,\tau}{{\mathcal{E}}}_{{\cal O}}^{\ \kappa},\hspace{.4cm}
[T_{\rho\sigma},s\bar{{\mathcal{E}}}_{{\cal O}}^{\ \tau}]_1\ =- s\sigma(\rho\sigma)_{\kappa}^{\,\,\tau}\bar{{\mathcal{E}}}_{{\cal O}}^{\ \kappa},\ \\
\,[ T_{\rho\sigma},\bar{q}{{\mathcal{V}}}_{{\cal O}}]_1&=&0,\hspace{2.4cm}
[ T_{\rho\sigma},q\bar{{\mathcal{V}}}_{{\cal O}}]_1\ =\ 0,\hspace{1.8cm}\\
\,[ {T}_{\rho},\bar{s}{{\mathcal{E}}}_{{\cal O}}^{\ \tau}]_1&=&0,\hspace{2.5cm}
\,[\hat{T}_{\rho},s\bar{{\mathcal{E}}}_{{\cal O}}^{\ \tau}]_1\ =\ 0,\hspace{1.8cm}\\
\,[\hat{T}_{\rho},\bar{s}{{\mathcal{E}}}_{{\cal O}}^{\ \tau}]_1&=&\sqrt{2}\ip \epsilon \bar{q}\delta_{\rho}^{\ \tau}{{\mathcal{V}}}_{{\cal O}},\hspace{0.8cm}
\,[{T}_{\rho},s\bar{{\mathcal{E}}}_{{\cal O}}^{\ \tau}]_1\ =\ \sqrt{2}\ip \epsilon {q}\delta_{\rho}^{\ \tau}\bar{{\mathcal{V}}}_{{\cal O}},\ \label{ThatrhoEkappaO}\\
\,[ {T}_{\rho},\bar{q}{{\mathcal{V}}}_{{\cal O}}]_1&=&-\sqrt{2}\ip  \bar{s}g_{\rho\kappa}{{\mathcal{E}}}_{{\cal O}}^{\ \kappa},\hspace{0.6cm}
\,[\hat{T}_{\rho},q\bar{{\mathcal{V}}}_{{\cal O}}]_1\ =-\sqrt{2}\ip  {s}g_{\rho\kappa}\bar{{\mathcal{E}}}_{{\cal O}}^{\ \kappa},\ \label{TrhoVO}\\
\,[ \hat{T}_{\rho},\bar{q}{{\mathcal{V}}}_{{\cal O}}]_1&=&0,\hspace{2.6cm}
[T_{\rho},q\bar{{\mathcal{V}}}_{{\cal O}}]_1\ =\ 0.\hspace{1.7cm}
\eeaa
Dimension 1 generators acting on fields associated to dimension 1 generators:
\beaa
\,[\hat{P}_{\tau},\bar{s}{{\mathcal{E}}}_{{\cal O}}^{\ \kappa}]_1&=&0,\hspace{2.6cm}[\hat{P}^{ *}_{\tau},s\bar{{\mathcal{E}}}_{{\cal O}}^{\ \kappa}]_1\ =\ 0,\\
\,[\hat{P}_{\tau},s\bar{{\mathcal{E}}}_{{\cal O}}^{\ \kappa}]_1&=&\!+5\hat{h} \delta^{\ \kappa}_{\tau} \hat{B}_{{\cal O}}-\!\ip r'\sigma[\rho\sigma]^{\ \kappa}_{\tau}B_{{\cal O}}^{\, \, [\rho\sigma]}+r\sigma(\rho\sigma)^{\ \kappa}_{\tau}\!B_{{\cal O}}^{\, \, (\rho\sigma)}\!\!\!,\nn\\
\,[\hat{P}^{*}_{\tau},\bar{s}{{\mathcal{E}}}_{{\cal O}}^{\ \kappa}]_1&=&\!-5\hat{h} \delta^{\ \kappa}_{\tau} \hat{B}_{{\cal O}}-\!\ip r'\sigma[\rho\sigma]^{\ \kappa}_{\tau}B_{{\cal O}}^{\, \, [\rho\sigma]}-r\sigma(\rho\sigma)^{\ \kappa}_{\tau}\!B_{{\cal O}}^{\, \, (\rho\sigma)},\\ 
\,[\hat{P}_{\tau},\bar{q}{{\mathcal{V}}}_{{\cal O}}]_1&=&0,\hspace{2.7cm}[\hat{P}_{\tau}^{ *},q\bar{{\mathcal{V}}}_{{\cal O}}]_1\ =\ 0,\\
\,[\hat{P}_{\tau},q\bar{{\mathcal{V}}}_{{\cal O}}]_1&=&\sqrt{2}\ip {\ell}g_{\tau\kappa}\phi_{{\cal O}}^{\ \kappa},\hspace{1.1cm}[\hat{P}_{\tau}^{ *},\bar{q}{{\mathcal{V}}}_{{\cal O}}]_1\ =\ \sqrt{2}\ip \ell g_{\tau\kappa} \hat{\phi}_{{\cal O}}^{\ \kappa},\\
\,[\Phi,\bar{s}{{\mathcal{E}}}_{{\cal O}}^{\ \kappa}]_1&=&0,\hspace{2.9cm}[\hat{\Phi},s\bar{{\mathcal{E}}}_{{\cal O}}^{\ \kappa}]_1\ =\ 0,\\
\,[\Phi,s\bar{{\mathcal{E}}}_{{\cal O}}^{\ \kappa}]_1&=& \!-\sqrt{2}\ip {\epsilon\ell}\hat{\phi}_{{\cal O}}^{\ \kappa},\hspace{1.4cm}[\bar{\Phi},\bar{s}{{\mathcal{E}}}_{{\cal O}}^{\ \kappa}]_1\ =-\sqrt{2}\ip {\epsilon\ell }{\phi}_{{\cal O}}^{\ \kappa},\ \\
\,[\Phi,\bar{q}{{\mathcal{V}}}_{{\cal O}}]_1&=&0,\hspace{2.9cm}[\bar{\Phi},q\bar{{\mathcal{V}}}_{{\cal O}}]_1\ =\ 0,\\
\,[\Phi,q\bar{{\mathcal{V}}}_{{\cal O}}]_1&=&-h\, B_{{\cal O}}+4\hat{h}\,\hat{B}_{{\cal O}},\hspace{0.7cm}[\bar{\Phi},\bar{q}{{\mathcal{V}}}_{{\cal O}}]_1\ =\ h\, B_{{\cal O}}-4\hat{h}\,\hat{B}_{{\cal O}}.
\eeaa 
Dimension 1 generators  acting on fields associated to dimensionless generators:
\beaa
\,[ \hat{P}_{\tau},hB_{{\cal O}}]_1\!&=&\!\!0,\hspace{2.4cm}[\hat{P}_{\tau}^{*},hB_{{\cal O}}]_1\ =\ \!0,\\
\,[\Phi,h{B}_{{\cal O}}]_1\!&=&\!\!\left(\!+\frac{12\epsilon  \bar{q}}{5R^2}g_{44}\!\right)\!{{\mathcal{V}}}_{{\cal O}},\hspace{.4cm}[\bar{\Phi},h{B}_{{\cal O}}]_1\ =\ \left(\!-\frac{12\epsilon  q}{5R^2}g_{44}\!\right)\!\bar{{\mathcal{V}}}_{{\cal O}},\\
\!\![\hat{P}_{\tau}\!,\hat{h}\hat{B}_{{\cal O}}]_1\!\!&=\!\!&\!\!\!\left(\!-\frac{  \bar{s}}{2R^2}g_{44}g_{\tau\kappa}\!\right)\!{{\mathcal{E}}}_{{\cal O}}^{\ \kappa},\hspace{.1cm}[ \hat{P}_{\tau}^{*}\!,\hat{h}\hat{B}_{{\cal O}}]_1\!=\!\left(\!\frac{  s}{2R^2}g_{44}g_{\tau\kappa}\!\right)\!\bar{{\mathcal{E}}}_{{\cal O}}^{\ \kappa},\ \ \\
\,[\Phi,\hat{h}\hat{B}_{{\cal O}}]_1&=&\!\left(\!-\frac{2\epsilon  \bar{q}}{5R^2}g_{44}\!\right)\!{{\mathcal{V}}}_{{\cal O}},\hspace{.5cm}[\bar{\Phi},\hat{h}\hat{B}_{{\cal O}}]_1=\left(\!+\frac{2\epsilon  q}{5R^2}g_{44}\!\right)\!\bar{{\mathcal{V}}}_{{\cal O}},
\eeaa
\beaa
\!\!\![\hat{P}_{\gamma},r'B_{{\cal O}}^{\,  [\tau\kappa]}]_1\!\!&\!=&\!\left(\!+\frac{\ip  \bar{s}g_{44}}{
	2R^2}(\sigma[\gamma\delta]^{\ \kappa}_{\rho}g^{\rho\tau})\!\right)\!{{\mathcal{E}}}_{{\cal O}}^{\ \delta},\nn\\
&&\!\!\![\hat{P}_{\gamma}^{*},r'B_{{\cal O}}^{\, \, [\tau\kappa]}]_1= \!\left(\!+\frac{\ip  sg_{44}}{
	2R^2}(\sigma[\gamma\delta]^{\ \kappa}_{\rho}g^{\rho\tau})\!\right)\!\bar{{\mathcal{E}}}_{{\cal O}}^{\ \delta},\ \\
\,[\Phi,r'B_{{\cal O}}^{\, \, [\tau\kappa]}]_1&=&0,\hspace{2.0cm}[\bar{\Phi},r'B_{{\cal O}}^{\, \, [\tau\kappa]}]_1\ =\ 0,\\
\!\!\![ \hat{P}_{\gamma}\!,rB_{{\cal O}}^{\,  (\tau\kappa)}]_1\!\!&\!=&\!\!\left(\!-\frac{\bar{s}g_{44}}{2R^2}(\sigma(\gamma\delta)^{\ \kappa}_{\rho}g^{\rho\tau})\!\right)\!{{\mathcal{E}}}_{{\cal O}}^{\ \delta},\nn\\
&&\,[\hat{P}_{\gamma}^{*},rB_{{\cal O}}^{\, \, (\tau\kappa)}]_1=\! \left(\!+\frac{sg_{44}}{
	2R^2}(\sigma(\gamma\delta)^{\ \kappa}_{\rho}g^{\rho\tau})\!\right)\!\bar{{\mathcal{E}}}_{{\cal O}}^{\ \delta},\\
\,[\Phi,rB_{{\cal O}}^{\, \, (\tau\kappa)}]_1&=&0,\hspace{2.0cm}[\bar{\Phi},rB_{{\cal O}}^{\, \, (\tau\kappa)}]_1\ =\ 0,\\
\,[\hat{P}_{\tau},\ell\,{\phi}_{{\cal O}}^{\, \, \rho}]_1\!&\!=\!&\!\!0,\hspace{2.3cm}[ \hat{P}_{\tau}^{ *},\ell\,\hat{\Phi}_{{\cal O}}^{\, \, \rho}]_1\ =\ \!0,
\eeaa
\beaa
\,[ \hat{P}_{\tau},\ell\,\hat{\phi}_{{\cal O}}^{\, \, \rho}]_1\!\!&=\!&\!\!\!\!\!\left(\!-\frac{\sqrt{2}\ip  \bar{q}}{R^2}g_{44}\delta_{\tau}^{\ \rho}\!\right)\!\!{{\mathcal{V}}}_{{\cal O}},\, \![\hat{P}_{\tau}^{*}\!,\ell\,{\phi}_{{\cal O}}^{\, \, \rho}]_1=\!\left(\!-\frac{\sqrt{2}\ip  {q}}{R^2}g_{44}\delta_{\tau}^{\ \rho}\!\right)\!\!\bar{{\mathcal{V}}}_{{\cal O}},\ \ \\
\,[\Phi,\ell\,{\phi}_{{\cal O}}^{\, \, \rho}]_1\!\!&=\!&\!\!\!\!\!\left(\!\frac{\sqrt{2}\ip  \bar{s}}{R^2}g_{44}\!\right)\!{{\mathcal{E}}}_{{\cal O}}^{\ \rho},\hspace{.7cm} [ \bar{\Phi},\ell\,\hat{\phi}_{{\cal O}}^{\, \, \rho}]_1=\left(\!\frac{\sqrt{2}\ip  {s}}{R^2}g_{44}\!\right)\!\bar{{\mathcal{E}}}_{{\cal O}}^{\ \rho},\ \ \\
\,[\Phi,\ell\,\hat{\phi}_{{\cal O}}^{\, \, \rho}]_1\!&\!=\!&\!\!0,\hspace{2.6cm}[ \bar{\Phi},\ell\,{\phi}_{{\cal O}}^{\, \, \rho}]_1\ =\ \!0.
\eeaa
While all actions of dimension 1 generators on fields associated with dimensionless generators vanish as $R\rightarrow\infty$ in the contraction procedure, these dimensionful couplings may have a connection to gravity coupling, which, although very small, does not completely vanish. The key point is that even though the standard contraction procedure leads to the Poincaré algebra, we might be moving towards a broken symmetry scenario where an effective global Poincaré-like symmetry emerges only when gravitational effects are neglected. The actual process of transitioning from a homogeneous symmetry group NPE $\simeq{\mathfrak{su}}(t_3,s_3)$ with $t_3+s_3=6$ to an inhomogeneous CPE $\simeq{\mathfrak{iu}}(t_2,s_2)$ with $t_2+s_2=5$ will be discussed further below.

Let's consider ${\cal O}$ and ${\cal O'}$ as scalars (for instance, ${\cal O},{\cal O'}\in\{\bar{\Phi},\Phi, T_{[o]}, \hat{T}_{[o]}\}$) and contemplate a Lagrangian {\bf quadratic in field strengths}. The terms in the last lines of (\ref{FOO'}) and (\ref{FhatOO'}) in the quadratic field strength expressions could help us understand the origin of the quartic complex scalar self-interaction terms for $\bar{{\mathcal{V}}}_{{\cal O}}$ and ${{\mathcal{V}}}_{{\cal O}}$. Additionally, the terms in the first lines of (\ref{SbarOO'}) and (\ref{SOO'}) might lead to quartic couplings between $\hat{B}_{{\cal O}}$ and ${B}_{{\cal O}}$. Furthermore, the terms linear in the fields in (\ref{FSTo}--\ref{FSThato}) and (\ref{FSPhi}--\ref{FSPhibar}) could give rise to mass-like terms when $C^{{\cal O}''}_{\ {\cal OO}'}\neq 0$. Note that the NPE, with a Lagrangian quadratic in field strengths, introduces this landscape of interactions and might contain the seeds for its own demise through symmetry breaking.

Let $(x)$ denote the coordinates in the adopted differential representation. We consider symmetry transformations that depend on $(x)$. We adopt the following transformation parameters: $\varepsilon(x)$, $\hat{\varepsilon}(x)$, $\hat{\varepsilon}^{\nu}(x)$, ${\varepsilon}^{\nu}(x)$, $\varepsilon^{[\tau\kappa]}(x)$, $\varepsilon^{(\tau\kappa)}(x)$, $\bar{\chi}_\nu(x)$, ${\chi}_\nu(x)$, $\bar{\chi}(x)$, and ${\chi}(x)$, associated respectively with the generators ${T_{[o]}}$, $\hat{T}_{[o]}$, $T_{\nu}$, $\hat{T}_{\nu}$, $M_{\tau\kappa}$, $T_{\tau\kappa}$, $\hat{P}^{\nu}$, $\hat{P}^{\nu *}$, ${\Phi}$, $\bar{\Phi}$. The transformations of fields and field strengths were stated in (\ref{NewFOTField}--\ref{NewFOTFieldSt}).

We now embark on the construction of a locally invariant Lagrangian utilizing Casimirs in the dual relationship between generators and homogeneously transforming field strengths (\ref{FieldStrengthsOp}), (\ref{FieldStrengthsOpPoin}). Local invariance, as well as the potential triggering of spontaneous symmetry breaking, motivates the construction of a Lagrangian quadratic in field strengths.

The field strengths ${\mathbb{F}}_{\ {\cal O}{\cal O}'}^{\hat{\cal O}}$ have an upper index $\hat{\cal O}$ that transforms as in (\ref{NewFOTField}), and its basis is the interaction basis, hence involving coupling constants. The lower indices ${\cal O}{\cal O}'$ are associated with covariant derivatives and are not in the interaction basis.

To construct an invariant, we contract the upper indices $\hat{\cal O}, \hat{\cal O}'$ using, in addition to the Minkowski metric for raising or lowering indices, the coefficients of $M_1^{-1}$ in (\ref{CasimirPEx}--\ref{CasimirPExM1}), completing a matrix ${\bf M_1}^{-1}_{\hat{\cal O}\hat{\cal O}'}$. For contracting lower indices ${\cal O},{\cal O}''$, we use the Minkowski metric for lowering or raising Lorentz indices, along with coefficients of $M_i$ in (\ref{CasimirPExM1}) or (\ref{CasimirPExM2}), completing a matrix ${\bf M_i}^{{\cal O}{\cal O}''}$. This depends on whether we decide to constrain the covariant derivatives for all the NPE generators or, after contraction, just those associated with the invariant subalgebra generated by dimension one generators.

The dual or adjoint basis $v^{\cal O}={\bf M_1}^{{\cal O}{\cal O}''}\!{\cal O}''$ transforms as $[\hat{\cal O},v^{\cal O}]=\!-\ip C_{\ \hat{\cal O}\tilde{\cal O}}^{{\cal O}}v^{\tilde{\cal O}}$ with the adjoint representation, similar to the field strength. The invariance of $v^{\cal O}{\bf M_1}^{-1}_{{\cal O}{\cal O}'}v^{{\cal O}'}$ implies $C^{\tilde{\cal O}}_{\ \hat{\cal O}{\cal O}}{\bf M_1}^{-1}_{\tilde{\cal O}{\cal O}'} =-{\bf M_1}^{-1}_{{\cal O}\tilde{\cal O}}C^{\tilde{\cal O}}_{\ \hat{\cal O}{\cal O}'}$, resulting in a totally antisymmetric array $C_{{\cal O}\hat{\cal O}{\cal O}'}$.

To ensure a quartic self--interaction potential bounded from below for scalar fields ${{\mathcal{V}}}_{{\cal O}}, \bar{{\mathcal{V}}}_{{\cal O}}$ (for ${\cal O}$ being a scalar such as $\Phi, \bar{\Phi}$) originating from the quadratic terms in (\ref{FSTo}--\ref{FSThato}) and (\ref{FOO'}-\ref{FhatOO'}), we select the overall sign of the Lagrangian quadratic in field strengths.

Utilizing the dual counterpart formalism, we can build a locally invariant quadratic Lagrangian ${{\mathbb L}}_i$ depending on the matrix ${\bf M_i}^{{\cal O}{\cal O}''}$ used to raise indices:
\beaa
\!\!\!\!\!\!4R^4{{\mathbb L}}_i\!=\!\ip\!\lambda^{\hat{\cal O}}{\mathbb {F}}_{\ {\cal O}{\cal O}'}^{\hat{\cal O}}{\bf M_1}^{-1}_{\hat{\cal O}\hat{\cal O}'}{\bf M_i}^{{\cal O}{\cal O}''}\!{\bf M_i}^{{\cal O}'{\cal O}'''}\!\ip\!\lambda^{\hat{\cal O}'}\!{\mathbb {F}}_{\ {\cal O}''{\cal O}'''}^{\hat{\cal O}'}\!=\!-\lambda^{\hat{\cal O}}{\mathbb {F}}_{\ {\cal O}{\cal O}'}^{\hat{\cal O}}\lambda_{\hat{\cal O}}{\mathbb {F}}^{\ {\cal O}{\cal O}'}_{\hat{\cal O}}\,&&\\
\!\!\!=\!\ip\![s{{\bar{{\mathbb R}}}}^{\ {\cal O}{\cal O}'}_{ \kappa}\,\! \bar{s}{{\mathbb R}}^{\ {\cal O}{\cal O}'}_{ \kappa}\,\!\! q{{\bar{{\mathbb Q}}}}^{{\cal O}{\cal O}'}\,\!\! \bar{q}{{\mathbb Q}}^{{\cal O}{\cal O}'}\,  \ell{{\bar{{\mathbb F}}}}^{\ {\cal O}{\cal O}'}_{\kappa}\,\! \ell{{\mathbb F}}^{\ {\cal O}{\cal O}'}_{\kappa}\,\! r'{{\mathbb G}}^{\ {\cal O}{\cal O}'}_{[\rho\tau]}\, r{{\mathbb H}}^{\ {\cal O}{\cal O}'}_{(\rho\tau)}\,\! h{{\mathbb F}}^{{\cal O}{\cal O}'}\! \,\hat{h}{{\hat{{\mathbb F}}}}^{{\cal O}{\cal O}'}]M_1^{-1}\!\cdot\!\!\!\!\!\!\!\!\!\!&&\nn\\
\!\!\! \! \ip[\bar{s}{{\mathbb R}}_{\ {\cal O}{\cal O}'}^{ \kappa}\ s{{\bar{{\mathbb R}}}}_{\ {\cal O}{\cal O}'}^{\kappa}\ \bar{q}{{\mathbb Q}}_{{\cal O}{\cal O}'}\ \,q{{\bar{{\mathbb Q}}}}_{{\cal O}{\cal O}'} \,\ell{{\mathbb F}}_{\ {\cal O}{\cal O}'}^{\kappa}\, \ell{{\bar{{\mathbb F}}}}_{\ {\cal O}{\cal O}'}^{\kappa}\, r'{{\mathbb G}}_{\ {\cal O}{\cal O}'}^{[\rho\tau]}\, r{{\mathbb H}}_{\ {\cal O}{\cal O}'}^{(\rho\tau)}\, h{{\mathbb F}}_{{\cal O}{\cal O}'}\,\bar{h}{{\hat{{\mathbb F}}}}_{{\cal O}{\cal O}'}]^{tr}\!\!\!\!\!\!\!&&\nn\\
= \left(-\frac{s\,\bar{s}\,g_{44}}{2R^2}\{ {{\bar{{\mathbb R}}}}^{\ {\cal O}{\cal O'}}_{ \kappa},{{\mathbb R}}_{\ {\cal O}{\cal O'}}^{ \kappa}\}
-\frac{q\,\bar{q}\,\epsilon\,g_{44}}{2\, R^2}\{ {{\bar{{\mathbb Q}}}}^{{\cal O}{\cal O'}},{{\mathbb Q}}_{{\cal O}{\cal O'}}\}\right.\hspace{2.5cm}&&\nn\\
-\frac{\ell^2\epsilon}{2}\{ {{\bar{{\mathbb F}}}}^{\ {\cal O}{\cal O'}}_{\kappa},{{\mathbb F}}_{\ {\cal O}{\cal O'}}^{\kappa}\}-{r'^2}{{\mathbb G}}^{\ {\cal O}{\cal O'}}_{[\rho\tau]}{{\mathbb G}}_{\ {\cal O}{\cal O'}}^{[\rho\tau]}- 
{r^2}{{\mathbb H}}^{\ {\cal O}{\cal O'}} _{ (\rho\tau)}{{\mathbb H}}_{\ {\cal O}{\cal O'}}^{ (\rho\tau)}\hspace{2cm}&&\nn\\
\left.- \frac{5h^2}{24}{{\mathbb F}}^{{\cal O}{\cal O'}}{{\mathbb F}}_{{\cal O}{\cal O'}}-   {5\hat{h}^2}{{\hat{{\mathbb F}}}}^{{\cal O}{\cal O'}}{\hat{{\mathbb F}}}_{{\cal O}{\cal O'}}\right),\hspace{3.5cm}\label{Li}&&
\eeaa
where $g_{44}, \epsilon \in \{1, -1\}$ will be assigned so that the action delivers the proper dynamics with the absence of ghosts. Although there is an appealing resemblance between the squared field strengths associated with dimensionless fields ${{\mathcal{E}}}_{{\cal O}}^{\ \tau}, \bar{{\mathcal{E}}}_{{\cal O}}^{\ \tau}, {{\mathcal{V}}}_{{\cal O}}, \bar{{\mathcal{V}}}_{{\cal O}}$ and the classical term corresponding to gravitation, with its dimensionful coupling, there are extra dimensionful contributions from the ${\bf M_i}$ arrays rising indices ${\cal O}, {\cal O'}$ in $\{ \hat{P}^{\nu *}, \hat{P}^{\nu}, \bar{\Phi}, \Phi\}$, each giving a factor of $g_{44}^{-1}R^2$.

Explorations on equations of motion and conserved currents goes along the same lines as with the LE in \cite{U1I}, so we devote our attention to model consistency. 

For scalars ${\cal O}$ and ${\cal O'}$, note that (\ref{TmuBO}--\ref{TmuBhatO}) and (\ref{TmuPhiO}) transform the scalar Higgs-like fields $B_{\cal O}$ and $\hat{B}_{\cal O}$ into vector fields ${\Phi}_{\cal O}^{\ \mu}$ and $\hat{\Phi}_{\cal O}^{\ \mu}$ (where $\mu$ is a Lorentz index), and vice versa. Similarly, with (\ref{ThatrhoEkappaO}) and (\ref{TrhoVO}), we transform the scalar Higgs-like fields $\bar{{\mathcal{V}}}_{{\cal O}}$ and ${{\mathcal{V}}}_{{\cal O}}$ into vector fields $\bar{{\mathcal{E}}}_{{\cal O}}^{\ \kappa}$ and ${{\mathcal{E}}}_{{\cal O}}^{\ \kappa}$, and vice versa. Additionally, the dimension--1 generators mix between $B_{\cal O}$, $\hat{B}_{\cal O}$, ${\Phi}_{\cal O}^{\ \mu}$, $\hat{\Phi}_{\cal O}^{\ \mu}$ and $\bar{{\mathcal{V}}}_{{\cal O}}$, ${{\mathcal{V}}}_{{\cal O}}$, $\bar{{\mathcal{E}}}_{{\cal O}}^{\ \kappa}$, ${{\mathcal{E}}}_{{\cal O}}^{\ \kappa}$. Such transformations seem to embody the leitmotif of mixing Higgs and gauge fields, as motivated in \cite{U1I}, in a rather unexpected way, beyond Kaluza--Klein constructions. The index $\mu$ in these non-scalar fields is a Lorentz index and is not associated with a covariant derivative (with an Einstein index) close to a translation.

To verify the correctness of the overall sign in the Lagrangian we study the terms giving rise to a quartic self--interaction potential for fields $\bar{{\mathcal{V}}}_{\bar{ \Phi}}, {{\mathcal{V}}}_{\Phi}, \bar{{\mathcal{V}}}_{\Phi}, {{\mathcal{V}}}_{\bar{ \Phi}}$:
\beaa
\ip\lambda^{T_{[o]}}{\cal F}_{{\cal O}{\cal O'}}M_1^{{\cal O}{\cal O''}}M_1^{{\cal O'}{\cal O'''}}{M_1}_{T_{[o]}T_{[o]}}^{-1}\ip\lambda^{T_{[o]}}{\cal F}_{{\cal O''}{\cal O'''}}/(4R^4)=\nn\\
=\cdots-\frac{(\lambda^{T_{[o]}})^2}{4R^4}\left(\!-\frac{12\epsilon q\bar{q}g_{44}}{5hR^2}\right)^2(\bar{{\mathcal{V}}}_{\bar{ \Phi}}{{\mathcal{V}}}_{\Phi}- \bar{{\mathcal{V}}}_{\Phi}{{\mathcal{V}}}_{\bar{ \Phi}})^2\left(\frac{2R^2}{\epsilon g_{44}}\right)^2\!\left(\frac{5}{24}\right).
\eeaa
This has the correct sign for a potential bounded from below (Lagrangian = Kinetic minus potential terms). A similar contribution comes from the field strength part $\hat{{\cal F}}_{{\cal O}{\cal O'}}$. Due to this interaction the complex scalar pairs $\bar{{\mathcal{V}}}_{\bar{ \Phi}}, {{\mathcal{V}}}_{\Phi}$ and $\bar{{\mathcal{V}}}_{\Phi}, {{\mathcal{V}}}_{\bar{ \Phi}}$ tend to have everywhere identical norms $\bar{{\mathcal{V}}}_{\bar{ \Phi}}{{\mathcal{V}}}_{\Phi},  \bar{{\mathcal{V}}}_{\Phi}{{\mathcal{V}}}_{\bar{ \Phi}}$.

In our exploration, three tasks remain to be accomplished. First, we need to verify the consistency of the novel non-commutativity of covariant derivatives for the NPE. This involves ensuring the consistency of kinetic-like terms associated with the translation generators, with the absence of ghost contributions for scalar and vector gauge fields. Additionally, we need to verify the correct sign for mass terms, as choices with a wrong mass sign might lead to instabilities, possibly triggering spontaneous symmetry--breaking mechanisms.

The second task is to explore possible changes in field variables that bring us closer to standard interactions with gravity. This exploration may also reveal potential actions leading to spontaneous symmetry--breaking mechanisms.

The third and more elusive task involves exploring scenarios where complex space-time coordinates can evolve into a real 4--dimensional space--time manifold combined with some internal spaces and broken leftovers.

\section{Consistency of Kinetic and mass terms}
Let's examine the consistency of the dynamics for pure scalar fields and pure vector fields, whether with a Lorentz index or an Einstein index. We begin with the signs of the kinetic terms associated with the covariant derivatives of the would-be translation generators $\hat{P}_\mu, \hat{P}^*_\mu$. The contributions from $-\ip\delta_{\hat{P}_\mu}, -\ip\delta_{\hat{P}^*_\mu}$ include both $-\ip\partial_\mu$ terms after contraction in 10D. Consider the scalar fields $B_{\cal O}, \hat{B}_{\cal O}, {{\mathcal{V}}}_{{\cal O}}, \bar{{\mathcal{V}}}_{{\cal O}}$ with ${\cal O}\in \{T_{[o]},\hat{T}_{[o]},\Phi,\bar{\Phi}\}$, which have the following forms:
  \beaa
   \!\!-h^2{\bf M}_1^{{\cal O}{\cal O}''}{\bf M}_1^{{\cal O}'{\cal O}'''}{{\bf M}_1}^{-1}_{T_{[o]}T_{[o]}}(\delta_{{\cal O}}({B}_{{\cal O}'})\!-\!\delta_{{\cal O}'}({B}_{{\cal O}}))(\delta_{{\cal O}''}({B}_{{\cal O}'''})\!-\!\delta_{{\cal O}'''}({B}_{{\cal O}''}))/(\!4R^4\!)\!\!\!\nn \\
   =\dots-\frac{4h^2}{4R^4}\!\left(\frac{2R^2 g^{\mu\nu}}{g_{44}}\right)\!\left(\frac{24}{5}\right)\!\left(\frac{5}{24}\right)
   (\delta_{\hat{P}_\mu}{B}_{T_{[o]}}\!-\!\delta_{T_{[o]}}{B}_{\hat{P}_\mu})(\delta_{\hat{P}^*_\nu}{B}_{T_{[o]}}\!-\!\delta_{T_{[o]}}{B}_{\hat{P}^*_\nu})\!\!\nn \\
    \ -\frac{4h^2}{4R^4}\!\left(\frac{2R^2 g^{\mu\nu}}{g_{44}}\right)\!\left(\frac{1}{5}\right)\!\left(\frac{5}{24}\right)
   (\delta_{\hat{P}_\mu}{B}_{\hat{T}_{[o]}}\!-\!\delta_{\hat{T}_{[o]}}{B}_{\hat{P}_\mu})(\delta_{\hat{P}^*_\nu}{B}_{\hat{T}_{[o]}}\!-\!\delta_{\hat{T}_{[o]}}{B}_{\hat{P}^*_\nu})\!\!\nn \\
 \!\!-\frac{2h^2}{4R^4}\!\left(\frac{2R^2 g^{\mu\nu}}{g_{44}}\right)\!\left(\frac{2R^2}{\epsilon g_{44}}\right)\!\left(\frac{5}{24}\right)\left\{
(\delta_{\hat{P}_\mu}{B}_{\Phi}\!-\!\delta_{\Phi}{B}_{\hat{P}_\mu})(\delta_{\hat{P}^*_\nu}{B}_{\bar{\Phi}}\!-\!\delta_{\bar{\Phi}}{B}_{\hat{P}^*_\nu})+(\Phi\!\leftrightarrow\!\bar{\Phi})\right\}\!,\!\!\!\!\!\nn 
\eeaa
\beaa
\!\!-\hat{h}^2{\bf M}_1^{{\cal O}{\cal O}''}{\bf M}_1^{{\cal O}'{\cal O}'''}{{\bf M}_1}^{-1}_{\hat{T}_{[o]}\hat{T}_{[o]}}(\delta_{{\cal O}}(\hat{B}_{{\cal O}'})\!-\!\delta_{{\cal O}'}(\hat{B}_{{\cal O}}))(\delta_{{\cal O}''}(\hat{B}_{{\cal O}'''})\!-\!\delta_{{\cal O}'''}(\hat{B}_{{\cal O}''}))/(\!4R^4\!)\!\!\!\nn \\
=\cdots-\frac{4\hat{h}^2}{4R^4}\!\left(\frac{2R^2 g^{\mu\nu}}{g_{44}}\right)\!\left(\frac{24}{5}\right)\!\left(\frac{5}{1}\right)
(\delta_{\hat{P}_\mu}\hat{B}_{T_{[o]}}\!-\!\delta_{T_{[o]}}\hat{B}_{\hat{P}_\mu})(\delta_{\hat{P}^*_\nu}\hat{B}_{T_{[o]}}\!-\!\delta_{T_{[o]}}\hat{B}_{\hat{P}^*_\nu})\!\!\nn \\
\ -\frac{4\hat{h}^2}{4R^4}\!\left(\frac{2R^2 g^{\mu\nu}}{g_{44}}\right)\!\left(\frac{1}{5}\right)\!\left(\frac{5}{1}\right)
(\delta_{\hat{P}_\mu}\hat{B}_{\hat{T}_{[o]}}\!-\!\delta_{\hat{T}_{[o]}}\hat{B}_{\hat{P}_\mu})(\delta_{\hat{P}^*_\nu}\hat{B}_{\hat{T}_{[o]}}\!-\!\delta_{\hat{T}_{[o]}}\hat{B}_{\hat{P}^*_\nu})\!\!\nn \\
\!\!-\frac{2\hat{h}^2}{4R^4}\!\left(\frac{2R^2 g^{\mu\nu}}{g_{44}}\right)\!\left(\frac{2R^2}{\epsilon g_{44}}\right)\!\left(\frac{5}{1}\right)\left\{
(\delta_{\hat{P}_\mu}\hat{B}_{\Phi}\!-\!\delta_{\Phi}\hat{B}_{\hat{P}_\mu})(\delta_{\hat{P}^*_\nu}\hat{B}_{\bar{\Phi}}\!-\!\delta_{\bar{\Phi}}\hat{B}_{\hat{P}^*_\nu})+(\Phi\!\leftrightarrow\!\bar{\Phi})\right\}\!,\!\!\!\!\!\nn
\eeaa
\beaa 
\!\!-q\bar{q}{\bf M}_1^{{\cal O}{\cal O}''}{\bf M}_1^{{\cal O}'{\cal O}'''}{{\bf M}_1}^{-1}_{\Phi\bar{ \Phi}}(\delta_{{\cal O}}(\bar{{\mathcal{V}}}_{{\cal O}'})\!-\!\delta_{{\cal O}'}(\bar{{\mathcal{V}}}_{{\cal O}}))(\delta_{{\cal O}''}({{\mathcal{V}}}_{{\cal O}'''})\!-\!\delta_{{\cal O}'''}({{\mathcal{V}}}_{{\cal O}''}))/(4R^4)\nn \\
=\cdots-\frac{2q\bar{q}}{4R^4}\!\left(\!\frac{2R^2 g^{\mu\nu}}{g_{44}\!}\right)\!\left(\frac{24}{5}\right)\!\left(\frac{\epsilon g_{44}}{2R^2}\right)\left\{
(\delta_{\hat{P}_\mu}\bar{{\mathcal{V}}}_{T_{[o]}}\!-\!\delta_{T_{[o]}}\bar{{\mathcal{V}}}_{\hat{P}_\mu})(\delta_{\hat{P}^*_\nu}{{\mathcal{V}}}_{T_{[o]}}\!-\!\delta_{T_{[o]}}{{\mathcal{V}}}_{\hat{P}^*_\nu})\right.\nn\\
\left.+(\bar{{\mathcal{V}}}_{T_{[o]}}\leftrightarrow {{\mathcal{V}}}_{T_{[o]}})\right\}\!\!\nn \\
-\frac{2q\bar{q}}{4R^4}\!\left(\frac{2R^2 g^{\mu\nu}}{g_{44}}\right)\!\left(\frac{1}{5}\right)\!\left(\frac{\epsilon g_{44}}{2R^2}\right)\left\{
(\delta_{\hat{P}_\mu}\bar{{\mathcal{V}}}_{\hat{T}_{[o]}}\!-\!\delta_{\hat{T}_{[o]}}\bar{{\mathcal{V}}}_{\hat{P}_\mu})(\delta_{\hat{P}^*_\nu}{{\mathcal{V}}}_{\hat{T}_{[o]}}\!-\!\delta_{\hat{T}_{[o]}}{{\mathcal{V}}}_{\hat{P}^*_\nu})\right.\nn\\
\left.+(\bar{{\mathcal{V}}}_{\hat{T}_{[o]}}\leftrightarrow {{\mathcal{V}}}_{\hat{T}_{[o]}})\right\}\!\!\nn \\
\!\!\!-\frac{2q\bar{q}}{4R^4}\!\!\left(\!\frac{2R^2 g^{\mu\nu}}{g_{44}\!}\right)\!\!\left(\frac{2R^2}{\epsilon g_{44}}\right)\!\!\left(\frac{\epsilon g_{44}}{2R^2}\right)\!\left\{
(\delta_{\hat{P}_\mu}\bar{{\mathcal{V}}}_{\Phi}\!-\!\delta_{\Phi}\bar{{\mathcal{V}}}_{\hat{P}_\mu})(\delta_{\hat{P}^*_\nu}{{\mathcal{V}}}_{\bar{\Phi}}\!-\!\delta_{\bar{\Phi}}{{\mathcal{V}}}_{\hat{P}^*_\nu})\!+\!({\hat{P}_\mu}\!\leftrightarrow\!{\hat{P}^*_\nu})\right\}\!\!\!\!\!\!\nn\\
\!\!\!-\frac{2q\bar{q}}{4R^4}\!\left(\!\frac{2R^2 g^{\mu\nu}}{g_{44}}\!\right)\!\!\left(\frac{2R^2}{\epsilon g_{44}}\right)\!\!\left(\frac{\epsilon g_{44}}{2R^2}\right)\!\!\left\{
(\delta_{\hat{P}_\mu}\bar{{\mathcal{V}}}_{\bar{\Phi}}\!-\!\delta_{\bar{\Phi}}\bar{{\mathcal{V}}}_{\hat{P}_\mu})(\delta_{\hat{P}^*_\nu}{{\mathcal{V}}}_{{\Phi}}\!-\!\delta_{{\Phi}}{{\mathcal{V}}}_{\hat{P}^*_\nu})\!+\!({\hat{P}_\mu}\!\leftrightarrow\!{\hat{P}^*_\nu})\right\}.\!\!\!\!\!\!\nn  
\eeaa
To ensure the absence of scalar ghosts, it is necessary to impose $g_{44} < 0$ and $\epsilon g_{44} > 0$. In other words, the coordinates $\chi_4$ and $\chi_5$ are of a space-like nature, excluding the conformal subalgebras (\ref{ConfAlg}--\ref{ConfAlg2}). A similar result appears to hold when introducing a central charge $\mathcal{D}$ as in (\ref{CCharge}, \ref{CCharge5D}, \ref{CCharge10D}) with the corresponding fields, Casimir, and invariant constructions.

The normalization of fields (indicated by a subindex $n$) is determined by the kinetic terms. Therefore, for the tree dynamics of these fields, the following normalizations are appropriate:
\beaa
 {B_n}_{T_{[o]}}\!=\!\left|\frac{8h^2}{4R^4}\!\left(\frac{2R^2}{g_{44}}\right)\right|^{1/2}\!{B}_{T_{[o]}},\ 
  {B_n}_{\hat{T}_{[o]}}\!=\!\left|\frac{8h^2}{4R^4}\!\left(\frac{2R^2 }{24g_{44}}\right)\right|^{1/2}\!{B}_{\hat{T}_{[o]}},\\
   {B_n}_{\Phi}\!=\!\left|\frac{4h^2}{4R^4}\!\left(\frac{20R^4}{12\epsilon g_{44}^2}\right)\right|^{1/2}\!{B}_{\Phi},\ 
  {B_n}_{\bar{\Phi}}\!=\!\left|\frac{4h^2}{4R^4}\!\left(\frac{20R^4}{12\epsilon g_{44}^2}\right)\right|^{1/2}\!{B}_{\bar{\Phi}},\\
  {\hat{B_n}}_{{T}_{[o]}}\!=\!\left|\frac{8\hat{h}^2}{4R^4}\!\left(\frac{48R^2}{g_{44}}\right)\right|^{-1/2}\!\hat{B}_{{T}_{[o]}},\ 
  {\hat{B_n}}_{\hat{T}_{[o]}}\!=\!\left|\frac{8\hat{h}^2}{4R^4}\!\left(\frac{2R^2 }{g_{44}}\right)\right|^{1/2}\!\hat{B}_{\hat{T}_{[o]}},
  \eeaa
  \beaa
  {\hat{B_n}}_{\Phi}\!=\!\left|\frac{4\hat{h}^2}{4R^4}\!\left(\frac{20R^4}{\epsilon g_{44}^2}\right)\right|^{1/2}\!\hat{B}_{\Phi},\ 
  {\hat{B_n}}_{\bar{\Phi}}\!=\!\left|\frac{4\hat{h}^2}{4R^4}\!\left(\frac{20R^4}{\epsilon g_{44}^2}\right)\right|^{1/2}\!\hat{B}_{\bar{\Phi}},\\
\bar{{\mathcal{V}_n}}_{T_{[o]}}\!=\!\left|\frac{4q\bar{q}}{4R^4}\!\left(\frac{48\epsilon}{10}\right)\right|^{1/2}\!\bar{{\mathcal{V}}}_{T_{[o]}},\ 
{{\mathcal{V}_n}}_{T_{[o]}}\!=\!\left|\frac{4q\bar{q}}{4R^4}\!\left(\frac{48\epsilon}{10}\right)\right|^{1/2}\!{{\mathcal{V}}}_{T_{[o]}},\\
\bar{\mathcal{V}}_{n\hat{T}_{[o]}}\!=\!\left|\frac{4q\bar{q}}{4R^4}\!\left(\frac{2\epsilon}{10}\right)\right|^{1/2}\!\bar{{\mathcal{V}}}_{\hat{T}_{[o]}},\ 
{{\mathcal{V}}}_{n\hat{T}_{[o]}}\!=\!\left|\frac{4q\bar{q}}{4R^4}\!\left(\frac{2\epsilon}{10}\right)\right|^{1/2}\!{{\mathcal{V}}}_{\hat{T}_{[o]}},\\
\bar{{\mathcal{V}}}_{n\Phi}\!=\!\left|\frac{4q\bar{q}}{4R^4}\!\left(\frac{2R^2}{g_{44}}\right)\right|^{1/2}\!\bar{{\mathcal{V}}}_{\Phi},\ 
{{\mathcal{V}}}_{n\bar{\Phi}}\!=\!\left|\frac{4q\bar{q}}{4R^4}\!\left(\frac{2R^2}{g_{44}}\right)\right|^{1/2}\!{{\mathcal{V}}}_{\bar{\Phi}},\\
\bar{{\mathcal{V}}}_{n\bar{\Phi}}\!=\!\left|\frac{4q\bar{q}}{4R^4}\!\left(\frac{2R^2}{g_{44}}\right)\right|^{1/2}\!\bar{{\mathcal{V}}}_{\bar{\Phi}},\ 
{{\mathcal{V}}}_{n{\Phi}}\!=\!\left|\frac{4q\bar{q}}{4R^4}\!\left(\frac{2R^2}{g_{44}}\right)\right|^{1/2}\!{{\mathcal{V}}}_{{\Phi}}.
\eeaa
These normalizations play a crucial role as mass terms can become either very large, rendering them irrelevant at lower energy, or extremely small, making them approximately massless at lower energy. Upon normalization, the standard kinetic terms are recovered after contraction. For instance:
 \beaa
 &&-\frac{4\hat{h}^2}{4R^4}\!\left(\frac{2R^2 g^{\mu\nu}}{g_{44}}\right)\!\left(\frac{24}{5}\right)\!\left(\frac{5}{1}\right)
 (\delta_{\hat{P}_\mu}\hat{B}_{T_{[o]}}\!-\!\delta_{T_{[o]}}\hat{B}_{\hat{P}_\mu})(\delta_{\hat{P}^*_\nu}\hat{B}_{T_{[o]}}\!-\!\delta_{T_{[o]}}\hat{B}_{\hat{P}^*_\nu})\ \nn\\
&& =(-Sign(g_{44}))\left(\half\partial_\mu B_{nT_{[o]}}
\partial^\mu B_{nT_{[o]}}+\half\hat{\partial}_\mu B_{nT_{[o]}}
\hat{\partial}^\mu B_{nT_{[o]}}\right)+\cdots,\ \ \\
&&-\frac{2q\bar{q}}{4R^4}\!\left(\!\frac{2R^2 g^{\mu\nu}}{g_{44}}\!\right)\!\!\left(\!\frac{2R^2}{\epsilon g_{44}}\!\right)\!\!\left(\!\frac{\epsilon g_{44}}{2R^2}\!\right)\!\left\{\!
(\delta_{\hat{P}_\mu}\!\bar{{\mathcal{V}}}_{\bar{\Phi}}\!-\!\delta_{\bar{\Phi}}\bar{{\mathcal{V}}}_{\hat{P}_\mu})(\delta_{\hat{P}^*_\nu}{{\mathcal{V}}}_{{\Phi}}\!-\!\delta_{{\Phi}}{{\mathcal{V}}}_{\hat{P}^*_\nu})\!+\!({\hat{P}_\mu}\!\leftrightarrow\!{\hat{P}^*_\nu})\!\right\}\!\!\nn\\
&&=(-Sign(\epsilon))\left(\partial_\mu \bar{{\mathcal{V}}}_{n\bar{\Phi}}
\partial^\mu {{\mathcal{V}}}_{n{\Phi}}+\hat{\partial}_\mu \bar{{\mathcal{V}}}_{n\bar{\Phi}}
\hat{\partial}^\mu {{\mathcal{V}}}_{n{\Phi}}\right)+\cdots.\ \ 
\eeaa
This analysis delves into the consistency of the process of eliminating the coordinates $\chi^4$ and $y^4$, exploring choices of the underlying metric that yield sensible models. The fundamental question arises whether the contraction is equivalent to some geometrical deformation of the Lie-group-like variety with extended space-time coordinates. Additionally, certain extra coordinates need to be ``integrated out," providing a potential avenue for obtaining effective theories. An alternative approach involves constructing an effective theory by considering contributions in the Lagrangian only up to certain powers of $1/R$. The normalizations introduced thus far are primarily aimed at achieving the proper appearance of kinetic terms when the differential operators are contracted. No contributions were considered from the actual process of ``integrating out" $\chi^4$ and $y^4$.

  Let us consider   mass--terms of purely scalar fields.  
 We consider first structure constants transforming scalar generators into scalar generators. This is the case in relations (\ref{ToPhi}--\ref{TohatPhi}) and (\ref{PhiPhiTs}) of the NPE, where
\beaa
C^\Phi_{\ T_{[o]}\Phi} =\ip,\   C^{\bar{\Phi}}_{\ T_{[o]}\bar{\Phi}} =-\ip,\ 
C^\Phi_{\ \hat{T}_{[o]}\Phi} =-4\ip,\ C^{\bar{\Phi}}_{\ \hat{T}_{[o]}\bar{\Phi}} =+4\ip,\nn\\
C^{T_{[o]}}_{\ {\Phi}\bar{\Phi}} =24\ip g_{44}\epsilon/(10R^2),\ \   C^{\hat{T}_{[o]}}_{\ {\Phi}\bar{\Phi}} =-4\ip g_{44}\epsilon/(10R^2).\ \ 
\eeaa
These structure constants lead to mass--term contributions:
\beaa
&&\!\!\!-4(C^\Phi_{\,T_{[o]}\Phi}(\lambda^{\hat{\cal O}}{\cal A}^{\hat{\cal O}}_{\ \Phi}))(C^{\bar{\Phi}}_{\,T_{[o]}\bar{\Phi}}(\lambda^{\hat{\cal O}'}{\cal A}^{\hat{\cal O}'}_{\ \bar{\Phi}})){\bf M}_1^{T_{[o]}T_{[o]}}{\bf M}_1^{\Phi\bar{\Phi}}{{\bf M}_1}^{-1}_{\hat{\cal O}\hat{\cal O}'}/(\!4R^4\!)\hspace{1.6cm}\nn \\
&&\!\!=\!-\frac{4h^2}{4R^4}\!\left(\frac{24}{5}\right)\left(\frac{2R^2 }{\epsilon g_{44}}\right)\!\left(\frac{5}{24}\right)
B_{\Phi}B_{\bar{\Phi}}\!
\ -\frac{4\hat{h}^2}{4R^4}\!\left(\frac{24}{5}\right)\left(\frac{2R^2 }{\epsilon g_{44}}\right)\!\!\left(\frac{5}{1}\right)\!
\hat{B}_{\Phi}\hat{B}_{\bar{\Phi}}\!\!\nn \\
&&-\frac{4\bar{q}q}{4R^4}\!\left(\frac{24}{5}\right)\!\left(\frac{2R^2 }{\epsilon g_{44}}\right)\!\left(\frac{\epsilon g_{44}}{2R^2}\right)\!
\bar{\mathcal{V}}_{\Phi}{\mathcal{V}}_{\bar{\Phi}}\!
-\frac{4\bar{q}q}{4R^4}\!\left(\frac{24}{5}\right)\left(\frac{2R^2 }{\epsilon g_{44}}\right)\!\left(\frac{\epsilon g_{44}}{2R^2}\right)
\bar{\mathcal{V}}_{\bar{\Phi}}{\mathcal{V}}_{\Phi},\ 
\eeaa
\beaa
&&\!\!-4(C^\Phi_{\,\hat{T}_{[o]}\Phi}(\lambda^{\hat{\cal O}}{\cal A}^{\hat{\cal O}}_{\ \Phi}))(C^{\bar{\Phi}}_{\,\hat{T}_{[o]}\bar{\Phi}}(\lambda^{\hat{\cal O}'}{\cal A}^{\hat{\cal O}'}_{\ \bar{\Phi}})){\bf M}_1^{\hat{T}_{[o]}\hat{T}_{[o]}}{\bf M}_1^{\Phi\bar{\Phi}}{{\bf M}_1}^{-1}_{\hat{\cal O}\hat{\cal O}'}/(\!4R^4\!)\hspace{1.5cm}\nn \\
&&\!\!=\!-\frac{4^3h^2}{4R^4}\!\left(\frac{1}{5}\right)\left(\frac{2R^2 }{\epsilon g_{44}}\right)\!\left(\frac{5}{24}\right)
B_{\Phi}B_{\bar{\Phi}}\!
\ -\frac{4^3\hat{h}^2}{4R^4}\!\left(\frac{1}{5}\right)\!\left(\frac{2R^2 }{\epsilon g_{44}}\right)\!\left(\frac{5}{1}\right)\!
\hat{B}_{\Phi}\hat{B}_{\bar{\Phi}}\!\nn \\
&&\!\! -\frac{4^3\bar{q}q}{4R^4}\!\left(\frac{1}{5}\right)\!\left(\frac{2R^2 }{\epsilon g_{44}}\right)\!\left(\frac{\epsilon g_{44}}{2R^2}\right)
\bar{\mathcal{V}}_{\Phi}{\mathcal{V}}_{\bar{\Phi}}\!
\ -\frac{4^3\bar{q}q}{4R^4}\!\left(\frac{1}{5}\right)\!\left(\frac{2R^2 }{\epsilon g_{44}}\right)\!\left(\frac{\epsilon g_{44}}{2R^2}\right)\!
\bar{\mathcal{V}}_{\bar{\Phi}}{\mathcal{V}}_{\Phi},
\eeaa
\beaa
&&\!\!-2(C^{T_{[o]}}_{\ \Phi\bar{\Phi}}(\lambda^{\hat{\cal O}}\!{\cal A}^{\hat{\cal O}}_{\ T_{[o]}}))(C^{T_{[o]}}_{\ \bar{\Phi}\Phi}(\lambda^{\hat{\cal O}'}\!{\cal A}^{\hat{\cal O}'}_{\ T_{[o]}})){\bf M}_1^{\Phi\bar{\Phi}}{\bf M}_1^{\bar{\Phi}\Phi}{{\bf M}_1}^{-1}_{\hat{\cal O}\hat{\cal O}'}/(\!4R^4\!)\hspace{1.6cm}\\
&&\!\!\!=\!-2\!\left(\!\frac{24g_{44}\epsilon}{10R^2}\!\right)^{\!2}\!\!\!\left(\!\frac{2R^2 }{\epsilon g_{44}}\!\right)^{\!2}\!\!\!\left(\!\frac{5}{24}\!\right)\!
h^2B_{T_{[o]}}B_{T_{[o]}}\!
\!\! -2\!\left(\!\frac{24g_{44}\epsilon}{10R^2}\!\right)^{\!2}\!\!\!\left(\!\frac{2R^2 }{\epsilon g_{44}}\!\right)^{\!2}\!\!\!\left(\!\frac{5}{1}\!\right)\!
\hat{h}^2\hat{B}_{T_{[o]}}\hat{B}_{T_{[o]}}\!\!\nn \\
&&\!\!\! -2\!\left(\!\frac{24g_{44}\epsilon}{10R^2}\!\right)^{\!2}\!\!\!\left(\!\frac{2R^2 }{\epsilon g_{44}}\!\right)^{\!2}\!\!\!\left(\!\frac{\epsilon g_{44}}{2R^2}\!\right)\!
\bar{q}q\bar{\mathcal{V}}_{T_{[o]}}{\mathcal{V}}_{T_{[o]}}\!
\!-2\!\left(\!\frac{24g_{44}\epsilon}{10R^2}\!\right)^{\!2}\!\!\!\left(\!\frac{2R^2 }{\epsilon g_{44}}\!\right)^{\!2}\!\!\!\left(\!\frac{\epsilon g_{44}}{2R^2}\!\right)\!
\bar{q}q{\mathcal{V}}_{T_{[o]}}\bar{\mathcal{V}}_{T_{[o]}},\nn
\eeaa
\beaa
&&\!\!-2(C^{\hat{T}_{[o]}}_{\ \Phi\bar{\Phi}}(\lambda^{\hat{\cal O}}{\cal A}^{\hat{\cal O}}_{\ \hat{T}_{[o]}}))(C^{\hat{T}_{[o]}}_{\ \bar{\Phi}\Phi}(\lambda^{\hat{\cal O}'}{\cal A}^{\hat{\cal O}'}_{\ \hat{T}_{[o]}})){\bf M}_1^{\Phi\bar{\Phi}}{\bf M}_1^{\bar{\Phi}\Phi}{{\bf M}_1}^{-1}_{\hat{\cal O}\hat{\cal O}'}/(\!4R^4\!)\hspace{1.6cm} \\
&&\!\!=\!-\frac{2h^2}{4R^4}\!\left(\!\frac{4g_{44}\epsilon}{10R^2}\!\right)^{\!2\!}\!\!\left(\!\frac{2R^2 }{\epsilon g_{44}}\right)^{\!\!2}\!\!\left(\!\frac{5}{24}\!\right)\!
B_{\hat{T}_{[o]}}B_{\hat{T}_{[o]}}\!
\! -\frac{2\hat{h}^2}{4R^4}\!\left(\!\frac{4g_{44}\epsilon}{10R^2}\!\right)^{\!2\!}\!\!\left(\!\frac{2R^2 }{\epsilon g_{44}}\!\right)^{\!2}\!\!\!\left(\!\frac{5}{1}\!\right)\!
\hat{B}_{\hat{T}_{[o]}}\hat{B}_{\hat{T}_{[o]}}\!\!\nn \\
&&\! -\frac{2\bar{q}q}{4R^4}\!\left(\!\frac{4g_{44}\epsilon}{10R^2}\!\right)^{\!\!2}\!\!\left(\!\frac{2R^2 }{\epsilon g_{44}}\!\right)^{\!\!2}\!\!\left(\!\frac{\epsilon g_{44}}{2R^2}\!\right)\!
\bar{\mathcal{V}}_{\hat{T}_{[o]}}{\mathcal{V}}_{\hat{T}_{[o]}}\!
-\frac{2\bar{q}q}{4R^4}\!\left(\!\frac{4g_{44}\epsilon}{10R^2}\!\right)^{\!\!2}\!\!\left(\!\frac{2R^2 }{\epsilon g_{44}}\!\right)^{\!\!2}\!\!\left(\!\frac{\epsilon g_{44}}{2R^2}\!\right)\!
{\mathcal{V}}_{\hat{T}_{[o]}}\!\bar{\mathcal{V}}_{\hat{T}_{[o]}}.\nn    
\eeaa
Upon renormalization, these gives valid mass terms for $\epsilon g_{44}>0$.

 We can gain also contributions to mass terms from structure constants transforming vector generators into vector generators in the NPE:
 \beaa
  C^{\bar{\Phi}}_{\ T_{\rho}\hat{P}^*_\tau} =\sqrt{2} g_{\rho\tau},\ \  C^\Phi_{\ \hat{T}_{\rho}\hat{P}_\tau} =\sqrt{2} g_{\rho\tau},\hspace{1cm}\nn\\
 C^{T_{[o]}}_{\ T_\mu \bar{T}_\nu} =-24\ip  g_{\mu\nu}/10,\ \   C^{\hat{T}_{[o]}}_{\ T_\mu \bar{T}_\nu} =-\ip g_{\mu\nu}/10,\\ 
C^{\hat{T}_{[o]}}_{\ \hat{P}_\rho \hat{P}^*_\tau} =-\ip g_{44} g_{\rho\tau}/(2R^2),\ \  C^{\hat{T}_{[o]}}_{\ \hat{P}^*_\rho \hat{P}_\tau} =\ip g_{44} g_{\rho\tau}/(2R^2).\ \ 
 \eeaa
They lead to mass--terms as follows:
\beaa
\!\!-4(C^{\bar{\Phi}}_{\,T_{\rho}\hat{P}^*_{\tau}}(\lambda^{\hat{\cal O}}{\cal A}^{\hat{\cal O}}_{\ \bar{\Phi}}))(C^{\Phi}_{\,\hat{T}_{\sigma}\hat{P}_{\kappa}}(\lambda^{\hat{\cal O}'}{\cal A}^{\hat{\cal O}'}_{\ \Phi})){\bf M}_1^{T_{\rho}\hat{T}_{\sigma}}{\bf M}_1^{\hat{P}^*_{\tau}\hat{P}_{\kappa}}{{\bf M}_1}^{-1}_{\hat{\cal O}\hat{\cal O}'}/(\!4R^4\!)\hspace{1.6cm}\\
\!\!=\!\cdots-\frac{4^2(\sqrt{2})^2h^2}{4R^4}\!\!\left(\!\frac{2}{\epsilon}\!\right)\!\left(\!\frac{R^2 }{g_{44}}\right)\!\left(\frac{5}{24}\!\right)\!
B_{\bar{\Phi}}B_{\Phi}\!
\ -\frac{4^2(\sqrt{2})^2\hat{h}^2}{4R^4}\!\left(\!\frac{2}{\epsilon}\!\right)\!\left(\!\frac{R^2 }{g_{44}}\!\right)\!\left(\!\frac{5}{1}\!\right)\!
\hat{B}_{\bar{\Phi}}\hat{B}_{\Phi}\!\!\nn \\
\ -\frac{4^2(\sqrt{2})^2\bar{q}q}{4R^4}\!\left(\!\frac{2}{\epsilon}\!\right)\!\left(\!\frac{R^2 }{g_{44}}\!\right)\!\left(\!\frac{\epsilon g_{44}}{2R^2}\!\right)\!
\bar{\mathcal{V}}_{\bar{\Phi}}{\mathcal{V}}_{\Phi}
-\frac{4^2(\sqrt{2})^2\bar{q}q}{4R^4}\!\left(\!\frac{2}{\epsilon}\!\right)\!\left(\!\frac{R^2 }{g_{44}}\!\right)\!\left(\!\frac{\epsilon g_{44}}{2R^2}\!\right)\!
{\mathcal{V}}_{\bar{\Phi}}\bar{\mathcal{V}}_{\Phi},\nn  
\eeaa
\beaa
&&\!\!\!-4(C^{\bar{\Phi}}_{\,T_{\rho}\hat{P}^*_{\tau}}(\lambda^{\hat{\cal O}}{\cal A}^{\hat{\cal O}}_{\ \bar{\Phi}}))(C^{\Phi}_{\,\hat{T}_{\sigma}\hat{P}_{\kappa}}(\lambda^{\hat{\cal O}'}{\cal A}^{\hat{\cal O}'}_{\ \Phi})){\bf M}_1^{T_{\rho}\hat{T}_{\sigma}}{\bf M}_1^{\hat{P}^*_{\tau}\hat{P}_{\kappa}}{{\bf M}_1}^{-1}_{\hat{\cal O}\hat{\cal O}'}/(\!4R^4\!)\hspace{1.6cm} \\
&&\!\!=\!\cdots-\frac{4^2(\sqrt{2})^2h^2}{4R^4}\!\left(\!\frac{2}{\epsilon}\!\right)\!\left(\!\frac{R^2 }{g_{44}}\!\right)\!\left(\!\frac{5}{24}\!\right)\!
B_{\bar{\Phi}}B_{\Phi}\!
\ -\frac{4^2(\sqrt{2})^2\hat{h}^2}{4R^4}\!\left(\!\frac{2}{\epsilon}\!\right)\!\left(\!\frac{R^2 }{g_{44}}\!\right)\!\left(\!\frac{5}{1}\!\right)\!
\hat{B}_{\bar{\Phi}}\hat{B}_{\Phi}\!\!\!\!\nn \\
&&\!\! -\frac{4^2(\sqrt{2})^2\bar{q}q}{4R^4}\!\left(\frac{2}{\epsilon}\right)\!\left(\frac{R^2 }{g_{44}}\right)\!\left(\frac{\epsilon g_{44}}{2R^2}\right)\!
\bar{\mathcal{V}}_{\bar{\Phi}}{\mathcal{V}}_{\Phi}
-\frac{4^2(\sqrt{2})^2\bar{q}q}{4R^4}\!\left(\frac{2}{\epsilon}\right)\!\left(\frac{R^2 }{g_{44}}\right)\!\left(\frac{\epsilon g_{44}}{2R^2}\right)\!
{\mathcal{V}}_{\bar{\Phi}}\bar{\mathcal{V}}_{\Phi},\!\!\!\!\nn 
\eeaa
\beaa
&&\!\!-2(C^{T_{[o]}}_{\,T_{\rho}\hat{T}_{\tau}}(\lambda^{\hat{\cal O}}{\cal A}^{\hat{\cal O}}_{\ T_{[o]}})) (C^{T_{[o]}}_{\,\hat{T}_{\sigma}{T}_{\kappa}}(\lambda^{\hat{\cal O}'}{\cal A}^{\hat{\cal O}'}_{\ T_{[o]}})){\bf M}_1^{T_{\rho}\hat{T}_{\sigma}}{\bf M}_1^{\hat{T}_{\tau}{T}_{\kappa}}{{\bf M}_1}^{-1}_{\hat{\cal O}\hat{\cal O}'}/(\!4R^4\!)\hspace{1.2cm}\nn \\
&&\!\!=\!\cdots-\frac{2\!\cdot\!4h^2}{4R^4}\!\left(\!\frac{-24\ip}{10}\!\right)\!\!\left(\!\frac{24\ip}{10}\!\right)\!\left(\!\frac{2}{\epsilon}\!\right)^{\!\!2}\!\!\!\left(\!\frac{5}{24}\!\right)\!
B_{T_{[o]}}^2\!
 \!-\frac{2\!\cdot\!4\hat{h}^2}{4R^4}\!\left(\!\frac{-24\ip}{10}\!\right)\!\left(\!\frac{24\ip}{10}\!\right)\!\left(\!\frac{2}{\epsilon}\!\right)^{\!\!2}\!\!\!\left(\!\frac{5}{1}\!\right)\!
\hat{B}_{T_{[o]}}^2\!\!\!\nn \\
&&-\frac{2\cdot2\cdot4\bar{q}q}{4R^4}\!\left(\!\frac{-24\ip}{10}\!\right)\!\left(\!\frac{+24\ip}{10}\!\right)\!\left(\frac{2}{\epsilon}\right)^{\!\!2}\!\!\left(\frac{\epsilon g_{44}}{2R^2}\right)\!
\bar{\mathcal{V}}_{T_{[o]}}{\mathcal{V}}_{T_{[o]}},\ 
\eeaa
\beaa 
&&\!\!-2(C^{\hat{T}_{[o]}}_{\,T_{\rho}\hat{T}_{\tau}}(\lambda^{\hat{\cal O}}{\cal A}^{\hat{\cal O}}_{\ \hat{T}_{[o]}})) (C^{\hat{T}_{[o]}}_{\,\hat{T}_{\sigma}{T}_{\kappa}}(\lambda^{\hat{\cal O}'}{\cal A}^{\hat{\cal O}'}_{\ \hat{T}_{[o]}})){\bf M}_1^{T_{\rho}\hat{T}_{\sigma}}{\bf M}_1^{\hat{T}_{\tau}{T}_{\kappa}}{{\bf M}_1}^{-1}_{\hat{\cal O}\hat{\cal O}'}/(\!4R^4\!)\hspace{1.4cm}\nn \\
&&\!\!=\!\cdots-\!\frac{2\cdot4{h}^2}{4R^4}\!\left(\!\frac{-\ip}{10}\!\right)\!\!\left(\!\frac{+\ip}{10}\!\right)\!\!\left(\frac{2}{\epsilon}\!\right)^{\!\!2}\!\!\left(\frac{5}{24}\right)\!
B_{\hat{T}_{[o]}}^2\!
 \!\!-\!\frac{2\cdot4\hat{h}^2}{4R^4}\!\left(\!\frac{-\ip}{10}\!\right)\!\!\left(\!\frac{+\ip}{10}\!\right)\!\!\left(\frac{2}{\epsilon}\right)^{\!\!2}\!\!\left(\frac{5}{1}\right)\!
\hat{B}_{\hat{T}_{[o]}}^2\!\!\!\nn \\
&&-\frac{2\cdot2\cdot4\bar{q}q}{4R^4}\!\left(\!\frac{-\ip}{10}\!\right)\!\left(\!\frac{+\ip}{10}\!\right)\!\left(\frac{2}{\epsilon}\right)^{\!\!2}\!\!\left(\frac{\epsilon g_{44}}{2R^2}\right)\!
\bar{\mathcal{V}}_{\hat{T}_{[o]}}{\mathcal{V}}_{\hat{T}_{[o]}},\ 
\eeaa
\beaa  
&&\!\!-4(C^{T_{[o]}}_{\,T_{\rho}\hat{T}_{\tau}}(\lambda^{\hat{\cal O}}{\cal A}^{\hat{\cal O}}_{\ T_{[o]}})) (C^{\hat{T}_{[o]}}_{\,\hat{T}_{\sigma}{T}_{\kappa}}(\lambda^{\hat{\cal O}'}{\cal A}^{\hat{\cal O}'}_{\ \hat{T}_{[o]}})){\bf M}_1^{T_{\rho}\hat{T}_{\sigma}}{\bf M}_1^{\hat{T}_{\tau}{T}_{\kappa}}{{\bf M}_1}^{-1}_{\hat{\cal O}\hat{\cal O}'}/(\!4R^4\!)\hspace{1.2cm}\nn \\
&&\!\!=\!\cdots-\frac{4\!\cdot\!4h^2}{4R^4}\!\left(\!\frac{-24\ip}{10}\!\right)\!\!\left(\!\frac{\ip}{10}\!\right)\!\!\left(\frac{2}{\epsilon}\right)^{\!\!2}\!\!\left(\frac{5}{24}\right)\!
B_{T_{[o]}}B_{\hat{T}_{[o]}}\nn\\
&&
\hspace{3.2cm} \!-\frac{4\!\cdot\!4\hat{h}^2}{4R^4}\!\left(\!\frac{-24\ip}{10}\!\right)\!\left(\!\frac{\ip}{10}\!\right)\!\left(\!\frac{2}{\epsilon}\!\right)^{\!\!2}\!\!\left(\!\frac{5}{1}\!\right)\!
\hat{B}_{T_{[o]}}\hat{B}_{\hat{T}_{[o]}}\!\!\!\nn \\
&&\hspace{2.0cm}-\frac{4\!\cdot4\bar{q}q}{4R^4}\!\left(\!\frac{-24\ip}{10}\!\right)\!\!\left(\!\frac{\ip}{10}\!\right)\!\!\left(\!\frac{2}{\epsilon}\!\right)^{\!\!2}\!\!\left(\!\frac{\epsilon g_{44}}{2R^2}\!\right)\!
\bar{\mathcal{V}}_{T_{[o]}}\!{\mathcal{V}}_{\hat{T}_{[o]}}\nn\\
&&\hspace{3.5cm}\!-\!\frac{4\cdot4\bar{q}q}{4R^4}\!\left(\!\frac{-24\ip}{10}\!\right)\!\left(\!\frac{\ip}{10}\!\right)\!\left(\!\frac{2}{\epsilon}\!\right)^{\!\!2}\!\!\left(\!\frac{\epsilon g_{44}}{2R^2}\!\right)\!
\bar{\mathcal{V}}_{\hat{T}_{[o]}}\!{\mathcal{V}}_{T_{[o]}}, 
\eeaa
\beaa
&&\!\!-2(C^{\hat{T}_{[o]}}_{\,\hat{P}_{\rho}\hat{P}^*_{\tau}}(\lambda^{\hat{\cal O}}{\cal A}^{\hat{\cal O}}_{\ \hat{T}_{[o]}})) (C^{\hat{T}_{[o]}}_{\,\hat{P}^*_{\sigma}\hat{P}_{\kappa}}(\lambda^{\hat{\cal O}'}{\cal A}^{\hat{\cal O}'}_{\ \hat{T}_{[o]}})){\bf M}_1^{\hat{P}_{\rho}\hat{P}^*_{\sigma}}{\bf M}_1^{\hat{P}^*_{\tau}\hat{P}_{\kappa}}{{\bf M}_1}^{-1}_{\hat{\cal O}\hat{\cal O}'}/(\!4R^4\!)\hspace{0.9cm}\nn \\
&&\!\!=\!\cdots-\frac{2\cdot4h^2}{4R^4}\!\left(\!\frac{-\ip g_{44}}{2R^2}\!\right)\!\left(\!\frac{+\ip g_{44}}{2R^2}\!\right)\!\left(\frac{R^2}{g_{44}}\right)^{\!\!2}\!\!\left(\frac{5}{24}\right)\!
B_{\hat{T}_{[o]}}^2\!\nn\\
&&\hspace{2.2cm}\!-\frac{2\cdot4\hat{h}^2}{4R^4}\!\left(\!\frac{-\ip g_{44}}{2R^2}\!\right)\!\left(\!\frac{+\ip g_{44}}{2R^2}\!\right)\!\left(\frac{R^2}{g_{44}}\right)^{\!\!2}\!\!\left(\frac{5}{1}\right)\!
\hat{B}_{\hat{T}_{[o]}}^2\!\!\!\nn \\
&&\hspace{2.0cm}-\frac{2\cdot2\cdot4\bar{q}q}{4R^4}\!\left(\!\frac{-\ip g_{44}}{2R^2}\!\right)\!\left(\!\frac{+\ip g_{44}}{2R^2}\!\right)\!\left(\frac{R^2}{g_{44}}\right)^{\!\!2}\!\!\left(\frac{\epsilon g_{44}}{2R^2}\right)\!
\bar{\mathcal{V}}_{\hat{T}_{[o]}}{\mathcal{V}}_{\hat{T}_{[o]}}.   
\eeaa
Upon renormalization, these gives valid mass terms for $\epsilon g_{44}>0$.

Thus, prior to contraction, the absence of ghost-like scalar terms necessitates  $\epsilon < 0, g_{44} < 0$, while correctly signed mass terms arise from $\epsilon g_{44} > 0$. Both conditions, $g_{44} < 0$ and $\epsilon < 0$, are associated with space-like coordinates. This, unfortunately, seems to leave aside the possibility of having a role for conformal subalgebras identified in (\ref{ConfAlg}--\ref{ConfAlg2}). It's noteworthy that, as anticipated, all mass terms vanish in the limit $R \to \infty$. Importantly, the possibility of having a wrongly signed mass term (that, together with a correctly signed kinetic term, could instigate self-induced symmetry breaking) is not feasible.
 
\section{Towards Symmetry Breaking}
We are conducting a preliminary investigation into the behavior of the considered models, examining aspects such as stability and symmetry breaking, under the influence of additional invariant multiplets with non-zero vacuum expectation values (VEV).

Initially, we focus on a scenario characterized by NPE symmetry. In standard symmetry--breaking mechanisms, a multiplet in the fundamental representation typically plays a crucial role. However, the identification of such fundamental irreps for the NPE remains an open question and will be the subject of future exploration. A plausible expectation is that such a multiplet might include Lorentz spinors, given that they constitute the fundamental representation of Lorentz symmetry.

Considering the possibility of an $SU(2)\times U(1)$--NPE scenario, the exploration of an adjoint representation becomes relevant for attempting a Higgs-like mechanism. We introduce a new adjoint multiplet $K^\mu, \bar{K}^\mu, K, \bar{K}, C^\mu, \bar{C}^\mu, C^{[\rho,\tau]}, C^{(\rho,\tau)}\!$, $C, \hat{C}$ corresponding to $\hat{P}^{*}_{\mu},\hat{P}_{\mu}, \bar{\Phi}, {\Phi}$, $\hat{T}_{\mu}, T_{\mu}, M_{\rho\sigma}, T_{\rho\sigma}, {T_{[o]}},{\hat{T}_{[o]}}$, respectively.

However, it is essential to note that all fields in the considered NPE-symmetric model with Lagrangian $L_i$ in (\ref{Li}) are massive. This is due to the presence of structure constants $C^{{\cal O}''}_{{\cal O O}'}\neq 0$ in (\ref{FSNew}) for each ${\cal O}''$, leading to mass terms. Consequently, there is no necessity to transfer additional degrees of freedom for already massive gauge fields. The fields acquiring extra mass terms are expected to play a role in breaking the symmetry.

An inverse process to standard symmetry breaking may be employed, directing some degrees of freedom associated with massive gauge fields toward fields in the new adjoint multiplet. This process allows the now massless gauge fields (such as $B_{\hat{P}^{*}_{\mu}},B_{\hat{P}_{\mu}}$, and perhaps $\hat{B}_{\hat{P}^{*}_{\mu}}, \hat{B}_{\hat{P}_{\mu}}$) to become precursors of gauge fields in the would-be Standard Model (SM). Additionally, the massless Goldstone bosons $C^{(\rho,\tau)}, K, \bar{K}$ could be related to the graviton and Higgs fields, respectively.

The Higgs--like multiplet ${\cal{C}}_2$, with quadratic invariant $({\cal{C}}_2)^2$, is endowed with a Higgs--like NPE-invariant potential $V_{pot}$:
\beaa
{\cal{C}}_2&=&K^\mu\otimes\hat{P}^*_\mu+\bar{K}^\mu\otimes\hat{P}_\mu+K\otimes\bar{\Phi}+\bar{K}\otimes{\Phi}\nn\\
&&+C^\mu\otimes\hat{T}_\mu+\bar{C}^\mu\otimes {T}_\mu+C\otimes T_{[o]}+\hat{C}\otimes\hat{T}_{[o]}\nn\\
&&+C^{[\rho\tau]}\otimes M_{\rho\tau}+C^{(\rho\tau)}\otimes T_{\rho\tau},\\
({\cal{C}}_2)^2&=&\frac{s\bar{s}g_{44}}{2R^2}\{K^\mu,\bar{K}_\mu\}+\frac{q\bar{q}\epsilon g_{44}}{2R^2}\{K,\bar{K}\}+\frac{\ell^2\epsilon}{2}\{C^\mu,\bar{C}_\mu\}\nn\\
&&+r^{\prime 2}C^{[\rho\tau]}C_{[\rho\tau]}+r^{ 2}C^{(\rho\tau)}C_{(\rho\tau)}+\frac{5h^2}{24}C^2+5\hat{h}^2\hat{C}^2,\\
-V_{pot}&=&-{\bf \mu}^2({\cal{C}}_2)^2-\lambda(({\cal{C}}_2)^2)^2.
\eeaa
The classical minimum under variation of any adjoint multiplet component gives the condition:
\beaa
<({\cal{C}}_2)^2>=-\mu^2/\lambda,\ {\rm with}\ -\mu^2>0.
\eeaa
To preserve Lorentz symmetry, we prohibit the existence of a vacuum expectation value (VEV) for non-scalar fields. We selectively assign a VEV, for instance, only to the field $\hat{C}$ (corresponding to the generator $\hat{T}_{[o]}$), leading to the acquisition of mass through its VEV:
\beaa
<\!{\hat{C}}\!>^2=\frac{-\mu^2}{10\lambda \hat{h}^2},\ {\rm with}\ -\mu^2>0,\ \ \hat{C}=<\!{\hat{C}}\!>\,+\hat{C}',
\eeaa
with $<{\hat{C}'}>=0$. In this case the fields $K^\mu\!, \bar{K}^\mu\!\!, K, \bar{K}, C^\mu\!, \bar{C}^\mu\!,  C^{(\rho,\tau)}\!\!,  C^{[\rho,\tau]}\!\!, C$ remain as goldstone bossons. Now, from the  NPE--covariant derivative and kinetic term we obtain
\beaa
\,[D_{{\cal O}}, {\cal{C}}_2]&=&\cdots+(\delta_{{\cal O}}\hat{C})\otimes \hat{T}_{[o]}+\ip\ell\phi_{{\cal O}}^{\ \tau}(+\hat{C})\otimes\hat{T}_\tau+\ip\ell\hat{\phi}_{{\cal O}}^{\ \tau}(-\hat{C})\otimes {T}_\tau\ \ \ \ \nn\\
&&\ \ \ +\ip \bar{s}\mathcal{E}_{{\cal O}}^{\ \tau}(+5\hat{C})\otimes\hat{P}^*_\tau+\ip {s}\bar{\mathcal{E}}_{{\cal O}}^{\ \tau}(-5\hat{C})\otimes\hat{P}_\tau\nn\\
&&\ \ \ +\ip \bar{q}\mathcal{V}_{{\cal O}}(+4\hat{C})\otimes\bar{\Phi}+\ip {q}\bar{\mathcal{V}}_{{\cal O}}(-4\hat{C})\otimes{\Phi},
\eeaa
\beaa
\,([D_{{\cal O}}, {\cal{C}}_2])^2&=&\cdots+\left(\frac{\epsilon\ell^2}{2}\{\phi_{{\cal O}}^{\ \tau},\hat{\phi}_{{\cal O}'}^{\ \kappa}\}g_{\tau\kappa}+\frac{25g_{44}s\bar{s}}{R^2}\{\mathcal{E}_{{\cal O}}^{\ \tau},\bar{\mathcal{E}}_{{\cal O}'}^{\ \kappa}\}g_{\tau\kappa}\right.\nn\\
&&\left.+\frac{16\epsilon g_{44}q\bar{q}}{2R^2}\{\mathcal{V}_{{\cal O}},\bar{\mathcal{V}}_{{\cal O}'}\}\right)(<\!{\hat{C}}\!>\,+\hat{C}')^2 (M_1)^{{\cal O O}'}.
\eeaa
From this, the gauge fields $\phi_{{\cal O}}^{\ \tau}$, $\hat{\phi}_{{\cal O}'}^{\ \kappa}$, $\mathcal{E}_{{\cal O}}^{\ \tau}$, $\bar{\mathcal{E}}_{{\cal O}'}^{\ \kappa}$, $\mathcal{V}_{{\cal O}}$, and $\bar{\mathcal{V}}_{{\cal O}'}$ acquire additional mass terms, beyond those inherent to the NPE-invariant Lagrangian, triggering symmetry breaking.

Note that $({\hat{P}_\tau}+{\hat{P}^*_\tau})^2+(-\ip({\hat{P}_\tau}-{\hat{P}^*_\tau}))^2=2\{\hat{P}^\tau,\hat{P}^*_\tau\}$. The kinetic terms of the new multiplet allow the mediation of some interchange of degrees of freedom, but this time in the opposite direction: moving degrees of freedom that made certain gauge fields massive to the new multiplet.

The vector $B_{\hat{P}_\tau}+B_{\hat{P}^*_\tau}$ and/or $-\ip(B_{\hat{P}_\tau}-B_{\hat{P}^*_\tau})$, now massless, would embody fields of the unbroken generator $T_{[o]}$ of the $U(1)$ symmetry. Additionally, some spin-2 combinations such as $\mathcal{E}_{\hat{P}_\tau}^{\ \kappa}+\mathcal{E}_{\hat{P}^*_\tau}^{\ \kappa}$, $\bar{\mathcal{E}}_{\hat{P}_\tau}^{\ \kappa}+\bar{\mathcal{E}}_{\hat{P}^*_\tau}^{\ \kappa}$, $\phi_{\hat{P}_\tau}^{\ \kappa}+\phi_{\hat{P}^*_\tau}^{\ \kappa}$, $\hat{\phi}_{\hat{P}_\tau}^{\ \kappa}+\hat{\phi}_{\hat{P}^*_\tau}^{\ \kappa}$, and/or $-\ip(\mathcal{E}_{\hat{P}_\tau}^{\ \kappa}-\mathcal{E}_{\hat{P}^*_\tau}^{\ \kappa})$, $-\ip(\bar{\mathcal{E}}_{\hat{P}_\tau}^{\ \kappa}-\bar{\mathcal{E}}_{\hat{P}^*_\tau}^{\ \kappa})$, $-\ip(\phi_{\hat{P}_\tau}^{\ \kappa}-\phi_{\hat{P}^*_\tau}^{\ \kappa})$, $-\ip(\hat{\phi}_{\hat{P}_\tau}^{\ \kappa}-\hat{\phi}_{\hat{P}^*_\tau}^{\ \kappa})$, together with perhaps $C^{(\rho\tau)}$, would embody a massless graviton associated with translations ${\hat{P}_\tau}+{\hat{P}^*_\tau}$ and/or $-\ip({\hat{P}_\tau}-{\hat{P}^*_\tau})$.

The transferred degrees of freedom would make most fields massive, leaving only the Goldstone bosons $K$ and $\bar{K}$ and some combination of $C^{(\rho\tau)}$ with spin-2 gauge fields massless. The $K$ and $\bar{K}$ would embody the Higgs field perhaps, alternatively, through some combination of spin-0 components of gauge fields, $K$ and $\bar{K}$ would  even acquire a negative mass term that might trigger a further stage of symmetry breaking.

First order gauge transformations such as
\beaa
(\mathcal{E}_{\hat{P}_\tau}^{\ \kappa}+\mathcal{E}_{\hat{P}^*_\tau}^{\ \kappa})'=\!(\mathcal{E}_{\hat{P}_\tau}^{\ \kappa}+\mathcal{E}_{\hat{P}^*_\tau}^{\ \kappa}) \!+\!\frac{1}{\bar{s}}(\delta_{\hat{P}_\tau}+\delta_{\hat{P}^*_\tau})\chi^\kappa(x)\!+\!\frac{\ip \hat{\varepsilon}(x)}{\hat{h}}(5\hat{h})(\mathcal{E}_{\hat{P}_\tau}^{\ \kappa}+\mathcal{E}_{\hat{P}^*_\tau}^{\ \kappa})\ \ &&\nn\\
\!+\!\frac{\ip {\varepsilon}^{[\mu\nu]}(x)}{r'}(+\ip r'\sigma[\mu\nu]_\rho^{\ \kappa})(\mathcal{E}_{\hat{P}_\tau}^{\ \rho}+\mathcal{E}_{\hat{P}^*_\tau}^{\ \rho})\!+\!\frac{\ip {\varepsilon}^{(\mu\nu)}(x)}{r}(+ r\sigma(\mu\nu)_\rho^{\ \kappa})(\mathcal{E}_{\hat{P}_\tau}^{\ \rho}+\mathcal{E}_{\hat{P}^*_\tau}^{\ \rho})\!\!\!&&\nn\\
+\frac{\ip {\varepsilon}^{\kappa}(x)}{\ell}\!\left(\frac{+\sqrt{2}\ip\epsilon\ell\bar{q}}{\bar{s}}\right)\!({\mathcal{V}}_{\hat{P}_\tau}+{\mathcal{V}}_{\hat{P}^*_\tau})+\frac{\ip \chi(x)}{\bar{q}}\left(\frac{-\sqrt{2}\ip\epsilon\ell\bar{q}}{\bar{s}}\right)(\hat{\phi}_{\hat{P}_\tau}^{\ \kappa}+\hat{\phi}_{\hat{P}^*_\tau}^{\ \kappa})&&\nn\\
+\frac{\ip {\chi}^\rho(x)}{\bar{s}}\left(-5\hat{h}\delta_\rho^{\ \kappa}(\hat{B}_{\hat{P}_\tau}\!+\!\hat{B}_{\hat{P}^*_\tau})\right.\hspace{5cm}&&\nn\\
\left.\!-\!\ip r'\sigma[\mu\nu]_\rho^{\ \kappa}(B^{\ [\mu\nu]}_{\hat{P}_\tau}\!+\!B^{\ [\mu\nu]}_{\hat{P}^*_\tau})\!-\! r\sigma(\mu\nu)_\rho^{\ \kappa}(B^{\ (\mu\nu)}_{\hat{P}_\tau}\!+\!B^{\ (\mu\nu)}_{\hat{P}^*_\tau})\right),\ &&
\eeaa 
\beaa 
(B_{\hat{P}_\tau}+B_{\hat{P}^*_\tau})'=(B_{\hat{P}_\tau}+B_{\hat{P}^*_\tau})+\frac{1}{h}(\delta_{\hat{P}_\tau}+\delta_{\hat{P}^*_\tau})\varepsilon(x)\hspace{3.5cm}&&\ \ \nn\\
\ \ \ +\frac{\ip \hat{\varepsilon}^\nu(x)}{\ell}\!\left(\frac{-12\epsilon\ell^2g_{\nu\kappa}}{5h}\right)\!(\phi_{\hat{P}_\tau}^{\ \kappa}+\phi_{\hat{P}^*_\tau}^{\ \kappa})
\!+\!\frac{\ip {\varepsilon}^\nu(x)}{\ell}\!\left(\frac{+12\epsilon\ell^2g_{\nu\kappa}}{5h}\right)\!(\hat{\phi}_{\hat{P}_\tau}^{\ \kappa}+\hat{\phi}_{\hat{P}^*_\tau}^{\ \kappa})\!\!\!\!&&\nn\\
\!+\!\frac{\ip \bar{\chi}(x)}{q}\!\left(\frac{+12\epsilon q\bar{q}g_{44}}{5R^2h}\right)\!({\mathcal{V}}_{\hat{P}_\tau}+{\mathcal{V}}_{\hat{P}^*_\tau})
\!+\!\frac{\ip {\chi}(x)}{\bar{q}}\!\left(\frac{-12\epsilon q\bar{q}g_{44}}{5R^2h}\right)\!(\bar{\mathcal{V}}_{\hat{P}_\tau}+\bar{\mathcal{V}}_{\hat{P}^*_\tau}),&&
\eeaa
and the such, will bridge the migration of degrees of freedom. Such procedure will be thoroughly examined for the $SU(2)\times U(1)$--NPE in a more realistic setting in a further contribution.

The outlined construction here aimed to deliver an emerging model which is effectively commutative,  i.e. the commutation of some translations effectively vanish. The emerging unbroken symmetry does not need to be the full CPE.

Notice that when constructing a CPE--invariant model from the outset, the multiplet obtained from a quartic Casimir could replace the adjoint multiplet. Using the transformation properties of the multiplet components of the quartic invariant in (\ref{QuartTransInic}-\ref{QuartTransEnd}), we can leverage the structure of an invariant Lagrangian involving only covariant derivatives. This constitutes the subject of further exploration.

\section{Transit Complex to Real Space--Time}

The consistency result for both $g_{44} < 0$ and $\epsilon < 0$ is noteworthy, as the challenge posed by multiple time-like coordinates is intricate. However, the presence of complexified space-time coordinates in the 10D representation introduces the time-like coordinate $y_0$ in the invariant $\bar{z}^{a}z_{a} = g_{\mu\nu}{\chi}^{\mu}{\chi}^{\nu} + g_{\mu\nu}{y}^{\mu}{y}^{\nu} + g_{44}{\chi}^{4}{\chi}^{4} + g_{44}{y}^{4}{y}^{4} + \epsilon{\chi}^{5}{\chi}^{5} + \epsilon{y}^{5}{y}^{5}$. We contemplate whether a symmetry breaking could lead to a scenario where, for instance, a pair of branes emerge, effectively decoupling the $\chi$'s and $y$'s of complex coordinates. Alternatively, we may consider a Kaluza--Klein approach in which the fields, after some unknown mechanism, do not depend on the $y$--coordinates. Nevertheless, the challange is how a local Lorentz symmetry might persist in an emerging  theory coupled to gravity in 4--dimensional space--time.

There is no compact realization of Lorentz symmetry in a miniature 4D--internal $y$-space-time manifold, so, some other setting is necessary. Various naive scenarios could be considered; for instance, $y^0$ evolving into a closed 1-sphere $S^1$ (loop) of radius $R_t$, while the remaining $y^1, y^2, y^3$ form a 3-sphere $S^3$ of radius $R_s$. However, such a scenario would evidently break the local Lorentz symmetry in the given differential representation.

On the other hand, there are scenarios where extra non-compact dimensions remain but become non-observable on the space--time brane manifold where we reside. In fact, the conserved quantity $\bar{z}^{a}z_{a}$ might transform to conserve, for instance, $g_{\mu\nu}{\chi}^{\mu}{\chi}^{\nu}$, $g_{\mu\nu}{y}^{\mu}{y}^{\nu}$, $g_{44}{\chi}^{4}{\chi}^{4}+g_{44}{y}^{4}{y}^{4}$, $\epsilon{\chi}^{5}{\chi}^{5}+\epsilon{y}^{5}{y}^{5}$ separately, exactly or approximately. In such a case, an $SU(1,5)$ manifold might evolve into, for example, a manifold with symmetry $SL(2)\times SL(2)\times U(1)\times U(1)$, in addition to the local translations corresponding to gravity.

Let's attempt to deduce the implications of the formalism when transitioning from NPE to CPE through contraction. The introduction of the parameter $R$ in the NPE suggests a pathway towards a CPE. For large values of $R$, the Lagrangian (\ref{Li}) would prefer covariant derivatives with labels ${\cal O}, {\cal O}'\in  \{\hat{P}_{\mu},\hat{P}^*_{\mu}, \Phi,\bar{\Phi}\}$ (or ${\cal O}, {\cal O}'\in  \{\hat{P}_{\mu},\hat{P}^*_{\mu}, \Phi,\bar{\Phi}, {\cal D}\}$ if the central charge is included). This effectively leads to the invariant Lie algebra subspace associated with $M_2$ (or some enlarged version to include ${\cal D}$).

The Lagrangian terms $\{{{\bar{{\mathbb R}}}}^{\ {\cal O}{\cal O'}}_{ \kappa},{{\mathbb R}}_{\ {\cal O}{\cal O'}}^\kappa\}$ and $\{{{\bar{{\mathbb Q}}}}^{{\cal O}{\cal O'}},{{\mathbb Q}}_{{\cal O}{\cal O'}}\}$, although accompanied by $R^{-2}$, might be related to the origin of gravity. However, the challenge remains to transition from this model to gravity, perhaps utilizing a Palatini-like formalism.

As indicated, a symmetry-breaking mechanism should be formulated so that only (the prior to symmetry breaking) standard model gauge fields, the graviton, and perhaps a scalar Higgs field become or remain massless. All other massive fields should be effectively suppressed or decoupled at lower energy regimes. The emerging symmetry might not be a complete CPE.

Is it conceivable to subsequently obtain a contribution to the kinetic term through the differential representation of generators $T_\mu, \bar{T}_\nu$? Could a combination, such as ${\hat{P}_\tau}+{\hat{P}^*_\tau}$ and ${T_\tau}+{\hat{T}_\tau}$, potentially emerge as the effective space-time translation? Notably, $T_\mu, \bar{T}_\nu$ carry $T_{[o]}$ (or U(1)) charge. However, this scenario does not seem to lead to a sensible model unless such a $U(1)$ symmetry is broken. In the case of larger gauge symmetries than $U(1)$, it is plausible that only the unbroken and chargeless components might contribute to the effective emerging space-time translations.

Replacing $T_\mu, \bar{T}_\nu$ with $RT_\mu, R\bar{T}_\nu$ in the NPE algebra (\ref{Dim0Start})--(\ref{PPMTs}) or in the Lorentz extension LE (\ref{Dim0Start})--(\ref{Dim0End}), we observe that commutators of generators in $\{\hat{P}_{\mu},\hat{P}^*_{\mu}, \Phi,\bar{\Phi}, T_\mu, \bar{T}_\nu\}$ in these modified algebras yield terms of order $1/R$ for the modified NPE. Conversely, they produce terms of order $1/R^2$ for the remaining generators of the Lorentz extension in both the modified NPE or LE. This indicates their significant suppression for large $R$. The mechanism governing the transition from a Non-contracted Poincaré extension (such as NPE) or from a modified Lorentz extension (such as LE) will determine whether a stage emerges where the CLE becomes prevalent.

Nevertheless, it's essential to note that we have a diverse landscape for attempting to discern the origin of the Standard Model (SM) Higgs field and the origin of gravity. On one hand, the scalar associated with the fields $\phi_b^{\,\,\rho}, \hat{\phi}_b^{\,\,\rho}$ has been discussed in \cite{U1I}. Additionally, there is a scalar contribution from the fields ${{\mathcal{E}}}_{b\tau}, \bar{{\mathcal{E}}}_{b\tau}$, both potentially combining with other scalar fields (like the fields $K, \bar{K}$ in an adjoint multiplet) to give rise to an emerging Higgs field.

On the other hand, the fields $\phi_b^{\,\,\rho}, \hat{\phi}_b^{\,\,\rho}$, as well as ${{\mathcal{E}}}_{b\tau}, \bar{{\mathcal{E}}}_{b\tau}$, might contribute to a vierbein, ultimately defining a metric.

It would be valuable to extend the provided formalism to include differential forms. Specifically, a combination of the one-forms ${{\mathcal{E}}}^{\ \rho}_{\hat{P}_{\mu}}dz^\mu, \bar{{\mathcal{E}}}^{\  \rho}_{\hat{P}^*_{\mu}}d\bar{z}^\mu$ might harbor some of the seeds for an emerging gravitational field. Similarly, a combination of $B^{\ [\rho,\sigma]}_{\hat{P}_{\mu}}dz^\mu, B^{\ [\rho,\sigma]}_{\hat{P}^*_{\mu}}d\bar{z}^\mu$ could contain the seeds for an emerging spin connection. Additionally, a combination of $B_{\hat{P}_{\mu}}dz^\mu, B_{\hat{P}^*_{\mu}}d\bar{z}^\mu$ might provide the seeds for the one-form associated with the U(1) symmetry.

In the next section, we delve into a model with local CPE symmetry, following the formalism for constructing gravity from the Poincaré algebra. This represents an intermediate stage towards a real space-time. It is anticipated that a mechanism, akin to those discussed in locally invariant NPE models, will bring about this stage.

\section{Transition Stage: Extended Space--Time GR}

The transition from the covariant derivative (\ref{CovDerNPE}) associated with the full Lie algebra structure of the NPE (\ref{Dim0Start})--(\ref{PPMTs}) to models closely resembling the Standard Model (SM), with the simplifications adopted here of just an $U(1)$ internal gauge group and coupling to General Relativity (GR), opens up various developmental pathways.

\subsection{CPE 4+1 Space-time and LE-connection Gravity}

The implementation of general diffeomorphisms in the space-time variety in General Relativity (GR) introduces tetrads, i.e., the gravitational field, which is invariant under local Lorentz transformations. Now, we embark on a similar procedure: considering a GR-like model of the CPE (\ref{Dim0Start})--(\ref{PPMTs}) in the limit $R\to\infty$. The analogue to the Lorentz symmetry in this context is the LE (\ref{Dim0Start})--(\ref{Dim0End}). The local Min\-kowski pseudo-metric $g_{\mu\nu}$ is extended to include the 5th complex dimension, becoming ${\rm diag}(1,-1,-1,-1, \epsilon)$ according to the CPE quadratic Casimir $J$ and the pseudo-metric $M_2^{-1}$ in (\ref{CasPEx2}--\ref{CasimirPExM2}). The choices of generators were made such that the pseudo-metric can be adopted in this manner and might also allow for contributions from $\{T_{\mu}, \hat{T}_{\mu}\}$ to future translations.

It's crucial to observe that the symmetric generators $T_{\mu\nu}, T_{[o]}, \hat{T}_{[o]}$ distinctly transform the elements in pairs such as $\{\hat{P}_{\mu}, \hat{P}_{\mu}^*\}, \{\Phi, \bar{\Phi}\}, \{T_{\mu}, \hat{T}_{\mu}\}$, and the formation of invariants requires their pairing. Additionally, when contracting Lorentz or Minkowski labels, there's a non-barred index with a barred index, the latter coming from a complex conjugate generator, except when one of the factors is a differential.

A Minkowski--like pseudo-metric $\eta_{\tilde{a}\tilde{b}}$ is adopted as the symmetric matrix that lowers and raises Minkowski indices, and contractions always involve a non-barred and a barred index except when  just one of the factors is a differential:

 \beaa
 &\eta_{\bar{\mu}\nu}=\eta_{\nu\bar{\mu}}:= g_{\mu\nu},\ \ \eta^{\bar{\mu}\nu}=\eta^{\nu\bar{\mu}}:=g^{\mu\nu},\\ 
 &\eta_{\bar{5}5}=\eta_{5\bar{5}}:=g_{55}=\epsilon,\ \ \eta^{\bar{5}5}=\eta^{5\bar{5}}:=g^{55}=\epsilon^{-1},\\
 &\eta_{\tilde{a}\tilde{b}}=0 \ {\rm if}\ \tilde{b}\neq\bar{\tilde{a}}\ {\rm or}\  \tilde{a}\neq\bar{\tilde{b}},\ \ 
 \eta^{\bar{\mu}\nu}\eta_{\bar{\nu}\rho}=\delta_{\rho}^{\ \bar{\mu}},\ \eta_{\bar{\mu}\nu}\eta^{\bar{\nu}\rho}=\delta^{\ \rho}_{\bar{\mu}},\
 V^{5}\eta_{\bar{5}5}=V_{5},\\  
 &\eta_{\bar{5}5}V^{\bar{5}}=V_{\bar{5}},\ \  V^{\mu}\eta_{\bar{\mu}\rho}=V_{\rho},\  
 \eta_{\bar{\mu}\rho}V^{\bar{\rho}}=V_{\bar{\mu}},\ V_{\mu}\eta^{\bar{\mu}\rho}=V^{\rho},\  
 \eta^{\bar{\mu}\rho}V_{\bar{\rho}}=V^{\bar{\mu}}.
 \eeaa
The lowering and rising of indices through $\eta_{\tilde{a}\tilde{b}}$ mostly change a sign but do not alter the non-barred or barred nature of the index.

Certain generators (and their corresponding fields) will be related through complex conjugation, while others might be real from the outset. The generator pairs $\{\hat{P}_{\mu}, \hat{P}_{\mu}^*\}, \{\Phi, \bar{\Phi}\}, \{T_{\mu}, \hat{T}_{\mu}\}$ (and their associated connections) will be related by conjugation. In the CPE, the first two pairs are translations, but the last pair is not (yet) a translation. The generators $M_{\mu\nu}, T_{\mu\nu}$ maintain their antisymmetry and symmetry, but to specify their indices, we can write them as $M_{\bar{\mu}\nu}, T_{\bar{\mu}\nu}$ or as $M_{\mu\bar{\nu}}, T_{\mu\bar{\nu}}$.

Notice that the pseudo--metric $\eta_{\tilde{a}\tilde{b}}$ combines non-barred with barred indices, and summations over repeated indices mostly occur under this constraint (unless they involve differentials). This might enforce a specific complex structure on the emerging space-time pseudo-metric. Furthermore, the arrays used in the CLE take the form in terms of the Minkowski--like metric $\eta_{\tilde{a}\tilde{b}}$:

 \beaa
 \sigma[\bar{\mu}\nu]_{\bar{\rho}}^{\ \sigma}=-\sigma[\bar{\nu}\mu]_{\bar{\rho}}^{\ \sigma}&:=&(\eta_{\bar{\rho}\mu}\delta_{\bar{\nu}}^{\,\,\sigma}-\eta_{\bar{\rho}\nu}\delta_{\bar{\mu}}^{\,\,\sigma}),\\
 \sigma[\bar{\mu}\nu]_{{\rho}}^{\  \bar{\sigma}}=-\sigma[\bar{\nu}\mu]_{{\rho}}^{\ \bar{\sigma}}&:=&(\eta_{\bar{\mu}\rho}\delta_{{\nu}}^{\,\,\bar{\sigma}}-\eta_{\bar{\nu}\rho}\delta_{{\mu}}^{\,\,\bar{\sigma}}),
 \eeaa
 \beaa
 \sigma(\bar{\mu}\nu)_{\bar{\rho}}^{\ \sigma}=\sigma(\bar{\nu}\mu)_{\bar{\rho}}^{\ \sigma}&:=&(\eta_{\bar{\rho}\mu}\delta_{\bar{\nu}}^{\,\,\sigma}+\eta_{\bar{\rho}\nu}\delta_{\bar{\mu}}^{\,\,\sigma}-(1/2)\eta_{\bar{\mu}\nu}\delta_{\bar{\rho}}^{\,\,\sigma}),\\
\sigma(\bar{\mu}\nu)_{{\rho}}^{\ \bar{\sigma}}=\sigma(\bar{\nu}\mu)_{{\rho}}^{\ \bar{\sigma}}&:=&(\eta_{\bar{\mu}\rho}\delta_{{\nu}}^{\,\,\bar{\sigma}}+\eta_{\bar{\nu}\rho}\delta_{{\mu}}^{\,\,\bar{\sigma}}-(1/2)\eta_{\bar{\mu}\nu}\delta_{{\rho}}^{\,\,\bar{\sigma}}).
 \eeaa

 The array $\eta_{\tilde{a}\tilde{b}}$ invites us to define its antisymmetric counterpart $\kappa_{\tilde{a}\tilde{b}}$. Both of them correspond to the flat complex structure limit:

\begin{center}
	\begin{tabular}{lr lr} $\eta_{\tilde{a}\tilde{b}}=$&
	\begin{tabular}{ r | cc}
		$\tilde{a}\backslash\tilde{b}$ &$0\cdots 5$&$\bar{0}\cdots\bar{5}$ \\ \hline
		\begin{tabular}{l}\\
			0\\
			:\\
			5\end{tabular} & O & $\eta_{a\bar{b}}$ \\ 	\begin{tabular}{l}\\
			$\bar{0}$\\
			:\\
			$\bar{5}$\end{tabular} & $\eta_{\bar{a}b}$ & O \\
	\end{tabular}&
		,\ \ $\kappa_{\tilde{a}\tilde{b}}=$&
		\begin{tabular}{ r | cc}
		$\tilde{a}\backslash\tilde{b}$ &$0\cdots 5$&$\bar{0}\cdots\bar{5}$ \\ \hline
		\begin{tabular}{l}\\
		0\\
		:\\
		5\end{tabular} & O & $-\eta_{a\bar{b}}$ \\ 	\begin{tabular}{l}\\
		$\bar{0}$\\
		:\\
		$\bar{5}$\end{tabular} & $\eta_{\bar{a}b}$ & O \\
		\end{tabular}
		\end{tabular}.
\end{center}
The indices $4$ and $\bar{4}$ do not appear after the contraction. The index 5 is sometimes hidden in the notation adopted.

In this CPE stage, only fields $\lambda^{{\cal O}}{{\cal A}}_{\hat{{\cal O}}}^{\ {\cal O}}$ for $\hat{{\cal O}}$ translations in $\{\hat{P}_{\varepsilon}, \hat{P}_{\varepsilon}^*, \Phi, \bar{\Phi}\}$ survive. The translation subalgebra is a largest abelian CPE subalgebra.

We simplify the upper indices in the fields $\lambda^{{\cal O}}{{\cal A}}_{\hat{{\cal O}}}^{\ {\cal O}}$ by writing $(\bar{\mu}, \mu, \bar{5}, 5)$ in place of $(\hat{P}_{\mu}, \hat{P}_{\mu}^*, \Phi, \bar{\Phi})$ respectively. From now on, the indices $\mu, \nu, \rho, \sigma$ will be reserved for space-time Minkowski or local Lorentz indices. We reserve ${a}, {b}, {c}, {e}$ for extended Minkowski indices in $\{\mu, 5\}$, $\bar{a}, \bar{b}, \bar{c}, \bar{e}$ for extended Minkowski indices in $\{\bar{\mu}, \bar{5}\}$, and $\tilde{a}, \tilde{b}, \tilde{c}, \tilde{e}$ for extended Minkowski indices in $\{\mu, 5, \bar{\mu}, \bar{5}\}$.

We simplify lower subindices in the fields $\lambda^{{\cal O}}{{\cal A}}_{\hat{{\cal O}}}^{\ {\cal O}}$ by writing $(\varepsilon, \bar{\varepsilon}, 5, \bar{5})$ in place of $(\hat{P}_{\varepsilon}, \hat{P}_{\varepsilon}^*, \Phi, \bar{\Phi})$ respectively. Hence, we write $\lambda^{\bar{\mu}}{{\cal A}}_{5}^{\ \bar{\mu}}$ in place of $\lambda^{\hat{P}_{\mu}}{{\cal A}}_{\Phi}^{\ \hat{P}_{\mu}}=s\bar{\mathcal{E}}_{\Phi}^{\ \mu}$ (in the notation of previous sections for NPE).

From now on, the use of $\varepsilon, \eta, \kappa, \tau$ will be reserved for space-time Einstein indices labeling covariant derivatives, and we reserve $\tilde{u}, \tilde{v}, \tilde{w}, \tilde{s}$ for extended Einstein indices in $\{\varepsilon, \bar{\varepsilon}, 5, \bar{5}\}$ that label covariant derivatives (associated with translations). Extended Einstein indices explicitly distinguish between non-conjugated and conjugated extended translations.  

Now, the \textbf{gravitational field components} $\lambda^{{\cal O}}{{\cal A}}_{\hat{{\cal O}}}^{\ {\cal O}}$ —also called \textbf{Viel\-beins}— are those where both ${\cal O}$ and $\hat{{\cal O}}$ are translations. They will be written $e_{\tilde{u}}^{\ \tilde{a}}$ for $\tilde{a}\in\{\mu, 5, \bar{\mu}, \bar{5}\}$ Minkowski, and $\tilde{u}\in \{\varepsilon, 5, \bar{\varepsilon}, \bar{5}\}$ Einstein indices.

The coupling constants associated with translations will be absorbed in the fields $e_{\tilde{u}}^{\ \tilde{a}}$ since the gravitational coupling will be placed separately, as in GR.

We define the gravitational field as the 1-form:
 \beaa
  e^{\tilde{a}}= e_{\tilde{u}}^{\ \tilde{a}}\,dz^{\tilde{u}}, \  \tilde{a}\in\{\mu, \bar{\mu},  5, \bar{5}\},\ \tilde{u}\in \{\varepsilon, \bar{\varepsilon},  5, \bar{5}\},\ dz^{\bar{\mu}}=d\bar{z}^{\mu}, dz^{\bar{5}}=d\bar{z}^{5},\\
 e_{\varepsilon}^{\ \bar{\mu}}=s\bar{\mathcal{E}}_{\hat{P}_{\varepsilon}}^{\ \mu},\ 
 e_{\bar{\varepsilon}}^{\ \bar{\mu}}=s\bar{\mathcal{E}}_{\hat{P}_{\varepsilon}^*}^{\ \mu},\ 
 e_{5}^{\ \bar{\mu}}=s\bar{\mathcal{E}}_{\Phi}^{\ \mu},\ 		
 e_{\bar{5}}^{\ \bar{\mu}}=s\bar{\mathcal{E}}_{\bar{\Phi}}^{\ \mu},\\
 e_{\varepsilon}^{\ {\mu}}=\bar{s}{\mathcal{E}}_{\hat{P}_{\varepsilon}}^{\ \mu},\ 
 e_{\bar{\varepsilon}}^{\ {\mu}}=\bar{s}{\mathcal{E}}_{\hat{P}_{\varepsilon}^*}^{\ \mu},\ 
 e_{5}^{\ {\mu}}=\bar{s}{\mathcal{E}}_{\Phi}^{\ \mu},\ 		
 e_{\bar{5}}^{\ {\mu}}=\bar{s}{\mathcal{E}}_{\bar{\Phi}}^{\ \mu},\\
 e_{\varepsilon}^{\ \bar{5}}=q\bar{\mathcal{V}}_{\hat{P}_{\varepsilon}},\ 
 e_{\bar{\varepsilon}}^{\ \bar{5}}=q\bar{\mathcal{V}}_{\hat{P}_{\varepsilon}^*},\ 
 e_{5}^{\ \bar{5}}=q\bar{\mathcal{V}}_{\Phi}^{\ \mu},\ 		
 e_{\bar{5}}^{\ \bar{5}}=q\bar{\mathcal{V}}_{\bar{\Phi}},\\	
e_{\varepsilon}^{\ {5}}=\bar{q}{\mathcal{V}}_{\hat{P}_{\varepsilon}},\ 
e_{\bar{\varepsilon}}^{\ {5}}=\bar{q}{\mathcal{V}}_{\hat{P}_{\varepsilon}^*},\ 
e_{5}^{\ {5}}=\bar{q}{\mathcal{V}}_{\Phi},\ 		
e_{\bar{5}}^{\ {5}}=\bar{q}{\mathcal{V}}_{\bar{\Phi}},
\eeaa
\beaa
{\rm with}\ \ e^{\ \bar{\tilde{a}}}_{\bar{\tilde{u}}}=\overline{(e_{\tilde{u}}^{\ \tilde{a}})},\ \  \bar{\bar{u}}=u,\ \ \bar{\bar{a}}=a.
\eeaa
 We define an Exterior Derivative $D$ associated with the LE--connection fields (and generalized spin--connection):
 \beaa
 & D(\cdot):=d(\cdot)+dz^{\tilde{u}}\left( \ip\ell(\hat{\phi}_{\tilde{u}}^{\ 5\bar{\mu}}[T_{\bar{5}\mu},\cdot]+{\phi}_{\tilde{u}}^{\ \mu\bar{5}}[\hat{T}_{\bar{\mu}5},\cdot]) +\ip h B_{\tilde{u}}[T_{[o]},\cdot] \right.\hspace{1cm}\nn\\
 &\hspace{1cm}+ \ip \hat{h} \hat{B}_{\tilde{u}}[\hat{T}_{[o]},\cdot]+\left.\ip r' B_{\tilde{u}}^{\ [\rho\bar{\sigma}]} [M_{\bar{\rho}\sigma},\cdot]+ \ip r {B}_{\tilde{u}}^{\ (\rho\bar{\sigma})}[{T}_{\bar{\rho}\sigma},\cdot]\right),\\
 &\ {\rm with}\ \ \hat{\phi}_{\bar{\tilde{u}}}^{\ 5\bar{\mu}}=\overline{{\phi}_{\tilde{u}}^{\ \mu\bar{5}}},
 \ \ B_{\bar{\tilde{u}}}=\overline{B_{\tilde{u}}},\ \ \hat{B}_{\bar{\tilde{u}}}=\overline{\hat{B}_{\tilde{u}}},\\ &B_{\bar{\tilde{u}}}^{\ [\sigma\bar{\rho}]}=-\overline{B_{\tilde{u}}^{\ [\rho\bar{\sigma}]}},\ \ {B}_{\bar{\tilde{u}}}^{\ (\sigma\bar{\rho})}=\overline{{B}_{\tilde{u}}^{\ (\rho\bar{\sigma})}}\!\!.\ \ \ \ \ 
   \eeaa
In order to make explicit hidden indices, we wrote $T_{\bar{5}\mu}, \hat{T}_{\bar{\mu}5}$ instead of $T_{\mu}, \hat{T}_{{\mu}}$, and $\hat{\phi}_{\tilde{\varepsilon}}^{\ 5\bar{\mu}}, {\phi}_{\tilde{\varepsilon}}^{\ \mu\bar{5}}$ instead of $\hat{\phi}_{\tilde{\varepsilon}}^{\ {\mu}}, {\phi}_{\tilde{\varepsilon}}^{\ \mu}$.

To make the LE-connection fields compatible with the extended gravitational field 1-form, we require the \textbf{First Cartan Structure Equation} (FCSE) \cite{QG}:
  \beaa
  D(e^{\tilde{a}})=0,\label{FCSE}
  \eeaa
Although the articulation of GR in terms of a pseudo-metric field alone is not able to deal with fermion fields, which will require Vielbeins (generalized tetrads), it provides a simple way to solve the FCSE. We consider a Covariant Derivative $D_{\tilde{u}}$ that acts on both Minkowski and Einstein indices. The action on Minkowski indices corresponds to the covariant derivative in equation (101) of \cite{U1I} for the Lorentz extension but along translation generators of the CPE. And $D_{\tilde{u}}$ acting on an Einstein vector $v^{\tilde{v}}$ takes the form:

  \beaa
  D_{\tilde{u}}(v^{\tilde{v}})=\delta_{\tilde{u}}(v^{\tilde{v}})+\Gamma^{\tilde{v}}_{\bar{\tilde{w}}\tilde{u}}v^{\tilde{w}}.
  \eeaa
  
The conjunction of connection fields in the external derivative $D$ and the connection $\Gamma^{\tilde{v}}_{\tilde{u}\tilde{w}}$ will give the action of the Covariant Derivative $D_{\tilde{u}}$ on general objects carrying both Minkowski and Einstein indices. Our covariant derivative $D_{\tilde{u}}$ will be defined by the Extended Levi--Civita Connection, leaving the $e_{\tilde{v}}^{\ \tilde{a}}$ invariant:
   \beaa
   D_{\tilde{u}}(e_{\tilde{v}}^{\ \tilde{a}}):=\ip(-\ip\delta_{\tilde{u}})(e_{\tilde{v}}^{\ \tilde{a}})\!+\!\left( \ip\ell(\hat{\phi}_{\tilde{u}}^{\ 5\bar{\nu}}[T_{\bar{5}\nu},e_{\tilde{v}}^{\ \tilde{a}}]\!+\!{\phi}_{\tilde{u}}^{\ \nu\bar{5}}[\hat{T}_{\bar{\nu}5},e_{\tilde{v}}^{\ \tilde{a}}]) 
   \! +\!\ip h B_{\tilde{u}}[T_{[o]},e_{\tilde{v}}^{\ \tilde{a}}]\right.\hspace{-.7cm}\nn\\
   + \ip \hat{h} \hat{B}_{\tilde{u}}[\hat{T}_{[o]},e_{\tilde{v}}^{\ \tilde{a}}]\!+\!\left.\ip r' B_{\tilde{u}}^{\ [\rho\bar{\sigma}]} [M_{\bar{\rho}\sigma},e_{\tilde{v}}^{\ \tilde{a}}]\!+\! \ip r {B}_{\tilde{u}}^{\ (\rho\bar{\sigma})}[{T}_{\bar{\rho}\sigma},e_{\tilde{v}}^{\ \tilde{a}}]\right)-\Gamma^{\bar{\tilde{w}}}_{\tilde{v}\tilde{u}}e_{\tilde{w}}^{\ \tilde{a}}=0.\ \!\label{CovDerVielb}
   \eeaa
In contrast to GR with just the Spin--Connection $\omega^{\ [\rho,\sigma]}_\varepsilon$ associated with the Local Lorentz algebra symmetry, here a variety of connections are associated with the LE generators of the Lie algebra of the group $U(t_1, s_1)$, where $s_1+t_1=5$ ($t_1=1$ if $\epsilon=-1$). We present some instances of (\ref{CovDerVielb}) for non-barred or barred Minkowski index $\tilde{a}\in\{\mu, \bar{\mu}, 5, \bar{5}\}$:
   \beaa
&D_{\tilde{u}}(e_{\tilde{v}}^{\ \mu}):=\ip(-\ip\delta_{\tilde{u}})(e_{\tilde{v}}^{\ \mu})\!+\!\left( \ip\ell({\phi}_{\tilde{u}}^{\ \nu\bar{5}}(\sqrt{2}\ip\eta_{5\bar{5}}\delta_{\bar{\nu}}^{\ \mu}e_{\tilde{v}}^{\ 5}))
\! +\!\ip \hat{h} \hat{B}_{\tilde{u}}(+5e_{\tilde{v}}^{\ \mu})\right.\hspace{1.9cm}\nn&\\
&~\hspace{-1cm}\!+\!\left.\ip r' B_{\tilde{u}}^{\ [\rho\bar{\sigma}]} (\ip\sigma[\bar{\rho}\sigma]_{\bar{\nu} }^{\ \mu}e_{\tilde{v}}^{\ \nu})\!+\! \ip r {B}_{\tilde{u}}^{\ (\rho\bar{\sigma})}(\sigma(\bar{\rho}\sigma)_{\bar{\nu}}^{\ \mu}e_{\tilde{v}}^{\ \nu})\right)-\Gamma^{\bar{\tilde{w}}}_{\tilde{v}\tilde{u}}e_{\tilde{w}}^{\ \mu}=0.\ \ \nn&\\
&D_{\tilde{u}}(e_{\tilde{v}}^{\ \bar{\mu}}):=\ip(-\ip\delta_{\tilde{u}})(e_{\tilde{v}}^{\ \bar{\mu}})\!+\!\left( \ip\ell(\hat{\phi}_{\tilde{u}}^{\ 5\bar{\nu}}(\sqrt{2}\ip\eta_{\bar{5}5}\delta_{{\nu}}^{\ \bar{\mu}}e_{\tilde{v}}^{\ \bar{5}}))
\! +\!\ip \hat{h} \hat{B}_{\tilde{u}}(-5e_{\tilde{v}}^{\ \bar{\mu}})\right.\hspace{1.9cm}\nn&\\
&~\hspace{-1cm}\!+\!\left.\ip r' B_{\tilde{u}}^{\ [\rho\bar{\sigma}]} (\ip\sigma[\bar{\rho}\sigma]_{{\nu}}^{\ \bar{\mu}}e_{\tilde{v}}^{\ \bar{\nu}})\!+\! \ip r {B}_{\tilde{u}}^{\ (\rho\bar{\sigma})}(-\sigma(\bar{\rho}\sigma)_{{\nu}}^{\ \bar{\mu}}e_{\tilde{v}}^{\ \bar{\nu}})\right)-\Gamma^{\bar{\tilde{w}}}_{\tilde{v}\tilde{u}}e_{\tilde{w}}^{\ \bar{\mu}}=0.\ \ \nn&\\
&\hspace{-.5cm}D_{\tilde{u}}(e_{\tilde{v}}^{\ 5}):=\!\ip(-\ip\delta_{\tilde{u}})(e_{\tilde{v}}^{\ 5})\!+\!\left(\! \ip\ell(\hat{\phi}_{\tilde{u}}^{\ 5\bar{\nu}}(-\!\sqrt{2}\ip\eta_{\nu\bar{\mu}}e_{\tilde{v}}^{\ \mu}))
\! +\!\ip {h} {B}_{\tilde{u}}(-e_{\tilde{v}}^{\ 5})\! +\!\ip \hat{h} \hat{B}_{\tilde{u}}(+4e_{\tilde{v}}^{\ 5})\!\right)\hspace{.3cm}\nn&\\
&\hspace{-1cm}-\Gamma^{\bar{\tilde{w}}}_{\tilde{v}\tilde{u}}e_{\tilde{w}}^{\ 5}=0.\ \ \nn&\\
&\hspace{-.5cm}D_{\tilde{u}}(e_{\tilde{v}}^{\ \bar{5}}):=\!\ip(-\!\ip\delta_{\tilde{u}})(e_{\tilde{v}}^{\ \bar{5}})\!+\!\left(\!\ip\ell({\phi}_{\tilde{u}}^{\ \nu\bar{5}}(-\sqrt{2}\ip\eta_{\bar{\nu}{\mu}}e_{\tilde{v}}^{\ \bar{\mu}}))
\! +\!\ip {h} {B}_{\tilde{u}}(+e_{\tilde{v}}^{\ \bar{5}})\! +\!\ip \hat{h} \hat{B}_{\tilde{u}}(-4e_{\tilde{v}}^{\ \bar{5}})\!\right)\hspace{.3cm}&\nn\\
&~\hspace{-1cm}-\Gamma^{\bar{\tilde{w}}}_{\tilde{v}\tilde{u}}e_{\tilde{w}}^{\ \bar{5}}=0.\ \ \nn&
\eeaa

We will depart from the construction in \cite{Chamseddine2001} for gravity in a complex Hermitian extended space--time. In order to find the Levi-Civita connection $\Gamma^{\tilde{v}}_{\tilde{u}\tilde{w}}$, we construct the extended space--time Einstein pseudo--metric $\mathcal{G}_{\tilde{u}\tilde{v}}$:
\beaa
&\mathcal{G}_{\tilde{u}\tilde{v}}:= e_{\tilde{u}}^{\ \bar{\tilde{a}}}e_{\tilde{v}}^{\ \bar{\tilde{b}}}\eta_{\tilde{b}\tilde{a}}= \left(e_{\tilde{u}}^{\ 0}e_{\tilde{v}}^{\ \bar{0}}	-e_{\tilde{u}}^{\ 1}e_{\tilde{v}}^{\ \bar{1}}-e_{\tilde{u}}^{\ 2}e_{\tilde{v}}^{\ \bar{2}}-e_{\tilde{u}}^{\ 3}e_{\tilde{v}}^{\ \bar{3}}+\epsilon e_{\tilde{u}}^{\ 5}e_{\tilde{v}}^{\ \bar{5}}\right.\\
&\left.+e_{\tilde{u}}^{\ \bar{0}}e_{\tilde{v}}^{\ {0}}	-e_{\tilde{u}}^{\ \bar{1}}e_{\tilde{v}}^{\ {1}}-e_{\tilde{u}}^{\ \bar{2}}e_{\tilde{v}}^{\ {2}}-e_{\tilde{u}}^{\ \bar{3}}e_{\tilde{v}}^{\ {3}}+\epsilon e_{\tilde{u}}^{\ \bar{5}}e_{\tilde{v}}^{\ {5}}\right).
\eeaa
This extended space--time pseudo--metric is also invariant by construction:
\beaa
D_{\tilde{w}}(\mathcal{G}_{\tilde{u}\tilde{v}})=\delta_{\tilde{w}}(\mathcal{G}_{\tilde{u}\tilde{v}})-\Gamma^{\tilde{s}}_{\tilde{u}\tilde{w}}\mathcal{G}_{\tilde{s}\tilde{v}}-\Gamma^{\tilde{s}}_{\tilde{v}\tilde{w}}\mathcal{G}_{\tilde{u}\tilde{s}}=0.
\eeaa
 On the other hand, we expect  a real differential interval:
\beaa
ds^2=\mathcal{G}_{\tilde{u}\tilde{v}}dz^{\tilde{u}}dz^{\tilde{v}}=\bar{\mathcal{G}}_{\tilde{u}\tilde{v}}dz^{\bar{\tilde{u}}}dz^{\bar{\tilde{v}}}.
\eeaa
In the limit of a Minkowskian flat metric for $\mathcal{G}_{\tilde{u}\tilde{v}}$, we expect $e_{\tilde{v}}^{\ \tilde{b}}=\delta_{\tilde{v}}^{\ \tilde{b}}$, which would perhaps enforce non-bared to bared indices to be the only non-vanishing contributions.

We expect thus the extended space--time  $M$ to have 5 complex dimensions, corresponding to a manifold with 10 real dimensions. Besides the symmetric pseudo--metric tensor $\mathcal{G}_{\tilde{u}\tilde{v}}$, we also define an anti--symmetric tensor $\mathcal{K}_{\tilde{u}\tilde{v}}$:
\begin{center}
	\begin{tabular}{lr lr} $\mathcal{G}_{\tilde{u}\tilde{v}}=$&
		\begin{tabular}{ r | cc}
			$\tilde{u}\backslash\tilde{v}$ &$0\cdots 5$&$\bar{0}\cdots\bar{5}$ \\ \hline
			\begin{tabular}{l}\\
				0\\
				:\\
				5\end{tabular} & O & $\mathcal{G}_{u\bar{v}}$ \\ 	\begin{tabular}{l}\\
				$\bar{0}$\\
				:\\
				$\bar{5}$\end{tabular} & $\mathcal{G}_{\bar{u}v}$ & O \\
		\end{tabular}&
		,\ \ $\mathcal{K}_{\tilde{u}\tilde{v}}=$&
		\begin{tabular}{ r | cc}
			$\tilde{u}\backslash\tilde{v}$ &$0\cdots 5$&$\bar{0}\cdots\bar{5}$ \\ \hline
			\begin{tabular}{l}\\
				0\\
				:\\
				5\end{tabular} & O & $-\mathcal{G}_{u\bar{v}}$ \\ 	\begin{tabular}{l}\\
				$\bar{0}$\\
				:\\
				$\bar{5}$\end{tabular} & $\mathcal{G}_{\bar{u}v}$ & O \\
		\end{tabular}
	\end{tabular}.
\end{center}
Again, the indices $4$ and $\bar{4}$ do not appear after the contraction. Clearly,
\beaa 
\mathcal{G}_{\tilde{u}\tilde{v}}=\overline{\mathcal{G}_{\bar{\tilde{v}}\bar{\tilde{u}}}},\ \ \ \mathcal{K}_{\tilde{u}\tilde{v}}=-\overline{\mathcal{K}_{\bar{\tilde{v}}\bar{\tilde{u}}}}.
\eeaa

We require that $\mathcal{G}_{\tilde{u}\tilde{v}}, \mathcal{K}_{\tilde{u}\tilde{v}}$ are locally invertible, just like the flat-limit arrays $\eta_{\tilde{a}\tilde{b}}, \kappa_{\tilde{a}\tilde{b}}$. The inverse of $\mathcal{G}_{\tilde{u}\tilde{v}}$ is denoted as $\mathcal{G}^{\bar{\tilde{v}}\bar{\tilde{u'}}}$, and it satisfies $\mathcal{G}_{\tilde{u}\tilde{v}} \mathcal{G}^{\bar{\tilde{v}}\bar{\tilde{w}}}=\delta_{\tilde{u}}^{\ \bar{\tilde{w}}}$. Furthermore, we demand that the gravitational field components $e_{\tilde{u}}^{\ \tilde{a}}$ are locally invertible (allowing us to find the local extended Lorentzian -Minkowskian- frame from the local extended Einstein frame), and they are only nonvanishing when carrying both non-barred and bared indices. In other words, only $e_{{u}}^{\ \bar{a}}, e_{\bar{u}}^{\ {a}}$ can be non-vanishing in consonance with their flat limits $\delta_{{u}}^{\ \bar{a}}, \delta_{\bar{u}}^{\ {a}}$. The vielbein arrays $e_{{u}}^{\ \bar{a}}, e_{\bar{u}}^{\ {a}}$ are invertible and fulfill $e_{{u}}^{\ \bar{a}}e^{\bar{u}}_{\ {a'}}=\delta_{{a'}}^{\ \bar{a}},\  e_{\bar{u}}^{\ {a}}e^{{u}}_{\ \bar{a'}}=\delta_{\bar{a'}}^{\ {a}}$,
$e_{{u}}^{\ \bar{a}}e^{\bar{u'}}_{\ {a}}=\delta_{{u}}^{\ \bar{u'}},\  e_{\bar{u}}^{\ {a}}e^{{u'}}_{\ \bar{a}}=\delta_{\bar{u}}^{\ {u'}}$.
Clearly,
\beaa
\mathcal{G}^{{\tilde{v}}{\tilde{u'}}}= e^{\tilde{u}}_{\ \bar{\tilde{a}}}e^{\tilde{v}}_{\ \bar{\tilde{b}}}\eta^{\tilde{b}\tilde{a}}.
\eeaa

 To find the gravitational  Levi--Civita connection we use the construction in \cite{Chamseddine2001}. The complex metric satisfies,
\beaa
D_w\mathcal{G}_{{u}\bar{v}}=\partial_{w}\mathcal{G}_{{u}\bar{v}}- \Gamma_{uw}^{\ \bar{u}'}\mathcal{G}_{{u'}\bar{v}}-\Gamma_{\bar{v}w}^{\ {v'}}\mathcal{G}_{{u}\bar{v}'}=0,\\
D_{\bar{w}}\mathcal{G}_{{v}\bar{u}}=\overline{D_w\mathcal{G}_{{u}\bar{v}}}=\partial_{\bar{w}}\mathcal{G}_{{v}\bar{u}}- \Gamma_{v\bar{w}}^{\ \bar{v}'}\mathcal{G}_{{v'}\bar{u}}-\Gamma_{\bar{u}\bar{w}}^{\ {u'}}\mathcal{G}_{{v}\bar{u}'}=0,\\
{\rm hence}\ \ \Gamma_{v\bar{w}}^{\ \bar{v}'}=\overline{\Gamma_{\bar{v}w}^{\ {v'}}},\ \ \Gamma_{\bar{u}\bar{w}}^{\ {u'}}=\overline{\Gamma_{uw}^{\ \bar{u}'}}.\label{CongGamma}
	\eeaa
A solution to this system compatible with the complex structure is the Chern--Connection which is given in terms of the torsionless connection $\mathring{\Gamma}_{\tilde{u}\tilde{w}}^{\ {\tilde{u}'}}$, and the torsion tensors $\mathcal{T}_{\tilde{u}w{\tilde{u}''}}$ and $\hat{\mathcal{T}}_{\tilde{u}\bar{w}{\tilde{u}''}}$:
\beaa
{\Gamma}_{\tilde{u}w}^{\ {\tilde{u}'}}&=&\mathring{\Gamma}_{\tilde{u}{w}}^{\ {\tilde{u}'}}+\half\mathcal{G}^{{\tilde{u}'}\bar{\tilde{u}}''}\mathcal{T}_{\tilde{u}w{\tilde{u}''}},\ {\rm where}\\
\mathring{\Gamma}_{\tilde{u}{w}}^{\ {\tilde{u}}'}&=&\half\mathcal{G}^{{\tilde{u}'}\bar{\tilde{u}}''}(\partial_w\mathcal{G}_{\tilde{u}\tilde{u}''}+\partial_{\tilde{u}}\mathcal{G}_{\tilde{u}''w}-\partial_{\tilde{u}''}\mathcal{G}_{w\tilde{u}}),\nn\\ \mathcal{T}_{\tilde{u}w{\tilde{u}''}}&=&(\partial_w\mathcal{K}_{\tilde{u}''\tilde{u}}+\partial_{\tilde{u}''}\mathcal{K}_{\tilde{u}w}+\partial_{\tilde{u}}\mathcal{K}_{w\tilde{u}''});\\
{\Gamma}_{\tilde{u}\bar{w}}^{\ {\tilde{u}'}}&=&\mathring{\Gamma}_{\tilde{u}\bar{w}}^{\ {\tilde{u}'}}+\half\mathcal{G}^{{\tilde{u}'}\bar{\tilde{u}}''}\hat{\mathcal{T}}_{\tilde{u}\bar{w}{\tilde{u}''}},\ {\rm where}\\
\mathring{\Gamma}_{\tilde{u}\bar{w}}^{\ {\tilde{u}}'}&=&\half\mathcal{G}^{{\tilde{u}'}\bar{\tilde{u}}''}(\partial_{\bar{w}}\mathcal{G}_{\tilde{u}''\tilde{u}}+\partial_{\tilde{u}}\mathcal{G}_{\bar{w}\tilde{u}''}-\partial_{\tilde{u}''}\mathcal{G}_{\tilde{u}\bar{w}}),\nn\\ \hat{\mathcal{T}}_{\tilde{u}\bar{w}{\tilde{u}''}}&=&(\partial_{\bar{w}}\mathcal{K}_{\tilde{u}\tilde{u}''}+\partial_{\tilde{u}''}\mathcal{K}_{\bar{w}\tilde{u}}+\partial_{\tilde{u}}\mathcal{K}_{\tilde{u}''\bar{w}})=\overline{\mathcal{T}_{\bar{\tilde{u}}w\bar{\tilde{u}}''}}.
\eeaa
Clearly $\mathring{\Gamma}_{\tilde{u}{w}}^{\ {\tilde{u}}'}=\mathring{\Gamma}_{{w}\tilde{u}}^{\ {\tilde{u}}'}=\overline{\mathring{\Gamma}_{\bar{\tilde{u}}\bar{w}}^{\ \bar{\tilde{u}}'}}$ is a torsionless part. From these we obtain:
\beaa
\Gamma_{uw}^{\ \bar{v}}=\mathcal{G}^{v'{\bar{v}}}\partial_w\mathcal{G}_{u{\bar{v}'}},\ \ \Gamma_{\bar{u}w}^{\ {v}}=0,\\
\Gamma_{\bar{u}\bar{w}}^{\ {v}}=\mathcal{G}^{\bar{v}'{{v}}}\partial_{\bar{w}}\mathcal{G}_{{{v}'}\bar{u}},\ \ \Gamma_{u\bar{w}}^{\ \bar{v}}=0.
\eeaa
This fulfills (\ref{CongGamma}). Now, there are 25 independent Vielbeins (not related by conjugation), and 10x25 independent equations (\ref{CovDerVielb}). The LE has 25 generators, and there are 10x25 LE connections. So, the 2nd. order gravity endeavor seems possible. Multiplying (\ref{CovDerVielb}) by $e^{\bar{\tilde{v}}}_{\ \tilde{b}}$ and projecting into the trace, antisymmetric, and symmetric parts, we obtain the 2nd. order gravity relations:
\beaa
\ell{\phi}_{{u}}^{\ \nu\bar{5}}=\frac{1}{\sqrt{2}}\eta^{\bar{5}5}e^v_{\ \bar{5}}\partial_u e_{\bar{v}}^{\ \nu}=\overline{\ell\hat{\phi}_{\bar{u}}^{\ 5\bar{\nu}}},\ \ \ell{\phi}_{\bar{u}}^{\ \nu\bar{5}}=-\frac{1}{\sqrt{2}}\eta^{\bar{\mu}\nu}e^{\bar{v}}_{\ \mu}\partial_{\bar{u}} e_{{v}}^{\ \bar{5}}=\overline{\ell\hat{\phi}_{{u}}^{\ 5\bar{\nu}}},\ \label{2ndOrderStart}\\
\hat{h}\hat{B}_{{u}}=\frac{\ip}{20}e^v_{\ \bar{\mu}}\partial_u e_{\bar{v}}^{\ \mu}=\overline{\hat{h}\hat{B}_{\bar{u}}},\ \ {h}{B}_{{u}}=-\ip e^v_{\ \bar{5}}\partial_u e_{\bar{v}}^{\ 5}\!+\!\frac{\ip}{5}e^v_{\ \bar{\mu}}\partial_u e_{\bar{v}}^{\ \mu}=\overline{{h}{B}_{\bar{u}}},\\
r'B^{\ [\sigma\bar{\rho}]}_{u}=
\!\frac{1}{4}\eta^{\bar{\sigma}'\sigma}\eta^{\bar{\rho}\rho'}\sigma[\bar{\rho}'\sigma']_{\bar{\mu}}^{\ \nu}e^{{v}}_{\ \bar{\nu}}\partial_{{u}} e_{\bar{v}}^{\ {\mu}}\hspace{5cm}\nn\\
\!=\!-\frac{1}{4}(\eta^{\bar{\rho}\rho'}\delta_{\sigma'}^{\ \sigma}-\eta^{\bar{\sigma}\sigma'}\delta_{\sigma'}^{\ \rho})e^{{v}}_{\ \bar{\rho}'}\partial_{{u}} e_{\bar{v}}^{\ {\sigma'}}\!
=\overline{r' B_{\bar{u}}^{\ [\sigma\bar{\rho}]}},\\
r {B}_{{u}}^{\ (\sigma\bar{\rho})}=\!\frac{\ip}{4}\eta^{\bar{\sigma}'\sigma}\eta^{\bar{\rho}\rho'}\sigma(\bar{\rho}'\sigma')_{\bar{\mu}}^{\ \nu}e^{{v}}_{\ \bar{\nu}}\partial_{{u}} e_{\bar{v}}^{\ {\mu}}\hspace{5cm}\nn\\
\!=\!\frac{\ip}{4}(\eta^{\bar{\rho}\rho'}\delta_{\bar{\sigma}'}^{\ \sigma}+\eta^{\bar{\sigma}\rho'}\delta_{\bar{\sigma}'}^{\ \rho}-\half\eta^{\bar{\rho}\sigma}\delta_{\bar{\sigma}'}^{\ \rho'})e^{{v}}_{\ \bar{\rho}'}\partial_{{u}} e_{\bar{v}}^{\ {\sigma}'}
=\overline{r B_{\bar{u}}^{\ (\sigma\bar{\rho})}}.\label{2ndOrderEnd} 
\eeaa
The Lagrangian associated to the equations of motion can be devised as in \cite{Chamseddine2001}. Let us explore symmetry breaking using an adjoint translation multiplet and its invariant in (\ref{CasPEx2}-\ref{CasimirPExM2}) prior to 2nd. order gravity:
\beaa
{\cal C}'_2&=&K^\mu\otimes\hat{P}^*_\mu+\bar{K}^{\bar{\mu}}\otimes\hat{P}_\mu+K^5\otimes\bar{\Phi}+\bar{K}^{\bar{5}}\otimes{\Phi},\\
({\cal C}'_2)^2&=&\frac{g_{44}}{2}\{K^\mu,\bar{K}_{\bar{\mu}}\}+\frac{g_{44}\epsilon}{2}\{K^5,\bar{K}_{\bar{5}}\},\\
{\cal L}'_{K}&=&g_{44}D_{\tilde{u}}(K^\mu)D_{\tilde{v}}(\bar{K}^{\bar{\mu}'}){\cal G}^{\bar{\tilde{u}}\bar{\tilde{v}} }\eta_{\bar{\mu}\mu'}+
g_{44}\epsilon D_{\tilde{u}}(K^5)D_{\tilde{v}}(\bar{K}^{\bar{5}}){\cal G}^{\bar{\tilde{u}}\bar{\tilde{v}} }\nn \\
&\ &-{\mu}^2 ({\cal C}'_2)^2-\lambda(({\cal C}'_2)^2)^2.
\eeaa
The non--trivial extended Lorentz transformation of the adjoint multiplet $K^\mu, \bar{K}^{\bar{\mu}}$, $K^5, \bar{K}^{\bar{5}}$ are given by:
\beaa
\,[T_{[o]}, K^5]_1&=&-K^5,\ \ \ \ \ \  [T_{[o]}, \bar{K}^{\bar{5}}]_1=+\bar{K}^{\bar{5}},\\
\,[\hat{T}_{[o]}, K^\mu]_1&=&+5K^\mu,\ \ \ \ [\hat{T}_{[o]}, \bar{K}^{\bar{\mu}}]_1=-5\bar{K}^{\bar{\mu}},\\
\,[\hat{T}_{[o]}, K^5]_1&=&+4K^5,\ \ \ \ [\hat{T}_{[o]}, \bar{K}^{\bar{5}}]_1=-4\bar{K}^{\bar{5}},\\
\,[M_{\bar{\rho}\sigma}, K^\mu]_1&=&\ip\sigma[\bar{\rho}\sigma]_{\bar{\nu}}^{\ \mu} K^\nu,\ \ \ [M_{\bar{\rho}\sigma}, \bar{K}^{\bar{\mu}}]_1=\ip\sigma[\bar{\rho}\sigma]_{\nu}^{\ \bar{\mu}}\bar{K}^{\bar{\nu}},\\
\,[T_{\bar{\rho}\sigma}, K^\mu]_1&=&+\sigma(\bar{\rho}\sigma)_{\bar{\nu}}^{\ \mu} K^\nu,\ \ [T_{\bar{\rho}\sigma}, \bar{K}^{\bar{\mu}}]_1=-\sigma(\bar{\rho}\sigma)_{\nu}^{\ \bar{\mu}}\bar{K}^{\bar{\nu}},\\
\,[\hat{T}_{\bar{\rho}5}, K^\mu]_1&=&\sqrt{2}\ip\epsilon\delta_{\bar{\rho}}^{\ \mu} K^5,\ \ [T_{\bar{5}\rho}, \bar{K}^{\bar{\mu}}]_1=\sqrt{2}\ip\epsilon\delta_\rho^{\ \bar{\mu}}\bar{K}^{\bar{5}},\\
\,[{T}_{\bar{5}\rho}, K^5]_1&=&-\sqrt{2}\ip\eta_{\rho\bar{\mu}} K^\mu,\ \ [\hat{T}_{\bar{\rho}5}, \bar{K}^{\bar{5}}]_1=-\sqrt{2}\ip\eta_{\mu\bar{\rho}}\bar{K}^{\bar{\mu}}.
\eeaa
To preserve Lorentz symmetry, the extreme of the potential for $-\mu^2>0$ gives VEV only to the scalar field:
\beaa
<K^5><\bar{K}^{\bar{5}}>=|<K^5>|^2=\frac{-\mu^2}{2\lambda g_{44}\epsilon},\ K^5=K'^5+<K^5>.
\eeaa
The kinetic terms give rise to the following mass--term contributions:
\beaa g_{44}D_{\tilde{u}}(K^\mu)D_{\tilde{v}}(\bar{K}^{\bar{\mu}'}){\cal G}^{\bar{\tilde{u}}\bar{\tilde{v}} }\eta_{\bar{\mu}\mu'}+
g_{44}\epsilon D_{\tilde{u}}(K^5)D_{\tilde{v}}(\bar{K}^{\bar{5}}){\cal G}^{\bar{\tilde{u}}\bar{\tilde{v}} }\nn \\
=\cdots+(-g_{44})\left(\ell\phi_{\tilde{u}}^{\mu\bar{5}}\ell\hat{\phi}_{\tilde{v}}^{5\bar{\mu}'}\epsilon\eta_{\mu'\bar{\mu}}{\cal G}^{\bar{\tilde{u}}\bar{\tilde{v}} }\right)(|<K^5>|^2 \epsilon)\nn\\
+(g_{44})\left((hB_{\tilde{u}}-4\hat{h}\hat{B}_{\tilde{u}})(hB_{\tilde{v}}-4\hat{h}\hat{B}_{\tilde{v}}){\cal G}^{\bar{\tilde{u}}\bar{\tilde{v}} }\right)(|<K^5>|^2 \epsilon).
\eeaa
The introduction of such mass terms involves transferring degrees of freedom (DOF) from the adjoint multiplets $K^\mu, \bar{K}^{\bar{\mu}}$, $K^5, \bar{K}^{\bar{5}}$ to the gauge fields, but there may not be sufficient DOF available for this purpose. It's possible that the (complex) degrees of freedom of $e_5^{\ \bar{5}}, e_{\bar{5}}^{\ 5}$ and of $e_5^{\ \bar{\mu}}, e_{\bar{5}}^{\ \mu}$ also need to be sacrificed to render these fields massive. It would be desirable if additionally the DOF from $B_u^{\ (\sigma\rho)}$ could be transferred to the new massive fields. Alternatively, a larger Lorentz scalar multiplet might be required. For instance, one could involve an adjoint multiplet of the LE. Such considerations will be addressed in more realistic models when dealing with $SU(2)\times U(1)$ or even larger internal symmetry groups.

  \section{Conclusions} 
 An inquiry into transformations mixing gauge and Higgs--like scalar fields in  \cite{U1I} lead to a Lie algebraic symmetry, called the Lorentz Extension (LE). It involved  an internal $U(1)$ as well as space--time symmetries such as Lorentz transformations. Now, an extension,  called Noncontracted Poincar\'e Extension (NPE), has been devised involving  translation precursors prior to a contraction procedure. The contracted lie algebra was called the Contracted Poincar\'e Extension (CPE). A NPE--invariant gauge field model was constructed, and several features of its consistency were examined. The model is locally invariant under the NPE  that includes internal as well as space--time symmetries on a manifold that recreates itself the symmetry. These exploration with just an internal $U(1)$ symmetry paves the way for more realistic extensions, with say $S(2)\times U(1)$, or greater internal symmetries. Briefly, a preliminary approach to symmetry breaking was discussed here.  The equations of motion and conserved charges obtained the LE model in \cite{U1I} can be easily extended for the NPE--model. 
 
 In some sense, the NPE model addresses non--commutative geometry since there even the translations do not commute among themselves. Most masses turn out to be in direct correspondence with such non--commutativity, and the symmetry breaking procedure had the task of making the SM gauge bossons massless again, and to move the corresponding degrees of freedom to make some goldstone bossons massive.

As indicated in \cite{U1I}, the space--time complexification aligns in part to several  early developments in \cite{EinsteinStrauss}\cite{ESchr},  and to manifold constructions in \cite{TorresdC}\cite{Hughston}\cite{Witten}\cite{Ashtekar}\cite{PST}\cite{Chamseddine2001}\cite{GORS}
\cite{Chamseddine2004}\cite{Valtancoli}
\cite{Esposito}\cite{Bogos}\cite{Chamseddine2006}\cite{LiSotiriouBarrow}\cite{SotiriouLieBarrow}\cite{HeckmanVerline}\cite{ChamConnes}\cite{MMZ}\cite{Yang}\cite{Tanaka}\cite{CMJR}\cite{RS1}\cite{RS2}. Also, recent measurements of branching rations of Higgs decaying into gauge bosons \cite{Aadetal}. Also, several Beyond Standard Models (BSM) have been examined  to cope with phenomenological deviations to SM predictions  such as \cite{CompH}\cite{Caoetal}\cite{NMSSM}\cite{GHU}\cite{Lowetal}.  The low energy phenomenological models resulting from the models studied here and further ones involving Electroweak internal symmetries might provide  BSM enhancing certain Higgs  decay chanels.

Clearly, there are multiple challenges to substantiate the nature of eons where  non--commutative dynamic is centered around a Lie Manifold before a transition to the present 4--dimensional era. One step in this direction would be to have gravitation in a complex space--time with extended Lorentz symmetry LE, like the one whose $2^{nd.}$ order gravity has been devised here. There, an intimate mixing between external and would--be internal symmetries takes place.  This  plants the seeds that  might render naturality to some SM structures once the transition to  real 4 real space--time is found. We hope, they could bring insight into the origin and nature of intriguing portions such as particle--antiparticle asymmetry, dark energy and matter. Furthermore, the imposed symmetry  might enrich the graviton field to address quantization challenges.
 
\section{Acknowledgments} Initial developments  towards this exploration  took place at the  Departamento de F\'{\i}sica Te\'orica y del Cosmos, Universidad de Granada, Spain, and CIMAT, Guanajuato, M\'exico. Financial support of the research period in Spain by the Fundaci\'on Carolina, Spain, is acknowledged. Finantial support by CIMAT for the research period in M\'exico is acknowledged. The Universidad Nacional de Colombia, Sede Medell\'{\i}n, is aknowleged for its support of this research.

\appendix
\section{Quartic CPE invariant}
Besides the CPE quadratic invariant (\ref{CasPEx2}-\ref{CasimirPExM2}), we find a CPE quartic invariant departing from the constructions in \cite{Invariants}:
\beaa
A_1&=&\sum_{A\in\{\mu,5\}}\tilde{M}^{A}_{\ A}=-\tilde{M}^{4}_{\ 4},\ \ A_2=\sum_{A,B\in\{\mu,5\}}\tilde{M}^{AB}\tilde{M}_{BA},
\eeaa
\beaa
\mathbb{S}&=&8(A_1^2-A_2)=-8\!\!\sum_{A,B\in\{\mu,5\}}\!\!(\tilde{M}_{\stackrel{A}{.}}^{\ B}\tilde{M}_{\stackrel{B}{.}}^{\ A}-\tilde{M}_{B}^{\ B}\tilde{M}_{A}^{\ A}),\\
B_2&=&\sum_{A\in\{\mu,5\}}\tilde{M}_4^{\ A}\tilde{M}_{A4}=-\frac{R^2}{8}(2 \{\hat{P}^{\mu},\hat{P}^*_{\mu}\}+\frac{2}{\epsilon} \{\Phi, \bar{\Phi}\}).
\eeaa
The combination $\mathbb{S}=8(A_1^2-A_2)$ is an anti--symmetrized expression over the under-dotted indices, and it is LE--invariant. $B_2$ is clearly a CPE-invariant. Further anti--symmetrized quadratic expressions —involving only anti-commutators $\{\cdot,\cdot\}$ with indices in $\{\mu,5\}$ where at most 2 are not repeated— are:
\beaa
\mathbb{S}&=&-4\!\!\sum_{A,B\in\{\nu,5\}}\!\!(\{\tilde{M}_{\stackrel{A}{.}}^{\ B},\tilde{M}_{\stackrel{B}{.}}^{\ A}\}-\{\tilde{M}_{B}^{\ B},\tilde{M}_{A}^{\ A}\})=\{2\ip\tilde{M}_{\sigma}^{\ \rho},2\ip\tilde{M}_{\rho}^{\ \sigma}\}\nn\\
&&+2\left(
\!\frac{2}{\epsilon}\!\right)\!\{\hat{T}_\mu, T^{\mu}\}+\!4\{T_{[o]},T_{[o]}\}
\!-\{\frac{2}{5}(T_{[o]}-\hat{T}_{[o]}),\frac{2}{5}(T_{[o]}-\hat{T}_{[o]})\},\ \ \\
\mathbb{S}_{\mu\sigma}&=&-4\!\!\sum_{B\in\{\nu,5\}}\!\!(\{\tilde{M}_{\stackrel{\mu}{.}}^{\ B},\tilde{M}_{\stackrel{B\sigma}{.\ }}\}-\{\tilde{M}_{B}^{\ B},\tilde{M}_{\mu\sigma}\})=g^{\nu\rho}\{2\ip\tilde{M}_{\mu\nu},2\ip\tilde{M}_{\rho \sigma}\}\nn\\
&&+\left(\!\frac{2}{\epsilon}\!\right)\{\hat{T}_\mu, T_{\sigma}\}+\{\frac{2}{5}(T_{[o]}-\hat{T}_{[o]}),2\ip\tilde{M}_{\mu\sigma}\},\nn\\
\mathbb{S}_{5}^{\ 5}&=&-4\!\!\sum_{B\in\{\nu,5\}}\!\!(\{\tilde{M}_{\stackrel{5}{.}}^{\ B},\tilde{M}_{\stackrel{B}{. }}^{\ 5}\}-\{\tilde{M}_{B}^{\ B},\tilde{M}_{5}^{\ 5}\})\nn\\
&&=\left(
\!\frac{2}{\epsilon}\!\right)\{\hat{T}_\mu, T^{\mu}\}+4\{T_{[o]},T_{[o]}\}
-2\{\frac{2}{5}(T_{[o]}-\hat{T}_{[o]}),T_{[o]}\},
\eeaa
\beaa
\mathbb{V}_{5\mu}&=&2\sqrt{2}\!\!\sum_{B\in\{\nu,5\}}\!\!(\{\tilde{M}_{\stackrel{5}{.}}^{\ B},\tilde{M}_{\stackrel{B\mu}{.\ }}\}-\{\tilde{M}_{B}^{\ B},\tilde{M}_{5\mu}\})\nn\\
&&=\ip\{T^{\nu},2\ip\tilde{M}_{\nu \mu}\}-2\ip\{{T}_\mu, T_{[o]}\}+\ip\{\frac{2}{5}(T_{[o]}-\hat{T}_{[o]}),{T}_\mu\}.
\eeaa
The adjoint $\mathbb{V}_{\mu 5}^{\dagger}$ is $\mathbb{V}_{\mu 5}^{\dagger}=(\mathbb{V}_{5\mu})^\dagger$, $\mathbb{S}_{\mu\sigma}^\dagger=\mathbb{S}_{\sigma\mu}$, etc., with $\hat{T}_\mu=(T_\mu)^\dagger$, $\hat{P}^*_{\mu}=(\hat{P}_{\mu})^\dagger$, etc. Notice that adjunction inverts the order of the indices. The action of dimension 1 generators $\hat{P}_{\mu},\cdots, \bar{\Phi}$ over those above produces multiplet components:
\beaa
&&L_{4\mu}=\frac{4\ip}{R}\!\!\sum_{B\in\{\nu,5\}}\!\!(\{\tilde{M}_{\stackrel{4}{.}}^{\ B},\tilde{M}_{\stackrel{B\mu}{.\ }}\}-\{\tilde{M}_{B}^{\ B},\tilde{M}_{4\mu}\})\nn\\
&&\ \ \ \ \ \ =\{\sqrt{2}\hat{P}^{\nu},2\ip\tilde{M}_{\nu \mu}\}+\ip\left(
\!\frac{2}{\epsilon}\!\right)\{\Phi, T_{\mu}\}+\{\sqrt{2}\hat{P}^{\nu}, \frac{2}{5}(T_{[o]}-\hat{T}_{[o]})\},\ \ \\
&&\tilde{L}_{45}=-\frac{2\sqrt{2}\ip}{R}\!\!\sum_{B\in\{\nu,5\}}\!\!(\{\tilde{M}_{\stackrel{4}{.}}^{\ B},\tilde{M}_{\stackrel{B5}{.\ }}\}-\{\tilde{M}_{B}^{\ B},\tilde{M}_{45}\})\nn\\
&&\ \ \ \ \ \ =-\ip\{\sqrt{2}\hat{P}^{\nu},\hat{T}_{\nu }\}-2\{\Phi, T_{[o]}\}+\{\Phi, \frac{2}{5}(T_{[o]}-\hat{T}_{[o]})\},\\
&&{\hat{K}_{4\sigma}}\!\!~_{\mu\nu}=\frac{4\ip}{R}(\{\tilde{M}_{\stackrel{4\sigma}{\ .}},\tilde{M}_{\stackrel{\mu\nu}{\ . }}\}-\{\tilde{M}_{4\nu},\tilde{M}_{\mu\sigma}\})\nn\\
&&\ \ \ \ \ \ \ \ \ 
=\{\sqrt{2}\hat{P}_{\sigma},2\ip\tilde{M}_{\mu \nu}\}-\{\sqrt{2}\hat{P}_{\nu},2\ip\tilde{M}_{\mu \sigma}\},
\eeaa
\beaa
&{{\Pi}_{4\nu}}\!\!~_{5\sigma}=-\frac{2\sqrt{2}\ip}{R}(\{\tilde{M}_{\stackrel{4\nu}{\ .}},\tilde{M}_{\stackrel{5\sigma}{\ . }}\}\!-\!\{\tilde{M}_{4\sigma},\tilde{M}_{5\nu}\})=\!\sqrt{2}\ip(\{\hat{P}_{\nu},T_{\sigma}\}\!-\!\{\hat{P}_{\sigma},T_{\nu}\}),\ \ \\
&{\hat{W}_{4}^{\ 5}}\!\!~_{5\mu}=\frac{4\ip}{R}(\{\tilde{M}^{\ 5}_{\stackrel{4}{\ .}},\tilde{M}_{\stackrel{5\mu}{.\  }}\}\!-\!\{\tilde{M}_{5}^{\ 5},\tilde{M}_{4\mu}\})\!=\!\ip\!\left(
\!\frac{2}{\epsilon}\!\right)\!\!\{\Phi,T_{\mu}\}\!+2\{\sqrt{2}\hat{P}_{\mu},T_{[o]}\},\ \ \\
&{{K}_{45}}\!\!~_{\mu\nu}=-\frac{2\sqrt{2}\ip}{R}(\{\tilde{M}_{\stackrel{45}{\ .}},\tilde{M}_{\stackrel{\mu\nu}{\ . }}\}-\{\tilde{M}_{4\nu},\tilde{M}_{\mu5}\})\nn\\
&=\{\Phi,2\ip\tilde{M}_{\mu \nu}\}+\ip\{\sqrt{2}\hat{P}^{\nu},\hat{T}_\mu\}.
\eeaa
The adjoints $L^\dagger_{\mu 4}, \tilde{L}^\dagger_{54}, {\hat{K}^\dagger_{\nu\mu}\!\!~_{\sigma 4}}, {{\Pi}^\dagger_{\sigma 5}\!\!~_{\nu 4}}, {\hat{W}^\dagger}_{\mu 5}\!\!~_{\ 4}^{5}, {{K}^\dagger}_{\nu\mu}\!\!~_{54}$ are obtained correspondingly. Acting with the dimension 1 generators one obtains dimension 2 quadratic expressions:
\beaa
S&=&\!-\frac{4}{R^2}\!\sum_{B\in\{\nu,5\}}\!\!\{\tilde{M}_4^{\ B},\tilde{M}_{B4}\}=-\frac{8}{R^2}B_2=2\{\hat{P}^{\nu},\hat{P}^*_{\nu}\}\!+\!\frac{2}{\epsilon} \{\Phi, \bar{\Phi}\},\ \\
S^{\mu\sigma}&=&-\frac{4}{R^2}\{\tilde{M}_4^{\ \mu},\tilde{M}^\sigma_{\ 4}\}=2\{\hat{P}^{\mu},\hat{P}^{*\sigma}\},\\
S^5_{\ 5}&=&
-\frac{4}{R^2}\{\tilde{M}_{4}^{\ 5},\tilde{M}_{54}\}=\frac{2}{\epsilon} \{\Phi, \bar{\Phi}\},\\
\!V^\mu_{\ \ 5}&=\!&\!\!\!\!
\frac{2\sqrt{2}}{R^2}\{\tilde{M}_{4}^{\ \mu}\!,\tilde{M}_{54}\}\!=\!\{\sqrt{2}\hat{P}^{\mu}, \bar{\Phi}\},\\ 
\!V_5^{\dagger\,\mu}&=\!&
\frac{2\sqrt{2}}{R^2}\{\tilde{M}_{45},\tilde{M}^{\mu}_{\ 4}\}\!=\!\{ {\Phi},\sqrt{2}\hat{P}^{* \mu}\}.\ \ 
\eeaa
The quartic invariant of ${\mathfrak{iu}}(t_2,s_2)$ with $s_2>2,t_2>0$ and $s_2+t_2=5$, seemingly corresponding to an extension of the squared Pauli--Lubanski pseudovector, is given by:
\beaa
{\cal{I}}_4&=&-\half\{\mathbb{S},S\}+\{\mathbb{S}_{\mu\sigma},S^{\mu\sigma}\}+\{\mathbb{S}_5^{\ 5},S^{5}_{\ 5}\}
+\frac{2}{\epsilon}\{\mathbb{V}_{5\mu},V_5^{\dagger\,\mu}\}+\frac{2}{\epsilon}\{\mathbb{V}^\dagger_{\mu 5},V^\mu_{\ \ 5}\}\nn\\
&&\!\!+\{L_{4}^{\ \mu},L^\dagger_{\mu 4} \}+\!\frac{2}{\epsilon}\{\tilde{L}_{45},\tilde{L}^\dagger_{54}\}-\!\half\{\hat{K}_{4\sigma}\!\!~_{\mu\nu},\hat{K}^{\dagger\nu\mu}\!~^{\sigma}_{\ 4}\}-\!\half\!\left(\!\frac{2}{\epsilon}\!\right)\!\{{\Pi}_{4\nu}\!\!~_{5\sigma} ,{\Pi}^{\dagger\sigma}_{\ \  5}\!\!~^{\nu}_{\ 4}\}\nn\\
&& 
-\{\hat{W}_{4}^{\ 5}\!\!~_{5}^{\ \mu},\hat{W}^\dagger_{\mu 5}\!\!~_{\ 4}^{5}\}-\frac{2}{\epsilon}\{{K}_{45}\!\!~^{\mu\nu},{K}^\dagger_{\nu\mu}\!\!~_{54}\}.
\eeaa
The invariance of ${\cal{I}}_4$ under $M_{\mu\nu}$, $T_{\mu\nu}$, $T_{[o]}$, and $\hat{T}_{[o]}$ is easily tested. Invariance under dimension 1 generators and $T_\mu, \hat{T}_\mu$ follows from:
\beaa
\,[\hat{P}_\tau,\mathbb{S}]&=&-2\sqrt{2}L_{4\tau},\hspace{2.0cm} [\hat{P}^*_\tau,\mathbb{S}]=+2\sqrt{2}L^\dagger_{\tau 4},\label{QuartTransInic}\\
\,[\Phi,\mathbb{S}]&=&-4\tilde{L}_{45},\hspace{2.7cm} [\bar{\Phi},\mathbb{S}]=+4\tilde{L}^\dagger_{54},\\
\,[T_\tau,\mathbb{S}]&=&0,\hspace{3.4cm} [\hat{T}_\tau,\mathbb{S}]=0,\\
\,[\hat{P}_\tau,\mathbb{S}_{\mu\sigma}]\!\!&\!=\!\!&\!\!\!-\sqrt{2}g_{\mu\tau}L_{4\sigma}\!-\!\sqrt{2}\hat{K}_{4\sigma}\!\!~_{\mu\tau},\, [\hat{P}^*_\tau\!,\mathbb{S}_{\sigma\mu}]\!=\!\sqrt{2}g_{\tau\mu}L^\dagger_{\sigma 4}\!+\!\sqrt{2}\hat{K}^{\dagger}_{\tau\mu}\!\!~_{\sigma 4},\ \ \ \ \\
\,[\Phi,\mathbb{S}_{\mu\sigma}]&=&+2{K}_{45}\!\!~_{\mu\sigma},\hspace{2.0cm} [\bar{\Phi},\mathbb{S}_{\sigma\mu}]=-2{K}^\dagger_{\sigma\mu}\!\!~_{54},\\
\,[T_\tau,\mathbb{S}_{\mu\sigma}]&=&2\ip g_{\mu\tau}\mathbb{V}_{5\sigma},\hspace{1.8cm} [\hat{T}_\tau,\mathbb{S}_{\sigma\mu}]=2\ip g_{\tau\mu}\mathbb{V}^\dagger_{\sigma 5},\\
\,[\hat{P}_\tau,\mathbb{S}_{5}^{\ 5}]&=&
\!\!-\sqrt{2}\hat{W}_{4}^{\ 5}\!\!~_{5\tau},\hspace{1.7cm} [\hat{P}^*_\tau,\mathbb{S}_{5}^{\ 5}]=
\!+\sqrt{2}\hat{W}^\dagger_{\tau 5}\!\!~_{\ 4}^{5},\ \ \\
\,[\Phi,\mathbb{S}_{5}^{\ 5}]&=&-2\tilde{L}_{45},\hspace{2.5cm} [\bar{\Phi},\mathbb{S}_{5}^{\ 5}]=+2\tilde{L}^\dagger_{45},\\
\,[T_\tau,\mathbb{S}_{5}^{\ 5}]&=&-2\ip \mathbb{V}_{5\tau},\hspace{2.1cm} [\hat{T}_\tau,\mathbb{S}_{5}^{\ 5}]=-2\ip \mathbb{V}^\dagger_{\tau 5},\\
\,[\hat{P}_\tau,\mathbb{V}_{5\mu}]&=&
\!\!-\sqrt{2}{\Pi}_{4\mu}\!\!~_{5\tau},\hspace{1.1cm} [\hat{P}^*_\tau,\mathbb{V}_{5\mu }]=
\!+\sqrt{2}g_{\mu\tau}\tilde{L}^\dagger_{54}\!+\!\sqrt{2}{K}^\dagger_{\tau\mu}\!\!~_{54},\ \ \ \\
\,[\Phi,\mathbb{V}_{5\mu }]&=&-\epsilon L_{4\mu}+\epsilon\hat{W}_{4}^{\ 5}\!\!~_{5\mu},\hspace{0.6cm} [\bar{\Phi},\mathbb{V}_{5\mu }]=0,\\
\,[T_\tau,\mathbb{V}_{5\mu }]&=&0,\hspace{3.0cm} [\hat{T}_\tau,\mathbb{V}_{5\mu }]=-\ip\epsilon\mathbb{S}_{\tau\mu}+\ip\epsilon g_{\tau\mu} \mathbb{S}_{5}^{\ 5},\\
\eeaa
\beaa
\,[\hat{P}_\tau,L_{4\mu}]&=&
0,\hspace{3.0cm} [\hat{P}^*_\tau,L_{4\mu}]=
\!+\sqrt{2}g_{\mu\tau}S-\sqrt{2}S_{\mu\tau},\ \ \\
\,[\Phi,L_{4\mu}]&=&0,\hspace{3.2cm} [\bar{\Phi},L_{4\mu}]=-2V_{\mu 5},\\
\,[T_\tau,L_{4\mu}]&=&0,\hspace{3.1cm} [\hat{T}_\tau,L_{4\mu}]=2\ip g_{\mu\tau}\tilde{L}_{45},\\
\,[\hat{P}_\tau,\tilde{L}_{45}]&=&
0,\hspace{3.0cm} [\hat{P}^*_\tau,\tilde{L}_{45}]=
\!-\sqrt{2}V^\dagger_{5\tau},\ \ \\
\,[\Phi,\tilde{L}_{45}]&=&0,\hspace{3.2cm} [\bar{\Phi},\tilde{L}_{45}]=\epsilon S-\epsilon S^{5}_{\  5},\\
\,[T_\tau,\tilde{L}_{45}]&=&-\ip\epsilon L_{4\tau},\hspace{2.1cm} [\hat{T}_\tau,\tilde{L}_{45}]=0,\\
\,\!\![\hat{P}_\tau,\hat{K}_{4\sigma}\!\!~_{\mu\nu}]&=&
0,\hspace{2.2cm} [\hat{P}^*_\tau,\hat{K}_{4\sigma}\!\!~_{\mu\nu}]\!=
\!\sqrt{2}g_{\nu\tau}S_{\sigma\mu}\!-\!\sqrt{2}g_{\sigma\tau}S_{\nu\mu},\ \ \ \\
\,[\Phi,\hat{K}_{4\sigma}\!\!~_{\mu\nu}]&=&0,\hspace{2.8cm} [\bar{\Phi},\hat{K}_{4\sigma}\!\!~_{\mu\nu}]=0,\\
\,\!\![T_\tau,\hat{K}_{4\sigma}\!\!~_{\mu\nu}]&=&2\ip g_{\mu\tau}{\Pi}_{4\sigma}\!\!~_{5\nu},\;  [\hat{T}_\tau,\hat{K}_{4\sigma}\!\!~_{\mu\nu}]\!=\!2\ip\! g_{\sigma\tau}{K}_{45}\!\!~_{\mu\nu}\!-\!2\ip\! g_{\nu\tau}{K}_{45}\!\!~_{\mu\sigma},\ \ \\
\,[\hat{P}_\tau,{\Pi}_{4\nu}\!\!~_{5\sigma}]&=&
0,\hspace{2.2cm} [\hat{P}^*_\tau,{\Pi}_{4\nu}\!\!~_{5\sigma}]=\!\sqrt{2}g_{\sigma\tau}V_{\nu 5}\!-\!
\sqrt{2}g_{\nu\tau}V_{\sigma 5},\ \ \\
\,[\Phi,{\Pi}_{4\nu}\!\!~_{5\sigma}]&=&0,\hspace{2.9cm} [\bar{\Phi},{\Pi}_{4\nu}\!\!~_{5\sigma}]=0,\\
\,[T_\tau,{\Pi}_{4\nu}\!\!~_{5\sigma}]&=&0,\ \  [\hat{T}_\tau,{\Pi}_{4\nu}\!\!~_{5\sigma}]=\!-\ip\epsilon(\hat{K}_{4\nu}\!\!~_{\tau\sigma}\!+\! g_{\sigma\tau}\hat{W}_{4}^{\ 5}\!\!~_{5\nu}\!-\! g_{\nu\tau}\hat{W}_{4}^{\ 5}\!\!~_{5\sigma}),\ \ \label{TbarPi}
\eeaa
\beaa
\,\!\!\![\hat{P}_\tau,\hat{W}_{4}^{\ 5}\!\!~_{5\mu}]&=&
0,\hspace{2.7cm} [\hat{P}^*_\tau,\hat{W}_{4}^{\ 5}\!\!~_{5\mu}]=
\sqrt{2}g_{\mu\tau}S_{5}^{\ 5},\ \ \ \\
\,[\Phi,\hat{W}_{4}^{\ 5}\!\!~_{5\mu}]&=&0,\hspace{2.8cm} [\bar{\Phi},\hat{W}_{4}^{\ 5}\!\!~_{5\mu}]=-2V_{\mu 5},\\
\,[T_\tau,\hat{W}_{4}^{\ 5}\!\!~_{5\mu}]&=&-2\ip{\Pi}_{4\tau}\!\!~_{5\mu},\hspace{1.5cm}  [\hat{T}_\tau,\hat{W}_{4}^{\ 5}\!\!~_{5\mu}]
\!=\!-2\ip{K}_{45}\!~_{\tau\mu},\ \ \\
\,\!\!\![\hat{P}_\tau,{K}_{45}\!~_{\mu\nu}]&=&
0,\hspace{2.8cm} [\hat{P}^*_\tau,{K}_{45}\!~_{\mu\nu}]=
\!\sqrt{2}g_{\nu\tau}V^\dagger_{5\mu},\ \ \\
\,\!\![\Phi,{K}_{45}\!~_{\mu\nu}]&=&0,\hspace{2.9cm} [\bar{\Phi},{K}_{45}\!~_{\mu\nu}]=-\epsilon S_{\nu\mu},\\
\,\!\![T_\tau,{K}_{45}\!~_{\mu\nu}]&=&\!\!\!-\ip\epsilon(\hat{K}_{4\tau}\!\!~_{\mu\nu}\!-\! g_{\mu\tau}\hat{W}_{4}^{\ 5}\!\!~_{5\nu}\!),\, [\hat{T}_\tau,{K}_{45}\!~_{\mu\nu}]=0.
\eeaa
Clearly, $S$ is a CPE--invariant. The action of naive dimension 1 generators over the dimension 2 quadratic components obviously cancels, and: 
\beaa
\,[T_\tau,S_{\mu\sigma}]&=&2\ip g_{\sigma\tau} V_{\mu 5},\hspace{.8cm} [\hat{T}_\tau,S_{\sigma\mu}]=2\ip g_{\tau\sigma} V^\dagger_{5\mu},\\
\,[T_\tau,S_5^{\ 5}]&=&-2\ip V_{\tau 5},\hspace{.7cm} [\hat{T}_\tau,S_5^{\ 5}]=-2\ip V^\dagger_{5\tau},\\
\,[T_\tau,V_{\mu 5}]&=&0,\hspace{2cm} [\hat{T}_\tau,V_{\mu 5}]=-\ip\epsilon(S_{\mu\tau}-g_{\mu\tau}S_5^{\ 5}).\label{QuartTransEnd}
\eeaa
From the invariance of ${\cal{I}}_4$ under $T_\mu, \hat{T}_\nu$ and the trace and traceless parts of the Jacobi identity $0=[[T_\mu, \hat{T}_\nu],{\cal{I}}_4]$, we can also obtain the invariance under further LE-generators. Notice that the diverse components of the obtained CPE-multiplet are at most spin 2 under Lorentz transformations, giving hope for quantizability. For instance, ${\Pi}_{4\nu}\!\!~_{5\sigma}$ is an anti-symmetric vector. And from (\ref{TbarPi}), the multiplet component $\hat{K}_{4\nu}\!\!~_{\tau\sigma}$ is in $(\frac{1}{2},\frac{1}{2})\otimes((1,0)\oplus(0,1))$ $=(\frac{1}{2},\frac{1}{2})\oplus(\frac{1}{2},\frac{1}{2})\oplus((\frac{3}{2},\frac{1}{2})\oplus(\frac{1}{2},\frac{3}{2}))$ which has at most total spin 2.

\end{document}